\documentclass[10pt]{article}
\usepackage{graphicx}
\usepackage{amsmath}
\usepackage{amssymb}
\usepackage{caption2}
\begin{document}
\bibliographystyle{prsty}
\begin{center}
{\large {\bf \sc{Analysis of  the doubly heavy baryon states and pentaquark states with QCD sum rules }}} \\[2mm]
Zhi-Gang Wang \footnote{E-mail: zgwang@aliyun.com.  }     \\
 Department of Physics, North China Electric Power University, Baoding 071003, P. R. China
\end{center}

\begin{abstract}
In this article, we  study  the doubly heavy baryon states and  pentaquark states with the QCD sum rules by carrying out the operator product expansion   up to   the vacuum condensates of dimension $7$ and $13$ respectively  in a consistent way. In calculations, we separate  the contributions of the negative parity and positive parity hadron  states  unambiguously, and  study the masses and pole residues of the doubly heavy baryon states and  pentaquark states
 in details. The present predictions can be confronted to the experimental data in the future.
\end{abstract}



\section{Introduction}

In 2017, the LHCb collaboration observed the doubly charmed baryon state  $\Xi_{cc}^{++}$ in the $\Lambda_c^+ K^- \pi^+\pi^+$ mass spectrum, and obtained the mass $M_{\Xi_{cc}^{++}}=3621.40 \pm 0.72 \pm 0.27 \pm 0.14\, \rm{MeV}$   \cite{LHCb-Xicc}.
The observation of the $\Xi_{cc}^{++}$ provides the crucial experimental input on the strong correlation between the two charm quarks, which may shed light on the
  spectroscopy of the doubly charmed  baryon states, tetraquark states and pentaquark states. The attractive interaction induced by  one-gluon exchange  favors  formation  of  the diquarks in  color antitriplet \cite{One-gluon},  the favored configurations are the scalar ($C\gamma_5$) and axialvector ($C\gamma_\mu$) diquark states from the QCD sum rules \cite{Dosch-Diquark-1989}. For the heavy-heavy  quark systems  $QQ$,  only the   axialvector  diquarks $\varepsilon^{ijk} Q^{T}_j C\gamma_\mu Q_k$ and tensor diquarks $\varepsilon^{ijk} Q^{T}_j C\sigma_{\mu\nu }Q_k$ survive due to the Fermi-Dirac  statistics,
the axialvector diquarks $\varepsilon^{ijk} Q^{T}_j C\gamma_{\mu} Q_k$ are more stable than the tensor diquarks $\varepsilon^{ijk} Q^{T}_j C\sigma_{\mu\nu} Q_k$,
we can take the axialvector diquarks $\varepsilon^{ijk}  Q^T_jC\gamma_\mu Q_k$ as basic constituents to study doubly heavy  baryon states \cite{Xicc-QCDSR,Xicc-QCDSR-Azizi,WangHHbaryon-4-P,WangH-HHbaryon-6-4-P,WangH-HHbaryon-6-5-N}, tetraquark states \cite{Nielsen-Lee,QQ-QCDSR-Wang2011,QQ-QCDSR-Chen,Wang-QQ-tetraquark} and pentaquark states with the QCD sum rules. The doubly heavy pentaquark states have not been studied with the QCD sum rules. In Ref.\cite{Zhu-QQ-penta},  the mass spectrum of the doubly heavy pentaquark states are studied in  a color-magnetic interaction model.

In 2015,  the  LHCb collaboration studied the $\Lambda_b^0\to J/\psi K^- p$ decays, and performed the amplitude analysis   on all relevant
masses and decay angles of the six-dimensional   data using the helicity formalism
and Breit-Wigner amplitudes to describe all resonances, and observed  two exotic states $P_c(4380)$ and $P_c(4450)$ in the $J/\psi p$ invariant mass distributions    with the significances of more than 9 standard deviations \cite{LHCb-4380}.    The $P_c(4380)$ and $P_c(4450)$ are excellent candidates for the hidden-charm pentaquark states \cite{Maiani1507,di-di-anti-penta,Wang1508-EPJC}. Up to now, no experimental candidates for the doubly charmed or doubly bottom pentaquark states  have been observed.

In Refs.\cite{WangHHbaryon-4-P,WangH-HHbaryon-6-4-P,WangH-HHbaryon-6-5-N,WangHbaryon-6-PN,WangHHHbaryon-4-PN},  we separate the contributions of the positive parity and negative parity baryon states explicitly, and study the    heavy, doubly-heavy and triply-heavy baryon states  with the QCD sum rules in a systematic way, the truncations of the operator product expansion are shown explicitly in Table 1. We carry out the operator product expansion up to the vacuum condensates of dimension 4 for the positive parity doubly heavy baryon states \cite{WangHHbaryon-4-P,WangH-HHbaryon-6-4-P}, another detailed studied including the contributions of the higher dimensional vacuum condensates are still needed. While in Ref.\cite{Xicc-QCDSR}, the contributions of the  positive parity and negative parity doubly heavy baryon states are not separated explicitly.

In Ref.\cite{Wang1508-EPJC}, we   construct    the  diquark-diquark-antiquark type  interpolating currents to  study  the $J^P={\frac{1}{2}}^\pm$, ${\frac{3}{2}}^\pm$,  ${\frac{5}{2}}^\pm$   hidden-charm pentaquark states with the QCD sum rules in a systematic way by taking into account the vacuum condensates up to dimension $10$ in the operator product expansion and separating the contributions of the positive parity and negative parity pentaquark states explicitly. In calculations, we take the energy scale formula $\mu=\sqrt{M^2_{P}-(2{\mathbb{M}}_c)^2}$ with the effective $c$-quark mass ${\mathbb{M}}_c$ to determine the ideal energy scales of the QCD spectral densities. In Ref.\cite{Wang-tri-H-penta}, we construct the diquark-diquark-antiquark type current to study the  ground state triply-charmed  pentaquark states with the QCD sum rules by carrying out the operator product expansion up to the vacuum condensates of dimension $10$, and take the energy scale formula $\mu=\sqrt{M^2_{P}-(3{\mathbb{M}}_c)^2}$ to determine the optimal energy scales of the QCD spectral densities.

In Ref.\cite{Wang-cssss}, we  study  the diquark-diquark-antiquark type charmed pentaquark states with $J^P={\frac{3}{2}}^\pm$
with the QCD sum rules by carrying out the operator product expansion   up to   the vacuum condensates of dimension $13$ in a consistent way to explore the possible assignments of the new excited $\Omega_c$ states as the pentaquark states. The new excited $\Omega_c$ states may also be the P-wave excitations of the ground state $\Omega_c$ \cite{P-wave-Omega-c}.
In Ref.\cite{Wang-Pc-mole}, we  study  the $\bar{D}\Sigma_c^*$,  $\bar{D}^{*}\Sigma_c$  and   $\bar{D}^{*}\Sigma_c^*$  pentaquark molecular states
with the QCD sum rules by carrying out the operator product expansion   up to   the vacuum condensates of dimension $13$ in a consistent way to explore the possible assignments of the $P_c(4380)$ and $P_c(4450)$ as the pentaquark molecular states. In calculations, we observe that the vacuum condensates of dimensions $11$ and $13$ play an important role in obtaining stable QCD sum rules \cite{Wang-cssss,Wang-Pc-mole}.

\begin{table}
\begin{center}
\begin{tabular}{|c|c|c|c|c|}\hline\hline
Baryons (Parity)               &OPE              &References       \\ \hline

$Qqq^\prime\,(\pm)$            &$6$              &\cite{WangH-HHbaryon-6-4-P,WangH-HHbaryon-6-5-N,WangHbaryon-6-PN}   \\  \hline
$QQq\,(+)$                     &$4$              &\cite{WangHHbaryon-4-P,WangH-HHbaryon-6-4-P}  \\  \hline
$QQq\,(-)$                     &$5$              &\cite{WangH-HHbaryon-6-5-N}  \\  \hline
$QQQ^\prime\,(\pm)$            &$4$              &\cite{WangHHHbaryon-4-PN}   \\      \hline \hline
\end{tabular}
\end{center}
\caption{ The truncations in the operator product expansion (OPE) in the QCD sum rules for the heavy, doubly-heavy and triply-heavy baryon states.}
\end{table}

 In this article, we extend our previous works \cite{WangHHbaryon-4-P,WangH-HHbaryon-6-4-P,WangH-HHbaryon-6-5-N,Wang-QQ-tetraquark,Wang1508-EPJC,WangHHHbaryon-4-PN,Wang-tri-H-penta,Wang-cssss,Wang-Pc-mole} to study  the doubly heavy baryon states and  pentaquark states with the QCD sum rules by carrying out the operator product expansion   up to   the vacuum condensates of dimension $7$ and $13$ respectively  in a consistent way.

 The article is arranged as follows:
  we derive the QCD sum rules for the masses and pole residues of the doubly heavy baryon states and pentaquark states in Sect.2;  in Sect.3, we present the numerical results and discussions; and Sect.4 is reserved for our
conclusion.

\section{QCD sum rules for  the doubly heavy baryon states and pentaquark states}

In the following, we write down  the two-point correlation functions $\Pi(p)$, $\Pi_{\mu\nu}(p)$ and $\Pi_{\mu\nu\alpha\beta}(p)$  in the QCD sum rules,
\begin{eqnarray}
\Pi(p)&=&i\int d^4x e^{ip \cdot x} \langle0|T\left\{J(x)\bar{J}(0)\right\}|0\rangle \, , \nonumber\\
\Pi_{\mu\nu}(p)&=&i\int d^4x e^{ip \cdot x} \langle0|T\left\{J_{\mu}(x)\bar{J}_{\nu}(0)\right\}|0\rangle \, , \nonumber\\
\Pi_{\mu\nu\alpha\beta}(p)&=&i\int d^4x e^{ip \cdot x} \langle0|T\left\{J_{\mu\nu}(x)\bar{J}_{\alpha\beta}(0)\right\}|0\rangle \, ,
\end{eqnarray}
where the currents $J(x)=J^{QQq}(x)$, $J^{QQs}(x)$, $J^{QQud\bar{q}}(x)$,  $J_\mu(x)= J_\mu^{QQq}(x)$, $J_\mu^{QQs}(x)$, $J_\mu^{QQud\bar{q}}(x)$, $J_{\mu\nu}(x)=J_{\mu\nu}^{QQud\bar{q}}(x)$,
\begin{eqnarray}
 J^{QQq}(x)&=&   \varepsilon^{ijk}  Q^T_i(x) C\gamma_\mu Q_j(x)\gamma_5\gamma^\mu q_{k}(x) \, ,\nonumber \\
 J^{QQs}(x)&=&   \varepsilon^{ijk}  Q^T_i(x) C\gamma_\mu Q_j(x)\gamma_5\gamma^\mu s_{k}(x) \, ,\nonumber \\
 J_\mu^{QQq}(x)&=&   \varepsilon^{ijk}  Q^T_i(x) C\gamma_\mu Q_j(x)\,  q_{k}(x) \, ,\nonumber \\
 J_\mu^{QQs}(x)&=&   \varepsilon^{ijk}  Q^T_i(x) C\gamma_\mu Q_j(x)\,  s_{k}(x) \, , \nonumber \\
 J^{QQud\bar{q}}(x)&=& \varepsilon^{ila}\varepsilon^{ijk} \varepsilon^{lmn} Q^T_j(x) C\gamma_\mu Q_k(x) u^T_m(x) C\gamma_5 d_n(x) \gamma_5\gamma^\mu C \bar{q}^T_a(x) \, ,\nonumber \\
 J_\mu^{QQud\bar{q}}(x)&=& \varepsilon^{ila}\varepsilon^{ijk} \varepsilon^{lmn} Q^T_j(x) C\gamma_\mu Q_k(x) u^T_m(x) C\gamma_5 d_n(x)   C \bar{q}^T_a(x) \, ,\nonumber \\
  J_{\mu\nu}^{QQud\bar{q}}(x)&=& \varepsilon^{ila}\varepsilon^{ijk} \varepsilon^{lmn} Q^T_j(x) C\gamma_\mu Q_k(x) u^T_m(x) C\gamma_\nu d_n(x)   C \bar{q}^T_a(x) \, ,
\end{eqnarray}
the $i$, $j$, $k$, $l$, $m$, $n$, $a$ are color indices, $Q=b,c$, $q=u,d$. In this article, we take the doubly heavy diquarks $ \varepsilon^{ijk}  Q^T_j C\gamma_\mu Q_k$ as basic constituents to construct  the currents  $J(x)$, $ J_{\mu}(x)$ and $J_{\mu\nu}(x)$ to interpolate the doubly heavy baryon states ($\Xi_{QQ}$,  $\Xi^*_{QQ}$, $\Omega_{QQ}$,  $\Omega^*_{QQ}$) and pentaquark  states with the spin $J={\frac{1}{2}}$, ${\frac{3}{2}}$ and ${\frac{5}{2}}$, respectively.
In Refs.\cite{QQ-QCDSR-Wang2011,Wang-QQ-tetraquark}, we take the doubly heavy diquarks $\varepsilon^{ijk} Q^{T}_jC\gamma_\mu Q_k$ as basic constituents to study the doubly heavy tetraquark states.  Recently, Azizi, Sarac and Sundu studied the meson-baryon type (or the color singlet-singlet type) doubly heavy pentaquark states with a hidden-charm (or hidden-bottom) quark pair using the QCD sum rules \cite{Azizi-Penta}, while in the present work, we study the diquark-diquark-antiquark type doubly heavy pentaquark states with two charmed (or bottom) quarks. In this article, we choose the famous Ioffe currents, for more general currents interpolating the doubly heavy baryon states, one can consult Ref.\cite{Xicc-QCDSR-Azizi}, the simple Ioffe currents have shortcomings,   more experimental data are still needed to select the best parameters in the more general currents.

The  three quark   currents $J(0)$ and $J_{\mu}(0)$ couple potentially to the ${\frac{1}{2}}^+$ and ${\frac{1}{2}}^-$,  ${\frac{3}{2}}^+$ doubly heavy baryon    states $B_{\frac{1}{2}}^{+}$ and $B_{\frac{1}{2}}^{-}$, $B_{\frac{3}{2}}^{+}$, respectively,
\begin{eqnarray}
\langle 0| J (0)|B_{\frac{1}{2}}^{+}(p)\rangle &=&\lambda^{+}_{\frac{1}{2}}\,  U^{+}(p,s) \, , \nonumber \\
\langle 0| J_{\mu} (0)|B_{\frac{1}{2}}^{-}(p)\rangle &=&f^{-}_{\frac{1}{2}}\,p_\mu \,U^{-}(p,s) \, , \nonumber \\
\langle 0| J_{\mu} (0)|B_{\frac{3}{2}}^{+}(p)\rangle &=&\lambda^{+}_{\frac{3}{2}}\, U^{+}_\mu(p,s) \, ,
\end{eqnarray}
the currents $J(0)$ and $J_{\mu}(0)$ also couple potentially to the ${\frac{1}{2}}^-$ and ${\frac{1}{2}}^+$,  ${\frac{3}{2}}^-$ doubly heavy baryon    states $B_{\frac{1}{2}}^{-}$ and $B_{\frac{1}{2}}^{+}$, $B_{\frac{3}{2}}^{-}$, respectively,
\begin{eqnarray}
\langle 0| J (0)|B_{\frac{1}{2}}^{-}(p)\rangle &=&\lambda^{-}_{\frac{1}{2}}\,i\gamma_5  U^{-}(p,s) \, , \nonumber \\
\langle 0| J_{\mu} (0)|B_{\frac{1}{2}}^{+}(p)\rangle &=&f^{+}_{\frac{1}{2}}\,p_\mu i\gamma_5\,U^{+}(p,s) \, , \nonumber \\
\langle 0| J_{\mu} (0)|B_{\frac{3}{2}}^{-}(p)\rangle &=&\lambda^{-}_{\frac{3}{2}}\,i\gamma_5 U^{-}_\mu(p,s) \, ,
\end{eqnarray}
because multiplying $i \gamma_{5}$ to the currents $J(x)$ and $J_{\mu}(x)$ changes their parity \cite{WangHHbaryon-4-P,WangH-HHbaryon-6-4-P,WangH-HHbaryon-6-5-N,WangHbaryon-6-PN,WangHHHbaryon-4-PN,Chung82,Oka96}.
The  five quark   currents $J(0)$, $J_{\mu}(0)$ and $J_{\mu\nu}$ couple potentially to the ${\frac{1}{2}}^-$, ${\frac{1}{2}}^+$,  ${\frac{3}{2}}^-$ and ${\frac{1}{2}}^-$,  ${\frac{3}{2}}^+$, ${\frac{5}{2}}^-$ doubly heavy pentaquark  states $P_{\frac{1}{2}}^{-}$, $P_{\frac{1}{2}}^{+}$, $P_{\frac{3}{2}}^{-}$ and $P_{\frac{1}{2}}^{-}$, $P_{\frac{3}{2}}^{+}$, $P_{\frac{5}{2}}^{-}$, respectively,
\begin{eqnarray}
\langle 0| J (0)|P_{\frac{1}{2}}^{-}(p)\rangle &=&\lambda^{-}_{\frac{1}{2}}\,  U^{-}(p,s) \, , \nonumber \\
\langle 0| J_{\mu} (0)|P_{\frac{1}{2}}^{+}(p)\rangle &=&f^{+}_{\frac{1}{2}}p_\mu\,  U^{+}(p,s) \, , \nonumber \\
\langle 0| J_{\mu} (0)|P_{\frac{3}{2}}^{-}(p)\rangle &=&\lambda^{-}_{\frac{3}{2}}\,  U^{-}_\mu(p,s) \, , \nonumber \\
\langle 0| J_{\mu\nu} (0)|P_{\frac{1}{2}}^{-}(p)\rangle &=&g^{-}_{\frac{1}{2}}p_\mu p_\nu\, U^{-}(p,s) \, , \nonumber\\
\langle 0| J_{\mu\nu} (0)|P_{\frac{3}{2}}^{+}(p)\rangle &=&f^{+}_{\frac{3}{2}}\, \left[p_\mu U^{+}_{\nu}(p,s)+p_\nu U^{+}_{\mu}(p,s)\right] \, , \nonumber\\
\langle 0| J_{\mu\nu} (0)|P_{\frac{5}{2}}^{-}(p)\rangle &=&\sqrt{2}\lambda^{-}_{\frac{5}{2}}\, U^{-}_{\mu\nu}(p,s) \, ,
\end{eqnarray}
the currents $J(0)$, $J_{\mu}(0)$ and $J_{\mu\nu}$ also couple potentially to the ${\frac{1}{2}}^+$, ${\frac{1}{2}}^-$,  ${\frac{3}{2}}^+$ and ${\frac{1}{2}}^+$,  ${\frac{3}{2}}^-$, ${\frac{5}{2}}^+$ doubly heavy pentaquark  states $P_{\frac{1}{2}}^{+}$, $P_{\frac{1}{2}}^{-}$, $P_{\frac{3}{2}}^{+}$ and $P_{\frac{1}{2}}^{+}$, $P_{\frac{3}{2}}^{-}$, $P_{\frac{5}{2}}^{+}$, respectively,
\begin{eqnarray}
\langle 0| J (0)|P_{\frac{1}{2}}^{+}(p)\rangle &=&\lambda^{+}_{\frac{1}{2}}\, i\gamma_5 U^{+}(p,s) \, , \nonumber \\
\langle 0| J_{\mu} (0)|P_{\frac{1}{2}}^{-}(p)\rangle &=&f^{-}_{\frac{1}{2}}p_\mu \, i\gamma_5 U^{-}(p,s) \, , \nonumber \\
\langle 0| J_{\mu} (0)|P_{\frac{3}{2}}^{+}(p)\rangle &=&\lambda^{+}_{\frac{3}{2}}\, i\gamma_5 U^{+}_\mu(p,s) \, , \nonumber \\
\langle 0| J_{\mu\nu} (0)|P_{\frac{1}{2}}^{+}(p)\rangle &=&g^{+}_{\frac{1}{2}}p_\mu p_\nu\,i\gamma_5 U^{+}(p,s) \, , \nonumber\\
\langle 0| J_{\mu\nu} (0)|P_{\frac{3}{2}}^{-}(p)\rangle &=&f^{-}_{\frac{3}{2}} \,i\gamma_5\left[p_\mu U^{-}_{\nu}(p,s)+p_\nu U^{-}_{\mu}(p,s)\right] \, , \nonumber\\
\langle 0| J_{\mu\nu} (0)|P_{\frac{5}{2}}^{+}(p)\rangle &=&\sqrt{2}\lambda^{+}_{\frac{5}{2}}\, i\gamma_5 U^{+}_{\mu\nu}(p,s) \, ,
\end{eqnarray}
because multiplying $i \gamma_{5}$ to the five quark currents $J(x)$, $J_{\mu}(x)$,  $J_{\mu\nu}(x)$ also changes their parity \cite{Wang1508-EPJC}.
The $\lambda^{\pm}_{\frac{1}{2}/\frac{3}{2}/\frac{5}{2}}$, $f^{\pm}_{\frac{1}{2}/\frac{3}{2}}$ and $g^{\pm}_{\frac{1}{2}}$ are the pole residues or the current-hadron coupling constants.
The spinors $U^\pm(p,s)$ satisfy the Dirac equations  $(\not\!\!p-M_{\pm})U^{\pm}(p)=0$, while the spinors $U^{\pm}_\mu(p,s)$ and $U^{\pm}_{\mu\nu}(p,s)$ satisfy the Rarita-Schwinger equations $(\not\!\!p-M_{\pm})U^{\pm}_\mu(p)=0$ and $(\not\!\!p-M_{\pm})U^{\pm}_{\mu\nu}(p)=0$,  and the relations $\gamma^\mu U^{\pm}_\mu(p,s)=0$,
$p^\mu U^{\pm}_\mu(p,s)=0$, $\gamma^\mu U^{\pm}_{\mu\nu}(p,s)=0$,
$p^\mu U^{\pm}_{\mu\nu}(p,s)=0$, $ U^{\pm}_{\mu\nu}(p,s)= U^{\pm}_{\nu\mu}(p,s)$, respectively.

 At the phenomenological side, we  insert  a complete set  of intermediate doubly heavy baryon states or pentaquark   states with the
same quantum numbers as the current operators $J (x)$,
$i\gamma_5 J (x)$, $J_\mu(x)$,
$i\gamma_5 J_\mu(x)$, $J_{\mu\nu}(x)$ and
$i\gamma_5 J_{\mu\nu}(x)$ into the correlation functions $\Pi(p)$,
$\Pi_{\mu\nu}(p)$ and $\Pi_{\mu\nu\alpha\beta}(p)$ to obtain the hadronic representation
\cite{SVZ79,PRT85}. After isolating the pole terms of the lowest
 states, we obtain the complex expressions:
 \begin{eqnarray}
 \Pi(p) & = & {\lambda^{+}_{\frac{1}{2}}}^2  {\!\not\!{p}+ M_{+} \over M_{+}^{2}-p^{2}  } +  {\lambda^{-}_{\frac{1}{2}}}^2  {\!\not\!{p}- M_{-} \over M_{-}^{2}-p^{2}  } +\cdots  \, ,  \\
 \Pi_{\mu\nu}(p) & = & {\lambda^{+}_{\frac{3}{2}}}^2  {\!\not\!{p}+ M_{+} \over M_{+}^{2}-p^{2}  } \left(- g_{\mu\nu}+\frac{\gamma_\mu\gamma_\nu}{3}+\frac{2p_\mu p_\nu}{3p^2}-\frac{p_\mu\gamma_\nu-p_\nu \gamma_\mu}{3\sqrt{p^2}}
\right)\nonumber\\
&&+  {\lambda^{-}_{\frac{3}{2}}}^2  {\!\not\!{p}- M_{-} \over M_{-}^{2}-p^{2}  } \left(- g_{\mu\nu}+\frac{\gamma_\mu\gamma_\nu}{3}+\frac{2p_\mu p_\nu}{3p^2}-\frac{p_\mu\gamma_\nu-p_\nu \gamma_\mu}{3\sqrt{p^2}}
\right)   \nonumber \\
& &+ {f^{-}_{\frac{1}{2}}}^2  {\!\not\!{p}+ M_{-} \over M_{-}^{2}-p^{2}  } p_\mu p_\nu+  {f^{+}_{\frac{1}{2}}}^2  {\!\not\!{p}- M_{+} \over M_{+}^{2}-p^{2}  } p_\mu p_\nu  +\cdots  \, ,
\end{eqnarray}
 for the doubly heavy baryon states, and
 \begin{eqnarray}
 \Pi(p) & = & {\lambda^{-}_{\frac{1}{2}}}^2  {\!\not\!{p}+ M_{-} \over M_{-}^{2}-p^{2}  } +  {\lambda^{+}_{\frac{1}{2}}}^2  {\!\not\!{p}- M_{+} \over M_{+}^{2}-p^{2}  } +\cdots  \, , \\
   \Pi_{\mu\nu}(p) & = & {\lambda^{-}_{\frac{3}{2}}}^2  {\!\not\!{p}+ M_{-} \over M_{-}^{2}-p^{2}  } \left(- g_{\mu\nu}+\frac{\gamma_\mu\gamma_\nu}{3}+\frac{2p_\mu p_\nu}{3p^2}-\frac{p_\mu\gamma_\nu-p_\nu \gamma_\mu}{3\sqrt{p^2}}
\right)\nonumber\\
&&+  {\lambda^{+}_{\frac{3}{2}}}^2  {\!\not\!{p}- M_{+} \over M_{+}^{2}-p^{2}  } \left(- g_{\mu\nu}+\frac{\gamma_\mu\gamma_\nu}{3}+\frac{2p_\mu p_\nu}{3p^2}-\frac{p_\mu\gamma_\nu-p_\nu \gamma_\mu}{3\sqrt{p^2}}
\right)   \nonumber \\
& &+ {f^{+}_{\frac{1}{2}}}^2  {\!\not\!{p}+ M_{+} \over M_{+}^{2}-p^{2}  } p_\mu p_\nu+  {f^{-}_{\frac{1}{2}}}^2  {\!\not\!{p}- M_{-} \over M_{-}^{2}-p^{2}  } p_\mu p_\nu  +\cdots  \, , \\
\Pi_{\mu\nu\alpha\beta}(p) & = & 2{\lambda^{-}_{\frac{5}{2}}}^2  {\!\not\!{p}+ M_{-} \over M_{-}^{2}-p^{2}  } \left[\frac{ \widetilde{g}_{\mu\alpha}\widetilde{g}_{\nu\beta}+\widetilde{g}_{\mu\beta}\widetilde{g}_{\nu\alpha}}{2}-\frac{\widetilde{g}_{\mu\nu}\widetilde{g}_{\alpha\beta}}{5}\right.\nonumber\\
&&-\frac{1}{10}\left( \gamma_{\mu}\gamma_{\alpha}+\frac{\gamma_{\mu}p_{\alpha}-\gamma_{\alpha}p_{\mu}}{\sqrt{p^2}}-\frac{p_{\mu}p_{\alpha}}{p^2}\right)\widetilde{g}_{\nu\beta}\nonumber\\
&&\left.-\frac{1}{10}\left( \gamma_{\nu}\gamma_{\alpha}+\frac{\gamma_{\nu}p_{\alpha}-\gamma_{\alpha}p_{\nu}}{\sqrt{p^2}}-\frac{p_{\nu}p_{\alpha}}{p^2}\right)\widetilde{g}_{\mu\beta}
+\cdots\right]\nonumber\\
&&+2 {\lambda^{+}_{\frac{5}{2}}}^2  {\!\not\!{p}- M_{+} \over M_{+}^{2}-p^{2}  } \left[\frac{ \widetilde{g}_{\mu\alpha}\widetilde{g}_{\nu\beta}+\widetilde{g}_{\mu\beta}\widetilde{g}_{\nu\alpha}}{2}
-\frac{\widetilde{g}_{\mu\nu}\widetilde{g}_{\alpha\beta}}{5}\right.\nonumber\\
&&-\frac{1}{10}\left( \gamma_{\mu}\gamma_{\alpha}+\frac{\gamma_{\mu}p_{\alpha}-\gamma_{\alpha}p_{\mu}}{\sqrt{p^2}}-\frac{p_{\mu}p_{\alpha}}{p^2}\right)\widetilde{g}_{\nu\beta}\nonumber\\
&&\left.
-\frac{1}{10}\left( \gamma_{\nu}\gamma_{\alpha}+\frac{\gamma_{\nu}p_{\alpha}-\gamma_{\alpha}p_{\nu}}{\sqrt{p^2}}-\frac{p_{\nu}p_{\alpha}}{p^2}\right)\widetilde{g}_{\mu\beta}
 +\cdots\right]   \nonumber\\
 && +{f^{+}_{\frac{3}{2}}}^2  {\!\not\!{p}+ M_{+} \over M_{+}^{2}-p^{2}  } \left[ p_\mu p_\alpha \left(- g_{\nu\beta}+\frac{\gamma_\nu\gamma_\beta}{3}+\frac{2p_\nu p_\beta}{3p^2}-\frac{p_\nu\gamma_\beta-p_\beta \gamma_\nu}{3\sqrt{p^2}}
\right)+\cdots \right]\nonumber\\
&&+  {f^{-}_{\frac{3}{2}}}^2  {\!\not\!{p}- M_{-} \over M_{-}^{2}-p^{2}  } \left[ p_\mu p_\alpha \left(- g_{\nu\beta}+\frac{\gamma_\nu\gamma_\beta}{3}+\frac{2p_\nu p_\beta}{3p^2}-\frac{p_\nu\gamma_\beta-p_\beta \gamma_\nu}{3\sqrt{p^2}}
\right)+\cdots \right]   \nonumber \\
& &+ {g^{-}_{\frac{1}{2}}}^2  {\!\not\!{p}+ M_{-} \over M_{-}^{2}-p^{2}  } p_\mu p_\nu p_\alpha p_\beta+  {g^{+}_{\frac{1}{2}}}^2  {\!\not\!{p}- M_{+} \over M_{+}^{2}-p^{2}  } p_\mu p_\nu p_\alpha p_\beta  +\cdots \, ,
\end{eqnarray}
for the doubly heavy pentaquark states,
where $\widetilde{g}_{\mu\nu}=g_{\mu\nu}-\frac{p_{\mu}p_{\nu}}{p^2}$.

We can rewrite the correlation functions $\Pi_{\mu\nu}(p)$ and $\Pi_{\mu\nu\alpha\beta}(p)$ into the following form according to Lorentz covariance,
\begin{eqnarray}
\Pi_{\mu\nu}(p)&=&\Pi_{\frac{3}{2}}(p^2)\,\left(- g_{\mu\nu}\right)+\Pi_{\frac{3}{2}}^1(p^2)\,\gamma_\mu \gamma_\nu+\Pi_{\frac{3}{2}}^2(p^2)\,\left(p_\mu\gamma_\nu-p_\nu \gamma_\mu\right) +\Pi_{\frac{1}{2},\frac{3}{2}}(p^2)\, p_\mu p_\nu\, , \\
\Pi_{\mu\nu\alpha\beta}(p)&=&\Pi_{\frac{5}{2}}(p^2)\,\left(g_{\mu\alpha}g_{\nu\beta}+g_{\mu\beta}g_{\nu\alpha} \right)+\Pi_{\frac{5}{2}}^1(p^2)\, g_{\mu\nu}g_{\alpha\beta}+\Pi_{\frac{5}{2}}^2(p^2)\, \left(g_{\mu\nu}p_{\alpha}p_{\beta}+g_{\alpha\beta}p_{\mu}p_{\nu}\right) \nonumber\\
&&+\Pi_{\frac{5}{2}}^3(p^2)\,\left(  g_{\mu \alpha} \gamma_\nu \gamma_\beta+ g_{\mu \beta} \gamma_\nu \gamma_\alpha+ g_{\nu \alpha} \gamma_\mu \gamma_\beta+ g_{\nu \beta} \gamma_\mu \gamma_\alpha \right) \nonumber\\
&&+\Pi_{\frac{5}{2}}^4(p^2)\,\left[  g_{\nu \beta}\left(\gamma_{\mu} p_{\alpha}- \gamma_{\alpha}p_{\mu}\right) +
g_{\nu \alpha}\left(\gamma_{\mu}p_{ \beta}-\gamma_{ \beta}p_{\mu}\right) + g_{\mu \beta}\left(\gamma_{\nu} p_{\alpha}- \gamma_{\alpha}p_{\nu}\right)\right.\nonumber\\
&&\left.+ g_{\mu \alpha} \left(\gamma_{\nu}p_{ \beta}-\gamma_{ \beta}p_{\nu} \right)\right]\nonumber\\
&&+\Pi_{\frac{3}{2},\frac{5}{2}}^1(p^2)\,\left(  g_{\mu \alpha} p_\nu p_\beta+ g_{\mu \beta} p_\nu p_\alpha+ g_{\nu \alpha} p_\mu p_\beta+ g_{\nu \beta} p_\mu p_\alpha \right) \nonumber\\
&&+\Pi_{\frac{3}{2},\frac{5}{2}}^2(p^2)\,\left(  \gamma_{\mu} \gamma_{\alpha} p_\nu p_\beta+ \gamma_{\mu}\gamma_{ \beta} p_\nu p_\alpha+ \gamma_{\nu} \gamma_{\alpha} p_\mu p_\beta+ \gamma_{\nu}\gamma_{ \beta} p_\mu p_\alpha \right) \nonumber\\
&&+\Pi_{\frac{3}{2},\frac{5}{2}}^3(p^2)\,\left[  \left(\gamma_{\mu} p_{\alpha}- \gamma_{\alpha}p_{\mu}\right) p_\nu p_\beta+
\left(\gamma_{\mu}p_{ \beta}-\gamma_{ \beta}p_{\mu}\right) p_\nu p_\alpha+ \left(\gamma_{\nu} p_{\alpha}- \gamma_{\alpha}p_{\nu}\right) p_\mu p_\beta\right.\nonumber\\
&&\left.+ \left(\gamma_{\nu}p_{ \beta}-\gamma_{ \beta}p_{\nu} \right)p_\mu p_\alpha \right] +\Pi_{\frac{1}{2},\frac{3}{2},\frac{5}{2}}(p^2)\,p_\mu p_\nu p_\alpha p_\beta \, ,
\end{eqnarray}
 the subscripts $\frac{1}{2}$, $\frac{3}{2}$ and $\frac{5}{2}$ in the components $\Pi_{\frac{3}{2}}(p^2)$, $\Pi_{\frac{3}{2}}^1(p^2)$, $\Pi_{\frac{3}{2}}^2(p^2)$,  $\Pi_{\frac{1}{2},\frac{3}{2}}(p^2)$, $\Pi_{\frac{5}{2}}(p^2)$, $\Pi_{\frac{5}{2}}^1(p^2)$, $\Pi_{\frac{5}{2}}^2(p^2)$, $\Pi_{\frac{5}{2}}^3(p^2)$, $\Pi_{\frac{5}{2}}^4(p^2)$, $\Pi_{\frac{3}{2},\frac{5}{2}}^1(p^2)$, $\Pi_{\frac{3}{2},\frac{5}{2}}^2(p^2)$, $\Pi_{\frac{3}{2},\frac{5}{2}}^3(p^2)$ and
$\Pi_{\frac{1}{2},\frac{3}{2},\frac{5}{2}}(p^2)$ denote the spins the pentaquark states, which means that  the pentaquark states with $J=\frac{1}{2}$, $\frac{3}{2}$ and $\frac{5}{2}$ have contributions. The components $\Pi_{\frac{1}{2},\frac{3}{2}}(p^2)$, $\Pi_{\frac{3}{2},\frac{5}{2}}^1(p^2)$, $\Pi_{\frac{3}{2},\frac{5}{2}}^2(p^2)$, $\Pi_{\frac{3}{2},\frac{5}{2}}^3(p^2)$ and $\Pi_{\frac{1}{2},\frac{3}{2},\frac{5}{2}}(p^2)$ receive contributions from more than one pentaquark state, so they can be neglected. We can rewrite $\gamma_\mu \gamma_\nu=g_{\mu\nu}-i\sigma_{\mu\nu}$, then the components $\Pi_{\frac{3}{2}}^1(p^2)$, $\Pi_{\frac{3}{2}}^2(p^2)$, $\Pi_{\frac{5}{2}}^3(p^2)$ and $\Pi_{\frac{5}{2}}^4(p^2)$ are associated with tensor structures which are antisymmetric in the Lorentz indexes  $\mu$, $\nu$, $\alpha$ or $\beta$. In calculations, we observe that such antisymmetric properties lead to smaller intervals of dimensions of the vacuum condensates, therefore worse QCD sum rules, so the components $\Pi_{\frac{3}{2}}^1(p^2)$, $\Pi_{\frac{3}{2}}^2(p^2)$, $\Pi_{\frac{5}{2}}^3(p^2)$ and $\Pi_{\frac{5}{2}}^4(p^2)$ can also be  neglected. If we take the replacement $J_{\mu\nu}(x)\to \widehat{J}_{\mu\nu}(x)=J_{\mu\nu}(x)-\frac{1}{4}g_{\mu\nu}J_\alpha{}^\alpha(x)$ to subtract the contributions of the $J=\frac{1}{2}$ pentaquark states, a lots of terms   $\propto g_{\mu\nu}$, $g_{\alpha\beta}$  disappear at the QCD side, and result in smaller intervals of dimensions of the vacuum condensates, so the components  $\Pi_{\frac{5}{2}}^1(p^2)$ and $\Pi_{\frac{5}{2}}^2(p^2)$ are not the optimal choices  to study the $J=\frac{5}{2}$ pentaquark states. Now only the components  $\Pi_{\frac{3}{2}}(p^2)$ and $\Pi_{\frac{5}{2}}(p^2)$ are left. We can obtain definite conclusion by studying the  QCD sum rules based on the components   $\Pi_{\frac{3}{2}}^1(p^2)$, $\Pi_{\frac{3}{2}}^2(p^2)$,   $\Pi_{\frac{5}{2}}^1(p^2)$, $\Pi_{\frac{5}{2}}^2(p^2)$, $\Pi_{\frac{5}{2}}^3(p^2)$ and $\Pi_{\frac{5}{2}}^4(p^2)$, this may be our next work.

In this article, we choose the tensor structures $1$, $g_{\mu\nu}$ and $g_{\mu\alpha}g_{\nu\beta}+g_{\mu\beta}g_{\nu\alpha}$ for the correlation functions $\Pi(p)$, $\Pi_{\mu\nu}(p)$ and $\Pi_{\mu\nu\alpha\beta}(p)$ respectively to study the $J^P={\frac{1}{2}}^+$, ${\frac{3}{2}}^+$ doubly heavy baryon states and the $J^P={\frac{1}{2}}^-$, ${\frac{3}{2}}^-$ and ${\frac{5}{2}}^-$ doubly heavy pentaquark   states to avoid contaminations,
\begin{eqnarray}
\Pi(p)&=&\Pi_{\frac{1}{2}}(p^2)+\cdots\, , \nonumber\\
\Pi_{\mu\nu}(p)&=&\Pi_{\frac{3}{2}}(p^2)\,\left(- g_{\mu\nu}\right)+\cdots\, , \nonumber\\
\Pi_{\mu\nu\alpha\beta}(p)&=&\Pi_{\frac{5}{2}}(p^2)\,\left( g_{\mu\alpha}g_{\nu\beta}+g_{\mu\beta}g_{\nu\alpha}\right)+ \cdots \, .
\end{eqnarray}

Now we obtain the hadron spectral densities at phenomenological side through the dispersion relation,
\begin{eqnarray}
\frac{{\rm Im}\Pi_{j}(s)}{\pi}&=&\!\not\!{p} \left[{\lambda^{+}_{j}}^2 \delta\left(s-M_{+}^2\right)+{\lambda^{-}_{j}}^2 \delta\left(s-M_{-}^2\right)\right] +M_{+}{\lambda^{+}_{j}}^2 \delta\left(s-M_{+}^2\right)-M_{-}{\lambda^{-}_{j}}^2 \delta\left(s-M_{-}^2\right)\, , \nonumber\\
&=&\!\not\!{p}\, \rho^1_{j,H}(s)+\rho^0_{j,H}(s) \, ,
\end{eqnarray}
where $j=\frac{1}{2}$, $\frac{3}{2}$ for the doubly heavy baryon states, and
\begin{eqnarray}
\frac{{\rm Im}\Pi_{j}(s)}{\pi}&=&\!\not\!{p} \left[{\lambda^{-}_{j}}^2 \delta\left(s-M_{-}^2\right)+{\lambda^{+}_{j}}^2 \delta\left(s-M_{+}^2\right)\right] +M_{-}{\lambda^{-}_{j}}^2 \delta\left(s-M_{-}^2\right)-M_{+}{\lambda^{+}_{j}}^2 \delta\left(s-M_{+}^2\right)\, , \nonumber\\
&=&\!\not\!{p}\, \rho^1_{j,H}(s)+\rho^0_{j,H}(s) \, ,
\end{eqnarray}
where $j=\frac{1}{2}$, $\frac{3}{2}$, $\frac{5}{2}$ for the doubly heavy pentaquark states,
we introduce  the subscript $H$ to denote  the hadron side. Then we introduce the weight function $\exp\left(-\frac{s}{T^2}\right)$ to obtain the QCD sum rules at the phenomenological side (or the hadron side),
\begin{eqnarray}
\int_{4m_Q^2}^{s_0}ds \left[\sqrt{s}\rho^1_{j,H}(s)+\rho^0_{j,H}(s)\right]\exp\left( -\frac{s}{T^2}\right)
&=&2M_{+}{\lambda^{+}_{j}}^2\exp\left( -\frac{M_{+}^2}{T^2}\right) \, ,
\end{eqnarray}
with $j=\frac{1}{2}$, $\frac{3}{2}$ for the doubly heavy baryon states,
\begin{eqnarray}
\int_{4m_Q^2}^{s_0}ds \left[\sqrt{s}\rho^1_{j,H}(s)+\rho^0_{j,H}(s)\right]\exp\left( -\frac{s}{T^2}\right)
&=&2M_{-}{\lambda^{-}_{j}}^2\exp\left( -\frac{M_{-}^2}{T^2}\right) \, ,
\end{eqnarray}
with $j=\frac{1}{2}$, $\frac{3}{2}$, $\frac{5}{2}$ for the doubly heavy pentaquark states,
where the $s_0$ are the continuum threshold parameters and the $T^2$ are the Borel parameters.
We separate the  contributions  of the negative parity hadron states from that of the positive parity hadron  states unambiguously. In Eqs.(17-18), we choose the special combinations  introduced in Ref.\cite{Wang1508-EPJC}  to obtain the QCD sum rules, which differ from the non-covariant approach in Refs.\cite{WangHHbaryon-4-P,WangH-HHbaryon-6-4-P,WangH-HHbaryon-6-5-N,WangHbaryon-6-PN,WangHHHbaryon-4-PN,Oka96}.

We carry out  the operator product expansion for the correlation functions $\Pi(p)$, $\Pi_{\mu\nu}(p)$ and $\Pi_{\mu\nu\alpha\beta}(p)$ up to the vacuum condensates of dimension $7$ for the doubly heavy baryon states and dimension $13$ for the doubly heavy pentaquark states, and assume vacuum saturation for the
 higher dimensional vacuum condensates.  In calculations, we take the full light quark and heavy quark propagators,
 \begin{eqnarray}
S^{ij}(x)&=& \frac{i\delta_{ij}\!\not\!{x}}{ 2\pi^2x^4}-\frac{\delta_{ij}\langle
\bar{q}q\rangle}{12} -\frac{\delta_{ij}x^2\langle \bar{q}g_s\sigma Gq\rangle}{192} -\frac{ig_sG^{a}_{\alpha\beta}t^a_{ij}(\!\not\!{x}
\sigma^{\alpha\beta}+\sigma^{\alpha\beta} \!\not\!{x})}{32\pi^2x^2} \nonumber\\
&&  -\frac{1}{8}\langle\bar{q}_j\sigma^{\mu\nu}q_i \rangle \sigma_{\mu\nu}+\cdots \, ,
\end{eqnarray}
\begin{eqnarray}
S^{ij}_s(x)&=& \frac{i\delta_{ij}\!\not\!{x}}{ 2\pi^2x^4}
-\frac{\delta_{ij}m_s}{4\pi^2x^2}-\frac{\delta_{ij}\langle
\bar{s}s\rangle}{12} +\frac{i\delta_{ij}\!\not\!{x}m_s
\langle\bar{s}s\rangle}{48}-\frac{\delta_{ij}x^2\langle \bar{s}g_s\sigma Gs\rangle}{192}+\frac{i\delta_{ij}x^2\!\not\!{x} m_s\langle \bar{s}g_s\sigma
 Gs\rangle }{1152}\nonumber\\
&& -\frac{ig_s G^{a}_{\alpha\beta}t^a_{ij}(\!\not\!{x}
\sigma^{\alpha\beta}+\sigma^{\alpha\beta} \!\not\!{x})}{32\pi^2x^2}  -\frac{\delta_{ij}x^4\langle \bar{s}s \rangle\langle g_s^2 GG\rangle}{27648}-\frac{1}{8}\langle\bar{s}_j\sigma^{\mu\nu}s_i \rangle \sigma_{\mu\nu}   +\cdots \, ,
\end{eqnarray}
\begin{eqnarray}
S_Q^{ij}(x)&=&\frac{i}{(2\pi)^4}\int d^4k e^{-ik \cdot x} \left\{
\frac{\delta_{ij}}{\!\not\!{k}-m_Q}
-\frac{g_sG^n_{\alpha\beta}t^n_{ij}}{4}\frac{\sigma^{\alpha\beta}(\!\not\!{k}+m_Q)+(\!\not\!{k}+m_Q)
\sigma^{\alpha\beta}}{(k^2-m_Q^2)^2}\right.\nonumber\\
&&\left. -\frac{g_s^2 (t^at^b)_{ij} G^a_{\alpha\beta}G^b_{\mu\nu}(f^{\alpha\beta\mu\nu}+f^{\alpha\mu\beta\nu}+f^{\alpha\mu\nu\beta}) }{4(k^2-m_Q^2)^5}+\cdots\right\} \, ,\nonumber\\
f^{\alpha\beta\mu\nu}&=&(\!\not\!{k}+m_Q)\gamma^\alpha(\!\not\!{k}+m_Q)\gamma^\beta(\!\not\!{k}+m_Q)\gamma^\mu(\!\not\!{k}+m_Q)\gamma^\nu(\!\not\!{k}+m_Q)\, ,
\end{eqnarray}
and  $t^n=\frac{\lambda^n}{2}$, the $\lambda^n$ is the Gell-Mann matrix   \cite{PRT85,Pascual-1984}.
In Eqs.(19-20), we retain the term $\langle\bar{q}_j\sigma_{\mu\nu}q_i \rangle$ ($\langle\bar{s}_j\sigma_{\mu\nu}s_i \rangle$ )  comes from the Fierz re-arrangement of the $\langle q_i \bar{q}_j\rangle$ ($\langle s_i \bar{s}_j\rangle$) to  absorb the gluons  emitted from other  quark lines to extract the mixed condensate  $\langle\bar{q}g_s\sigma G q\rangle$ ($\langle\bar{s}g_s\sigma G s\rangle$).

For the  correlation functions $\Pi(p)$
and $\Pi_{\mu\nu}(p)$ of the doubly heavy three-quark currents,  there are two heavy quark propagators and a light quark propagator, if each heavy quark line emits a gluon and each light quark line contributes  a quark pair, we obtain a operator $GG\bar{q}q$ (or $GG\bar{s}s$), which is of dimension $7$, for example,
\begin{eqnarray}
\Pi(p)&=&-2i\, \varepsilon^{ijk}  \varepsilon^{i^{\prime}j^{\prime}k^{\prime}}
 \int d^4x e^{ip\cdot x}\, \gamma_5 \gamma^{\mu}  S^{ii^{\prime}}(x) \gamma^\nu\gamma_5 \,  {\rm Tr}\left[\gamma_\mu S_Q^{kk^\prime}(x) \gamma_\nu C S_Q^{Tjj^\prime}(x)C\right] \nonumber\\
&\sim & \langle G^c_{\rho\sigma}G^b_{\alpha\beta}\rangle \langle \bar{q}{q}\rangle \, ,
\end{eqnarray}
with the simple replacements  $S^{ii^\prime}(x)\to -\frac{1}{12}\delta^{ii^\prime}\langle\bar{q}q\rangle$,  $ S_Q^{kk^\prime}(x) \to G^c_{\rho\sigma} t^c_{kk^\prime}$, $ S_Q^{Tjj^\prime}(x) \to G^b_{\alpha\beta} t^b_{jj^\prime}$,
we should take into account the vacuum condensates at least up to dimension 7.
For the  correlation functions $\Pi(p)$,
$\Pi_{\mu\nu}(p)$ and $\Pi_{\mu\nu\alpha\beta}(p)$ of the doubly heavy five-quark currents,  there are two heavy quark propagators and three light quark propagators, if each heavy quark line emits a gluon and each light quark line contributes  a quark pair, we obtain a operator $GG\bar{u}u\bar{d}d\bar{q}q$, which is of dimension 13, for example,
\begin{eqnarray}
\Pi(p)&=&-2i\,\varepsilon^{ila}\varepsilon^{ijk}\varepsilon^{lmn}\varepsilon^{i^{\prime}l^{\prime}a^{\prime}}\varepsilon^{i^{\prime}j^{\prime}k^{\prime}}
\varepsilon^{l^{\prime}m^{\prime}n^{\prime}}\int d^4x e^{ip\cdot x}\, \gamma_5 \gamma^{\mu}C S^{Ta^{\prime}a}(-x)C\gamma^\nu\gamma_5   \nonumber\\
&&{\rm  Tr} \left[\gamma_\mu S_Q^{kk^\prime}(x) \gamma_\nu C S_Q^{Tjj^\prime}(x)C\right] \,{\rm Tr}\left[\gamma_5 S^{nn^\prime}(x) \gamma_5 C S^{Tmm^\prime}(x)C\right] \nonumber\\
&\sim & \langle G^c_{\rho\sigma}G^b_{\alpha\beta}\rangle \langle \bar{q}{q}\rangle \langle \bar{q}{q}\rangle \langle \bar{q}{q}\rangle\, ,
\end{eqnarray}
 with the simple replacements $S^{Ta^{\prime}a}(-x)\to -\frac{1}{12}\delta^{aa^\prime}\langle\bar{q}q\rangle$, $S^{nn^\prime}(x)\to -\frac{1}{12}\delta^{nn^\prime}\langle\bar{q}q\rangle$,  $S^{Tmm^\prime}(x)\to -\frac{1}{12}\delta^{mm^\prime}\langle\bar{q}q\rangle$, $ S_Q^{kk^\prime}(x) \to G^c_{\rho\sigma} t^c_{kk^\prime}$, $ S_Q^{Tjj^\prime}(x) \to G^b_{\alpha\beta} t^b_{jj^\prime}$,  we should take into account the vacuum condensates at least up to dimension 13.  We can carry out the operator product expansion by taking into account the vacuum condensates beyond  dimension 7 or 13, however, it is a very difficult work.

  The higher dimensional vacuum condensates   play an important role in determining the Borel windows, as  there appear terms of the orders $\mathcal{O}\left(\frac{1}{T^2}\right)$, $\mathcal{O}\left(\frac{1}{T^4}\right)$, $\mathcal{O}\left(\frac{1}{T^6}\right)$ in the QCD spectral densities, which  manifest themselves at small values of the Borel parameter $T^2$, we have to choose large values of the $T^2$ to warrant convergence of the operator product expansion and appearance of the Borel platforms.  In this article, we take the truncations $n\leq 7(13)$ and $k\leq 1$ in a consistent way,
the operators of the orders $\mathcal{O}( \alpha_s^{k})$ with $k> 1$ are  discarded. For technical details in calculations, one can consult Refs.\cite{Wang-tetraquark-QCDSR,Wang-molecule-QCDSR}.

 Once the analytical QCD spectral densities  are obtained,  we can take the
quark-hadron duality below the continuum thresholds  $s_0$ and introduce the weight function $\exp\left(-\frac{s}{T^2}\right)$ to obtain  the following QCD sum rules:
\begin{eqnarray}
2M_{+}{\lambda^{+}_{j}}^2\exp\left( -\frac{M_{+}^2}{T^2}\right)
&=& \int_{4m_Q^2}^{s_0}ds \left[\sqrt{s}\rho^1_{j,QCD}(s)+\rho^0_{j,QCD}(s)\right]\exp\left( -\frac{s}{T^2}\right)\, ,\\
2M_{-}{\lambda^{-}_{j}}^2\exp\left( -\frac{M_{-}^2}{T^2}\right) &=&\int_{4m_Q^2}^{s_0}ds \left[\sqrt{s}\rho^1_{j,QCD}(s)+\rho^0_{j,QCD}(s)\right]\exp\left( -\frac{s}{T^2}\right)\, ,
\end{eqnarray}
where
\begin{eqnarray}
\rho^{1/0}_{j,QCD}(s)&=&\rho_0^{1/0}(s)+\rho_3^{1/0}(s)+\rho_4^{1/0}(s)+\rho_5^{1/0}(s)+\rho_7^{1/0}(s)\, ,\nonumber
\end{eqnarray}
with $j=\frac{1}{2}$, $\frac{3}{2}$ for the doubly heavy baryon states,
\begin{eqnarray}
\rho^{1/0}_{j,QCD}(s)&=&\rho_0^{1/0}(s)+\rho_3^{1/0}(s)+\rho_4^{1/0}(s)+\rho_5^{1/0}(s)+\rho_6^{1/0}(s)+\rho_7^{1/0}(s)+\rho_8^{1/0}(s)+\rho_9^{1/0}(s) \nonumber\\
&&+\rho_{10}^{1/0}(s)+\rho_{11}^{1/0}(s)+\rho_{13}^{1/0}(s)\, ,\nonumber
\end{eqnarray}
with $j=\frac{1}{2}$, $\frac{3}{2}$, $\frac{5}{2}$ for the doubly heavy pentaquark states,
the explicit expressions of the  QCD spectral densities are given  in the appendix.

We derive   Eqs.(24-25) with respect to  $\tau=\frac{1}{T^2}$, then eliminate the
 pole residues $\lambda^{\pm}_{j}$ and obtain the QCD sum rules for
 the masses of the doubly heavy baryon states and  pentaquark    states,
 \begin{eqnarray}
 M^2_{\pm} &=& \frac{-\frac{d}{d \tau}\int_{4m_Q^2}^{s_0}ds \,\left[\sqrt{s}\,\rho^1_{j,QCD}(s)+\,\rho^0_{j,QCD}(s)\right]\exp\left(- \tau s\right)}{\int_{4m_Q^2}^{s_0}ds \left[\sqrt{s}\,\rho_{j,QCD}^1(s)+\,\rho_{j,QCD}(s)\right]\exp\left( -\tau s\right)}\, .
\end{eqnarray}

\section{Numerical results and discussions}
We take  the standard values of the vacuum condensates $\langle
\bar{q}q \rangle=-(0.24\pm 0.01\, \rm{GeV})^3$,   $\langle
\bar{q}g_s\sigma G q \rangle=m_0^2\langle \bar{q}q \rangle$,
$m_0^2=(0.8 \pm 0.1)\,\rm{GeV}^2$, $\langle\bar{s}s \rangle=(0.8\pm0.1)\langle\bar{q}q \rangle$, $\langle\bar{s}g_s\sigma G s \rangle=m_0^2\langle \bar{s}s \rangle$,  $\langle \frac{\alpha_s
GG}{\pi}\rangle=(0.33\,\rm{GeV})^4 $    at the energy scale  $\mu=1\, \rm{GeV}$
\cite{SVZ79,PRT85,ColangeloReview}, and choose the $\overline{MS}$ masses $m_{c}(m_c)=(1.275\pm0.025)\,\rm{GeV}$, $m_{b}(m_b)=(4.18\pm0.03)\,\rm{GeV}$, $m_s(\mu=2\,\rm{GeV})=0.095^{+0.009}_{-0.003}\,\rm{GeV}$ from the Particle Data Group \cite{PDG}, and set $m_u=m_d=0$.
Furthermore, we take into account the energy-scale dependence of  the input parameters,
\begin{eqnarray}
\langle\bar{q}q \rangle(\mu)&=&\langle\bar{q}q \rangle(Q)\left[\frac{\alpha_{s}(Q)}{\alpha_{s}(\mu)}\right]^{\frac{12}{33-2n_f}}\, , \nonumber\\
 \langle\bar{s}s \rangle(\mu)&=&\langle\bar{s}s \rangle(Q)\left[\frac{\alpha_{s}(Q)}{\alpha_{s}(\mu)}\right]^{\frac{12}{33-2n_f}}\, , \nonumber\\
 \langle\bar{q}g_s \sigma Gq \rangle(\mu)&=&\langle\bar{q}g_s \sigma Gq \rangle(Q)\left[\frac{\alpha_{s}(Q)}{\alpha_{s}(\mu)}\right]^{\frac{2}{33-2n_f}}\, , \nonumber\\ \langle\bar{s}g_s \sigma Gs \rangle(\mu)&=&\langle\bar{s}g_s \sigma Gs \rangle(Q)\left[\frac{\alpha_{s}(Q)}{\alpha_{s}(\mu)}\right]^{\frac{2}{33-2n_f}}\, , \nonumber\\
m_c(\mu)&=&m_c(m_c)\left[\frac{\alpha_{s}(\mu)}{\alpha_{s}(m_c)}\right]^{\frac{12}{33-2n_f}} \, ,\nonumber\\
m_b(\mu)&=&m_b(m_b)\left[\frac{\alpha_{s}(\mu)}{\alpha_{s}(m_b)}\right]^{\frac{12}{33-2n_f}} \, ,\nonumber\\
m_s(\mu)&=&m_s({\rm 2GeV} )\left[\frac{\alpha_{s}(\mu)}{\alpha_{s}({\rm 2GeV})}\right]^{\frac{12}{33-2n_f}} \, ,\nonumber\\
\alpha_s(\mu)&=&\frac{1}{b_0t}\left[1-\frac{b_1}{b_0^2}\frac{\log t}{t} +\frac{b_1^2(\log^2{t}-\log{t}-1)+b_0b_2}{b_0^4t^2}\right]\, ,
\end{eqnarray}
   where $t=\log \frac{\mu^2}{\Lambda^2}$, $b_0=\frac{33-2n_f}{12\pi}$, $b_1=\frac{153-19n_f}{24\pi^2}$, $b_2=\frac{2857-\frac{5033}{9}n_f+\frac{325}{27}n_f^2}{128\pi^3}$,  $\Lambda=210\,\rm{MeV}$, $292\,\rm{MeV}$  and  $332\,\rm{MeV}$ for the flavors  $n_f=5$, $4$ and $3$, respectively  \cite{PDG,Narison-mix}, and evolve all the input parameters to the optimal energy scales   $\mu$ to extract the masses of the
   doubly heavy baryon states and pentaquark states.

In the article, we study the doubly heavy baryon states and pentaquark states, the two heavy quarks form a diquark state $\varepsilon^{ijk} Q^{T}_jC\gamma_\mu Q_k$,
 which  serves as a static well potential and combines with a light quark state in color triplet to form a compact baryon state or combine with a light diquark  and a light antiquark in  color antitriplet to form  a compact pentaquark state.
While in the hidden-charm or hidden-bottom pentaquark states, the heavy quark $Q$ serves as a static well potential and  combines with the light quark  to form a heavy diquark  in  color antitriplet,
 the heavy antiquark $\bar{Q}$ serves  as another static well potential and combines with the light diquark in color antitriplet  to form a heavy triquark in  color triplet, then the heavy diquark and heavy triquark combine together to form a hidden-charm or hidden-bottom tetraquark state. The quark structures of the doubly heavy pentaquark states and hidden-charm or hidden-bottom pentaquark states are quite different.

 The doubly heavy (or hidden-charm, hidden-bottom) tetraquark  states $X$, $Y$, $Z$ and pentaquark  states $P$
are characterized by the effective heavy quark masses ${\mathbb{M}}_Q$   and the virtuality
$V=\sqrt{M^2_{X/Y/Z/P}-(2{\mathbb{M}}_Q)^2}$. In Refs.\cite{Wang-tetraquark-QCDSR,Wang-molecule-QCDSR}, we study the acceptable energy scales of the QCD spectral densities  for the hidden-charm (or hidden-bottom) tetraquark states and molecular states   in the QCD sum rules in details  for the first time,  and suggest an energy scale formula  $\mu=V=\sqrt{M^2_{X/Y/Z}-(2{\mathbb{M}}_Q)^2}$ to determine  the optimal   energy scales.
The  energy scale formula also works well in studying the hidden-charm pentaquark states \cite{Wang1508-EPJC}. The updated values are ${\mathbb{M}}_c=1.82\,\rm{GeV}$ and ${\mathbb{M}}_b=5.17\,\rm{GeV}$ for the hidden-charm and hidden-bottom tetraquark states, respectively  \cite{Wang-EPJC-update}.

It is not necessary  for the ${\mathbb{M}}_Q$ in the doubly heavy tetraquark states and pentaquark states  to have the same values
 as the ones in the hidden-charm or hidden-bottom tetraquark states and pentaquark states. In Ref.\cite{Wang-QQ-tetraquark}, we observe that if we choose a slightly different value ${\mathbb{M}}_c=1.84\,\rm{GeV}$ for the doubly charmed tetraquark states, the criteria of the QCD sum rules can be satisfied more easily, while the value ${\mathbb{M}}_b=5.17\,\rm{GeV}$ survives for the doubly bottom tetraquark states.    In this article, we take the energy scale formula as a constraint to study the doubly heavy pentaquark states.

At the phenomenological side, we exclude the contaminations of the higher resonances by setting in the truncations $s_0$,
\begin{eqnarray}
\int_{4m_Q^2}^{s_0}ds \left[\sqrt{s}\rho^1_{j,H}(s)+\rho^0_{j,H}(s)\right]\exp\left( -\frac{s}{T^2}\right)\, . \nonumber
\end{eqnarray}

At the QCD side, there are terms of the Dirac $\delta$ function type,  $\delta(s-\overline{m}_{Q}^{2})$ and $\delta(s-\widetilde{m}_{Q}^{2})$, which are associated with
the higher dimensional vacuum condensates,
\begin{eqnarray}
\int_{4m_Q^2}^{s_0}\,ds \,\delta(s-\overline{m}_{Q}^{2})\exp\left( -\frac{s}{T^2}\right) &=&\int_{4m_Q^2}^{\infty}\,ds \,\delta(s-\overline{m}_{Q}^{2})\exp\left( -\frac{s}{T^2}\right)=\exp\left( -\frac{\overline{m}_{Q}^{2}}{T^2}\right)\, ,\nonumber\\
\int_{4m_Q^2}^{s_0}\,ds \,\delta(s-\widetilde{m}_{Q}^{2})\exp\left( -\frac{s}{T^2}\right) &=&\int_{4m_Q^2}^{\infty}\,ds \,\delta(s-\widetilde{m}_{Q}^{2})\exp\left( -\frac{s}{T^2}\right)=\exp\left( -\frac{\widetilde{m}_{Q}^{2}}{T^2}\right)\, ,
\end{eqnarray}
 the upper bounds of the integrals are  arbitrary for getting  the same values,  there may be some uncertainties originating from the higher resonances, as  the truncations $s_0$ cannot exclude the contaminations of the higher resonances rigourously.
 Firstly, we need good convergent behaviors in the operator product expansion to obtain solid predictions.

Now we write down the definition for the contributions of the different terms in the operator product expansion,
\begin{eqnarray}
D(n)&=& \frac{  \int_{4m_Q^2}^{s_0} ds\,\rho_{n}(s)\,\exp\left(-\frac{s}{T^2}\right)}{\int_{4m_Q^2}^{s_0} ds \,\rho(s)\,\exp\left(-\frac{s}{T^2}\right)}\, ,
\end{eqnarray}
in stead of
\begin{eqnarray}
D(n)&=& \frac{  \int_{4m_Q^2}^{\infty} ds\,\rho_{n}(s)\,\exp\left(-\frac{s}{T^2}\right)}{\int_{4m_Q^2}^{\infty} ds \,\rho(s)\,\exp\left(-\frac{s}{T^2}\right)}\, ,
\end{eqnarray}
where the $\rho_{n}(s)$ are the QCD spectral densities for the vacuum condensates of dimension $n$, and the total spectral densities
$\rho(s)=\sqrt{s}\rho^1_{QCD}(s)+ \rho^0_{QCD}(s)$. The definition in Eq.(29)
warrants  the contributions of the higher dimensional vacuum condensates play a less important role if the  operator product expansion is well convergent.

 The contributions of the perturbative terms $D(0)$ are usually small for the tetraquark states and pentaquark  states,
    we approximate the continuum contributions as $\rho(s)\Theta(s-s_0)$, and define
    the pole contributions ($\rm{PC}$) or ground state contributions as
   \begin{eqnarray}
{\rm PC}&=& \frac{  \int_{4m_Q^2}^{s_0} ds\,\rho(s)\,
\exp\left(-\frac{s}{T^2}\right)}{\int_{4m_Q^2}^{\infty} ds \,\rho(s)\,\exp\left(-\frac{s}{T^2}\right)}\, .
\end{eqnarray}
If the pole dominance is satisfied at the phenomenological side, the uncertainties originate from the higher dimensional vacuum condensates are greatly suppressed. So the QCD sum rules must satisfied the two criteria, pole dominance at the phenomenological side and convergence of the operator product expansion at the QCD side.

For the lowest doubly charmed baryon state $\Xi_{cc}$, $M_{\Xi_{cc}^{++}}=3621.40 \pm 0.72 \pm 0.27 \pm 0.14\, \rm{MeV}$ \cite{LHCb-Xicc}, which is smaller than $2 {\mathbb{M}}_c=3.68\,\rm{GeV}$, the energy scale formula $\mu=\sqrt{M^2_{\Xi^{++}_{cc}}-(2{\mathbb{M}}_c)^2}$ is failed to work, we can choose the typical energy scale
$\mu=1\,\rm{GeV}$ for the doubly charmed baryon states $\Xi_{cc}$,  $\Xi^*_{cc}$, $\Omega_{cc}$,  $\Omega^*_{cc}$, $2m_c({\rm 1\,GeV})=2.70\sim2.84\,{\rm{GeV}}< M_{\eta_c}<M_{\Xi^{++}_{cc}}$, the $\overline{MS}$ masses $m_{c}({\rm 1\,GeV})=1.35\sim 1.42\,\rm{GeV}$, the integrals $\int_{4m_c^2}^{s_0} ds \,\rho_{QCD}(s)\,\exp\left(-\frac{s}{T^2}\right)$ make sense for the conventional charmonium states and doubly charmed  baryon states.
In calculations, we observe that $2m_b({\rm 2.2\,GeV})=9.28\sim 9.44{\,\rm{GeV}}$, while the bottomonium masses $M_{\eta_b}=9.399\,\rm{GeV}$, $M_{\Upsilon}=9.46\,\rm{GeV}$ \cite{PDG}, the energy scale $\mu=2.2\,\rm{GeV}$ is the lowest energy scale or  marginal energy scale for the integrals $\int_{4m_b^2}^{s_0} ds \,\rho_{QCD}(s)\,\exp\left(-\frac{s}{T^2}\right)$ for the conventional bottomonium states. In this article, we choose $\mu=2.2\,\rm{GeV}$
 for the doubly bottom baryon states $\Xi_{bb}$,  $\Xi^*_{bb}$, $\Omega_{bb}$ and  $\Omega^*_{bb}$, which works well.

In Refs.\cite{WangHHbaryon-4-P,WangH-HHbaryon-6-4-P,WangH-HHbaryon-6-5-N,Wang1508-EPJC,WangHbaryon-6-PN,WangHHHbaryon-4-PN},  we separate the contributions of the positive parity and negative parity hadron states explicitly, and study the   heavy, doubly-heavy, triply-heavy baryon states and the hidden-charm pentaquark states with the QCD sum rules in a systematic way.
In calculations, we observe that the continuum threshold parameters $\sqrt{s_0}=M_{\rm{gr}}+ (0.6-0.8)\,\rm{GeV}$  works well,  where the subscript $\rm{gr}$ denotes the ground   states. In this article,  we can take the continuum threshold parameters as $\sqrt{s_0}< M_{B/P}+0.8\,\rm{GeV}$.

 In this  article, we choose the  Borel parameters $T^2$ and continuum threshold
parameters $s_0$  to satisfy   the  following four criteria:

$\bf C1.$ Pole dominance at the phenomenological side;

$\bf C2.$ Convergence of the operator product expansion;

$\bf C3.$ Appearance of the Borel platforms;

$\bf C4.$ Satisfying the energy scale formula only for the doubly heavy pentaquark states,\\
by try and error.

The resulting Borel parameters or Borel windows $T^2$, continuum threshold parameters $s_0$,  energy scales of the QCD spectral densities,  pole contributions of the ground states and contributions of the highest dimensional  vacuum condensates  are shown  explicitly in Table 2.
From the table, we can see that the pole dominance at the phenomenological side and the convergence of the operator product expansion at  the QCD side are satisfied, or the criteria $\bf C1$ and $\bf C2$ are satisfied.

\begin{figure}
 \centering
 \includegraphics[totalheight=5cm,width=7cm]{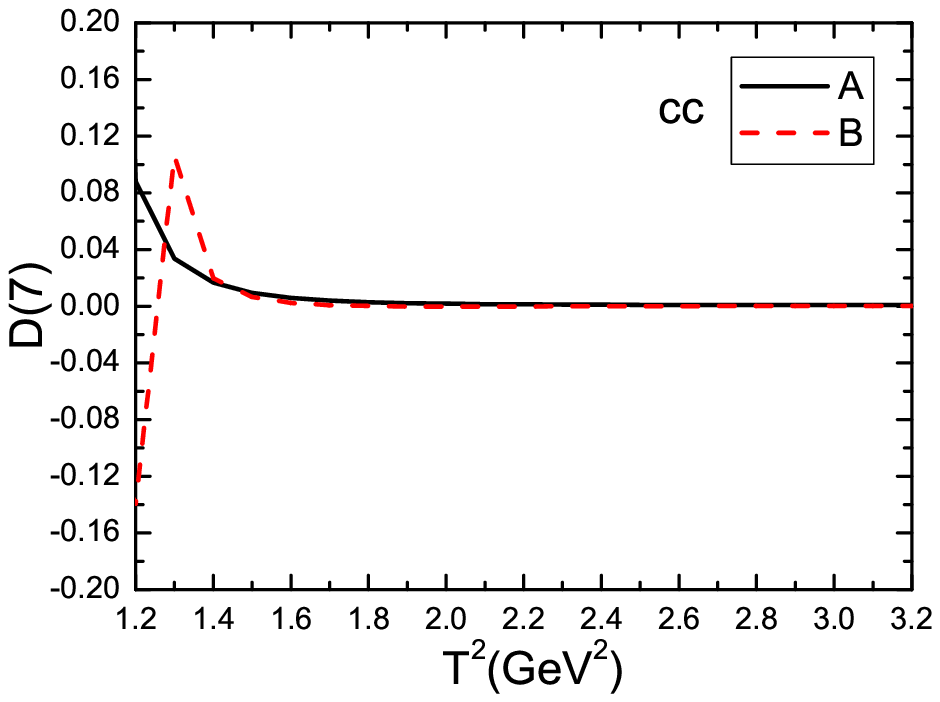}
 \includegraphics[totalheight=5cm,width=7cm]{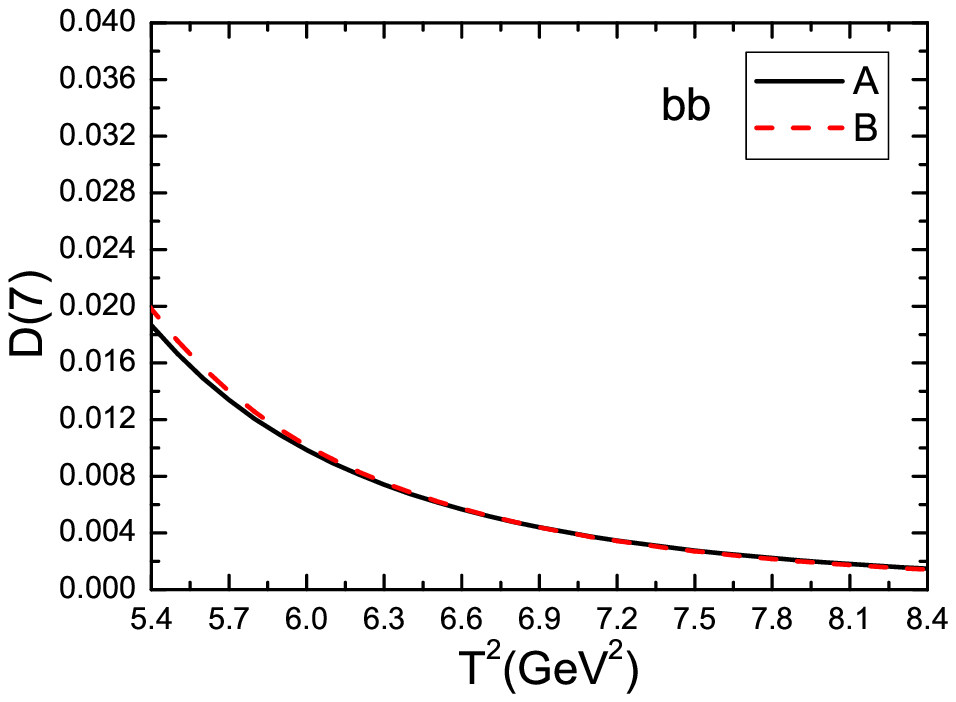}
 \caption{ The contributions  of the vacuum condensates of dimension $7$ in the operator product expansion, where $A$ and $B$ denote
 the  $\Xi_{QQ}$ baryon states with $J=\frac{1}{2}$ and $\frac{3}{2}$, respectively.  }
\end{figure}

\begin{figure}
 \centering
 \includegraphics[totalheight=5cm,width=7cm]{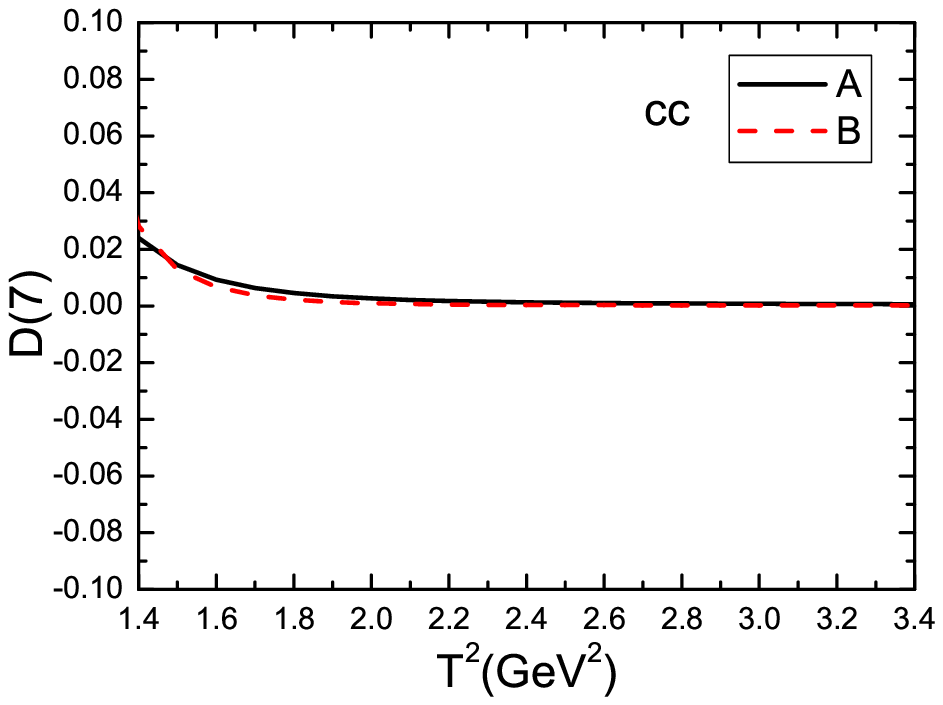}
 \includegraphics[totalheight=5cm,width=7cm]{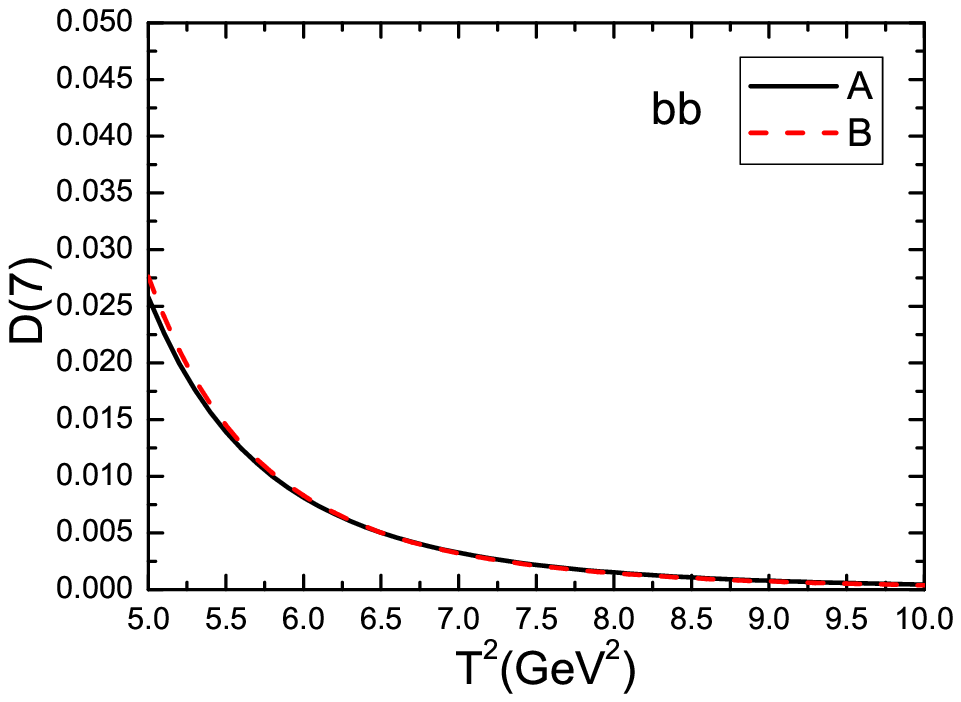}
 \caption{ The contributions  of the vacuum condensates of dimension $7$ in the operator product expansion, where $A$ and $B$ denote
 the  $\Omega_{QQ}$ baryon states with $J=\frac{1}{2}$ and $\frac{3}{2}$, respectively.  }
\end{figure}

\begin{figure}
 \centering
 \includegraphics[totalheight=5cm,width=7cm]{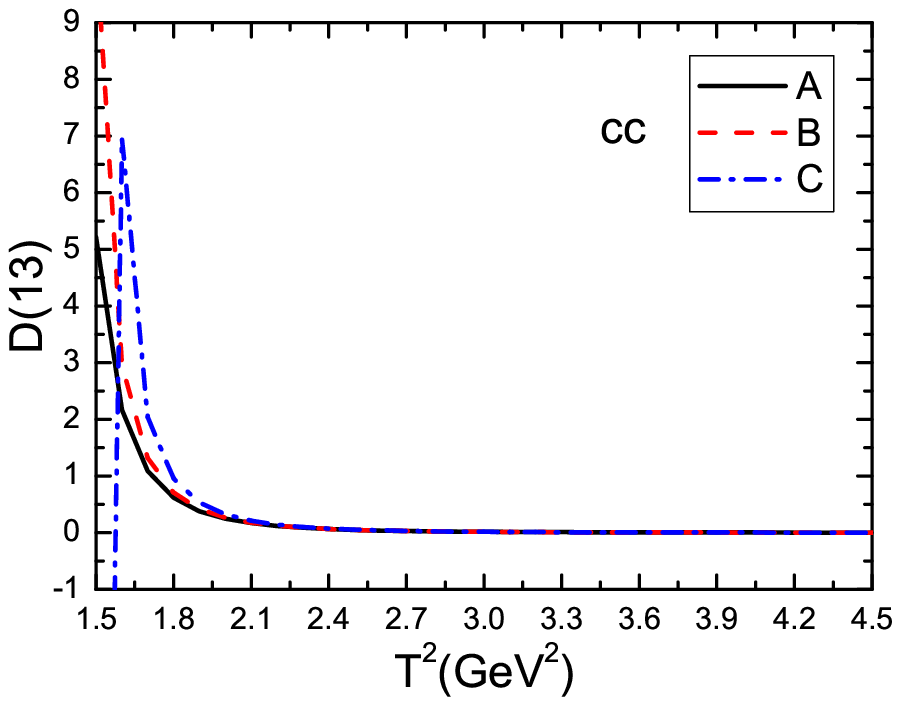}
 \includegraphics[totalheight=5cm,width=7cm]{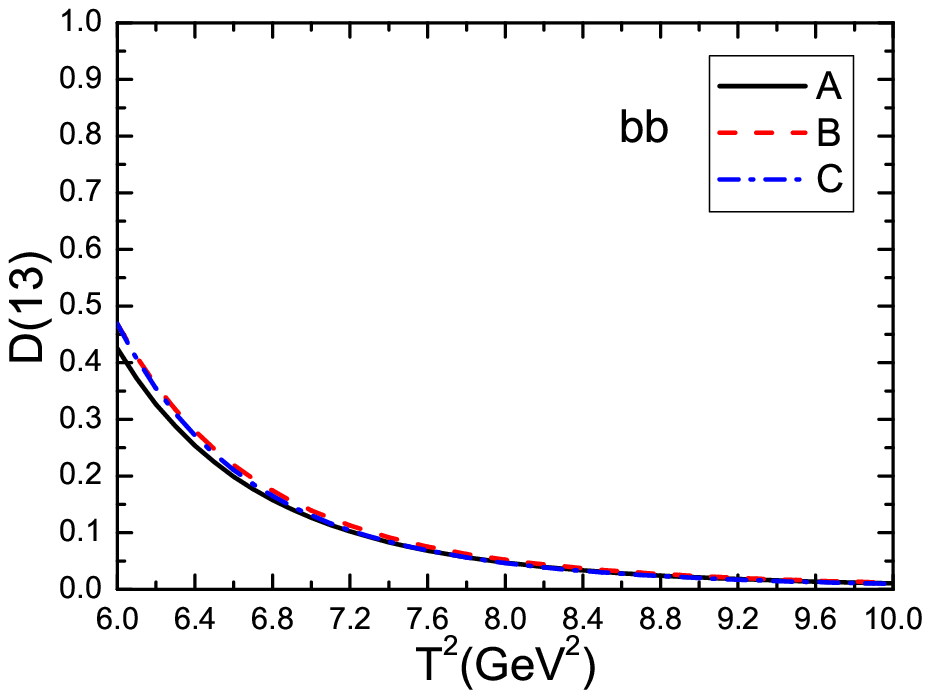}
 \caption{ The contributions  of the vacuum condensates of dimension $13$ in the operator product expansion, where $A$, $B$ and $C$ denote
 the  doubly heavy pentaquark states with $J=\frac{1}{2}$, $\frac{3}{2}$ and  $\frac{5}{2}$, respectively.  }
\end{figure}

In Figs.1-3,  we plot the contributions of the vacuum condensates of dimension $7$ and $13$ with variations of the Borel parameters $T^2$ for  the doubly heavy baryon states and pentaquark states, respectively. From the figures, we can see explicitly that the contributions of the highest dimensional vacuum condensates are tiny
 in the Borel windows for the doubly heavy baryon states and doubly-charmed pentaquark states, while the contributions of the vacuum condensates of dimension $13$ for the doubly bottom pentaquark states are somewhat larger, the smallest contributions are about $(5\sim 7)\%$  in the Borel windows, the operator product expansion is still convergent. In fact, for the doubly bottom pentaquark states, the $D(13)$ decrease  monotonously  and  quickly with the increase of the Borel parameter $T^2$, a slight larger Borel parameter can lead to much smaller contribution. In calculations, we observe that the predicted doubly heavy pentaquark masses are rather stable with variations of the Borel parameters  at the region $T^2\geq T_{min}^2$, where the $min$ denotes the minimal  values, the predictions survive for larger Borel parameters, the somewhat large contributions $D(13)$ cannot impair the predictive ability.

\begin{figure}
 \centering
 \includegraphics[totalheight=5cm,width=7cm]{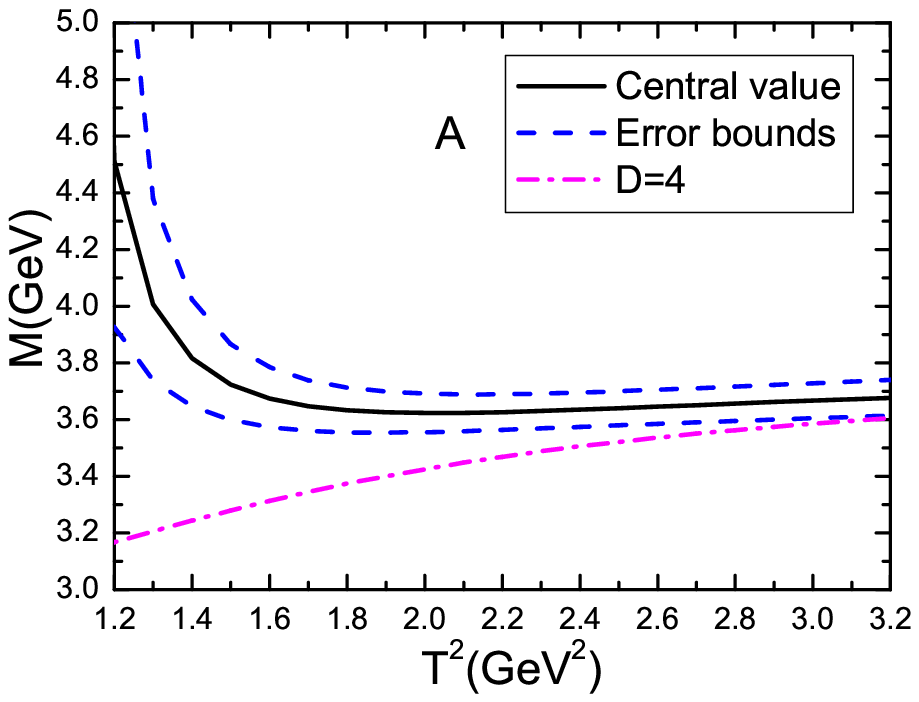}
 \includegraphics[totalheight=5cm,width=7cm]{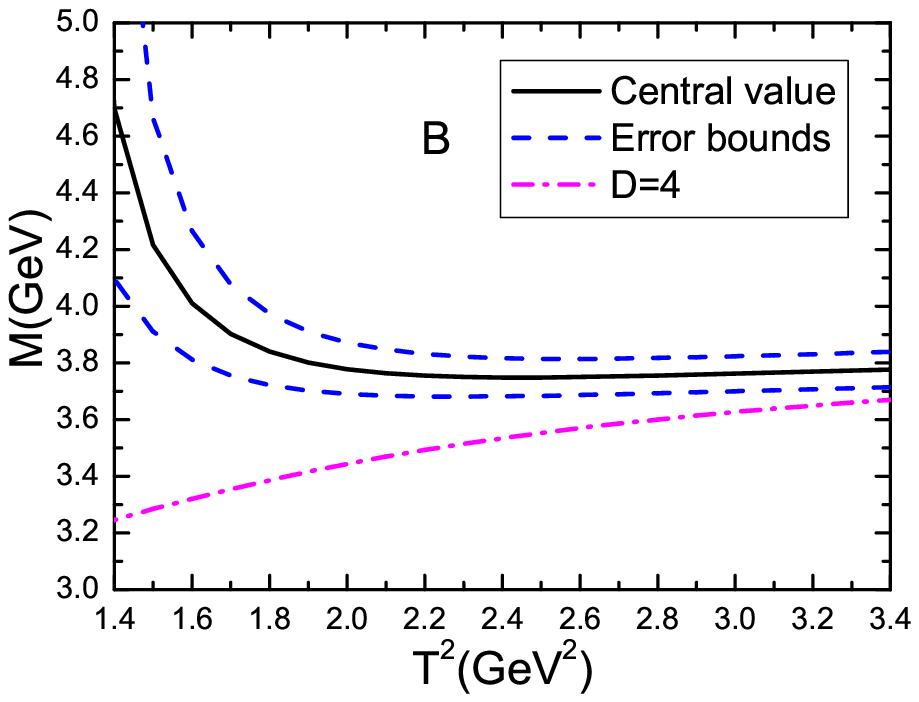}
 \includegraphics[totalheight=5cm,width=7cm]{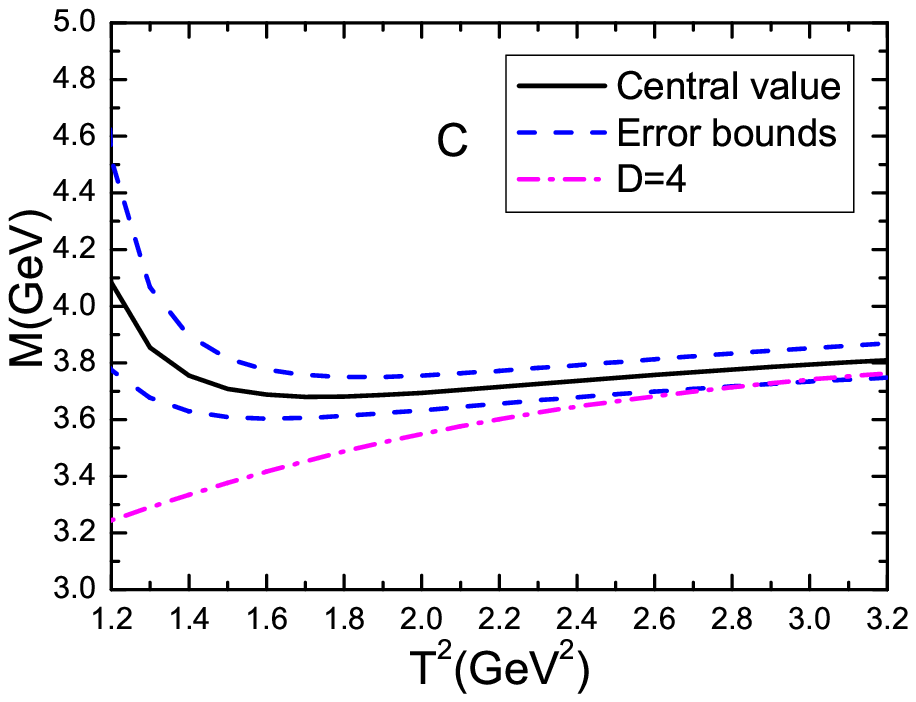}
 \includegraphics[totalheight=5cm,width=7cm]{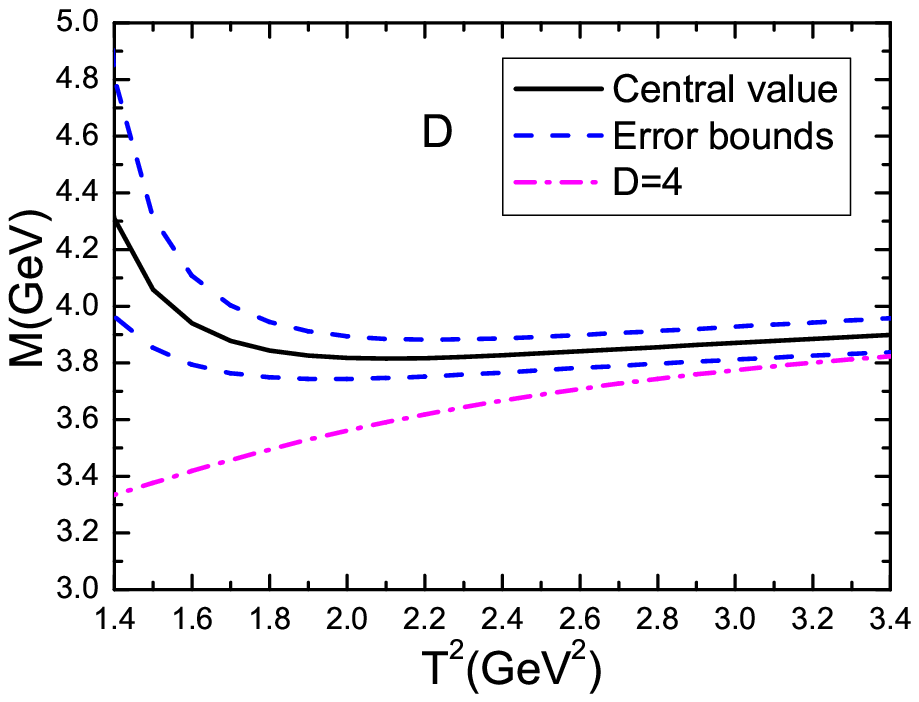}
 \includegraphics[totalheight=5cm,width=7cm]{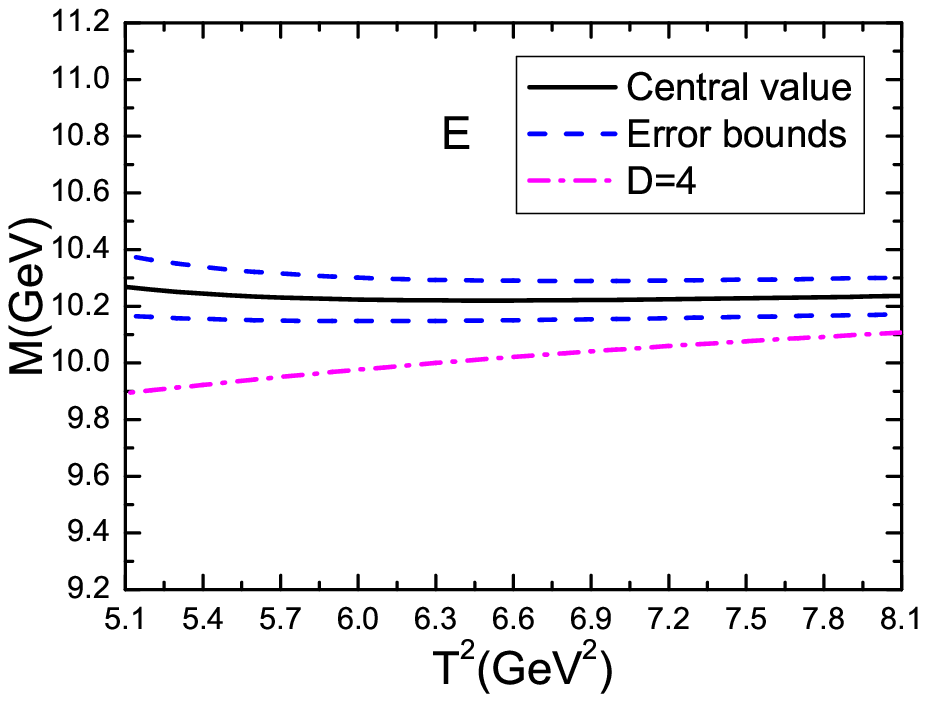}
 \includegraphics[totalheight=5cm,width=7cm]{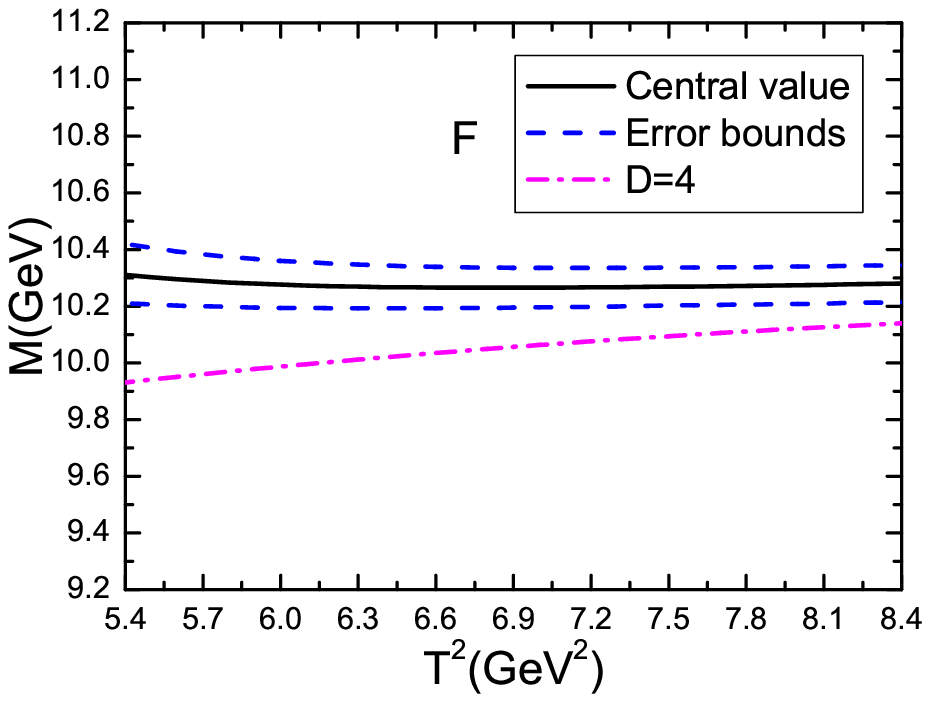}
 \includegraphics[totalheight=5cm,width=7cm]{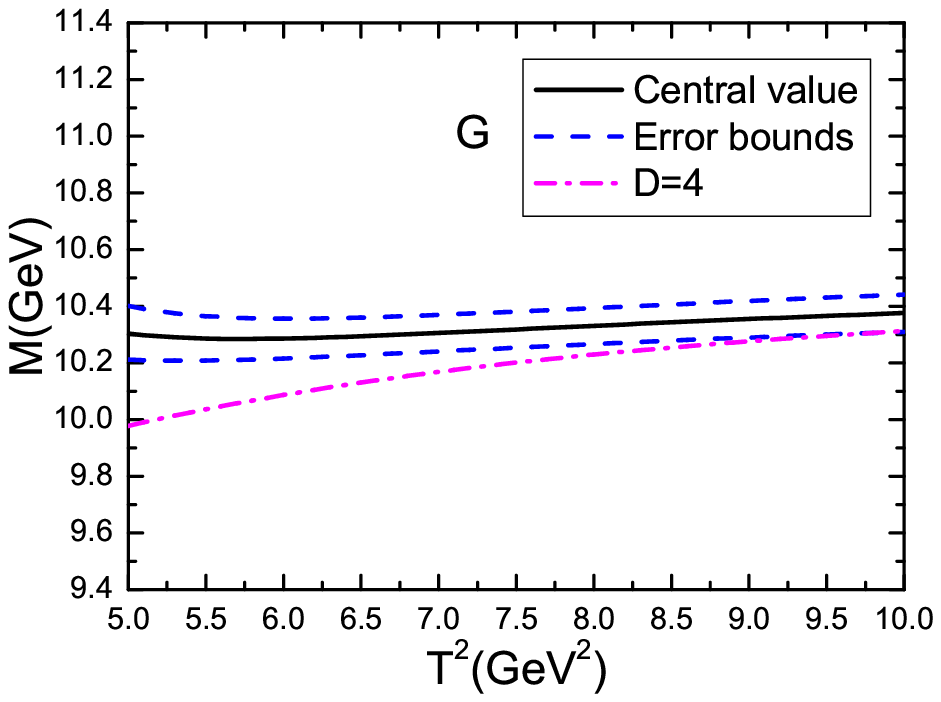}
 \includegraphics[totalheight=5cm,width=7cm]{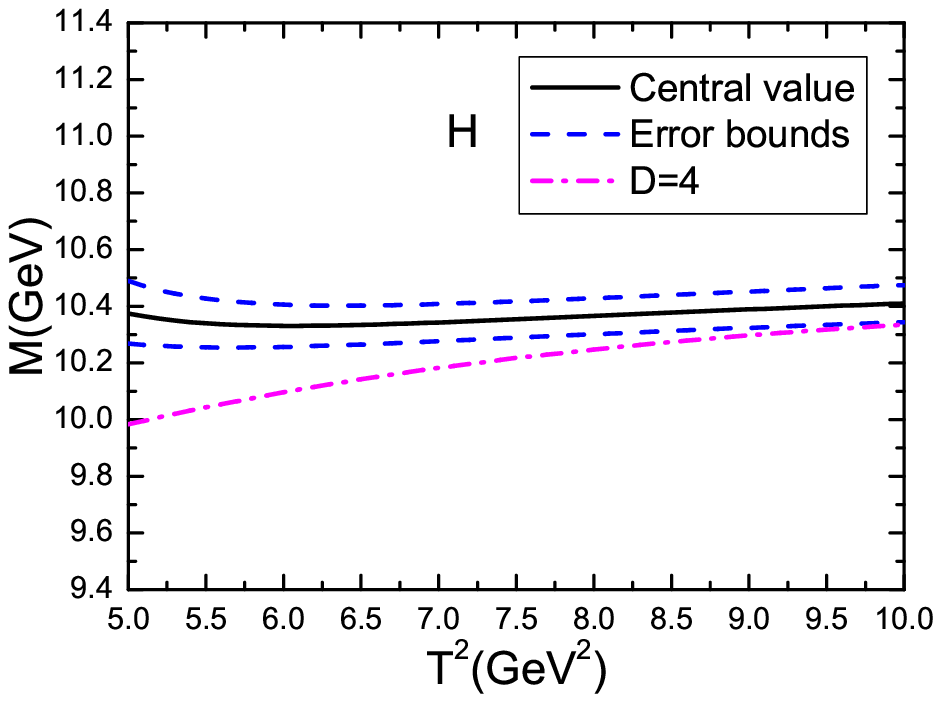}
  \caption{ The masses  of the doubly heavy baryon states  with variations of the Borel parameters $T^2$, where the $A$, $B$, $C$, $D$, $E$, $F$, $G$ and $H$ denote
  the $\Xi_{cc}$, $\Xi_{cc}^*$, $\Omega_{cc}$, $\Omega_{cc}^*$, $\Xi_{bb}$, $\Xi_{bb}^*$, $\Omega_{bb}$ and $\Omega_{bb}^*$,  respectively, the $D=4$ denotes the predictions based on the truncations  of the operator product expansion up to the vacuum condensates of dimension $4$.  }
\end{figure}

\begin{figure}
 \centering
 \includegraphics[totalheight=5cm,width=7cm]{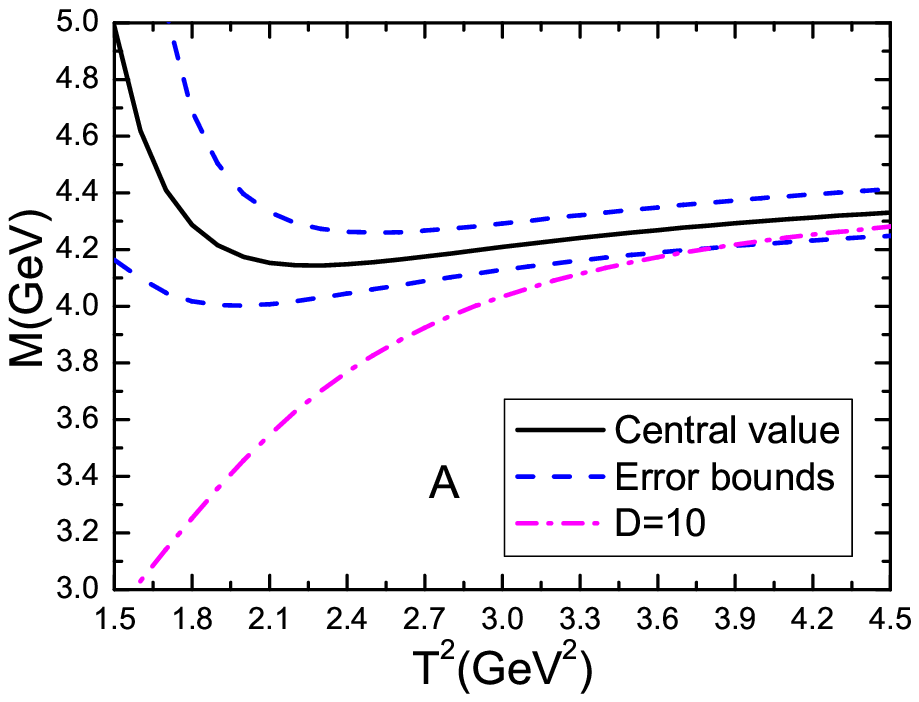}
 \includegraphics[totalheight=5cm,width=7cm]{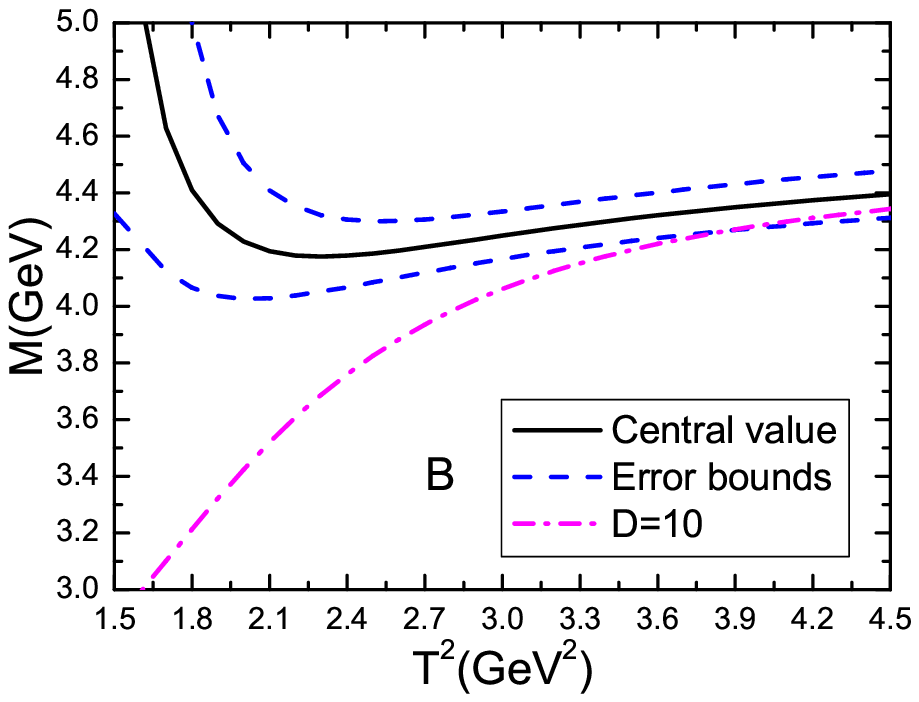}
 \includegraphics[totalheight=5cm,width=7cm]{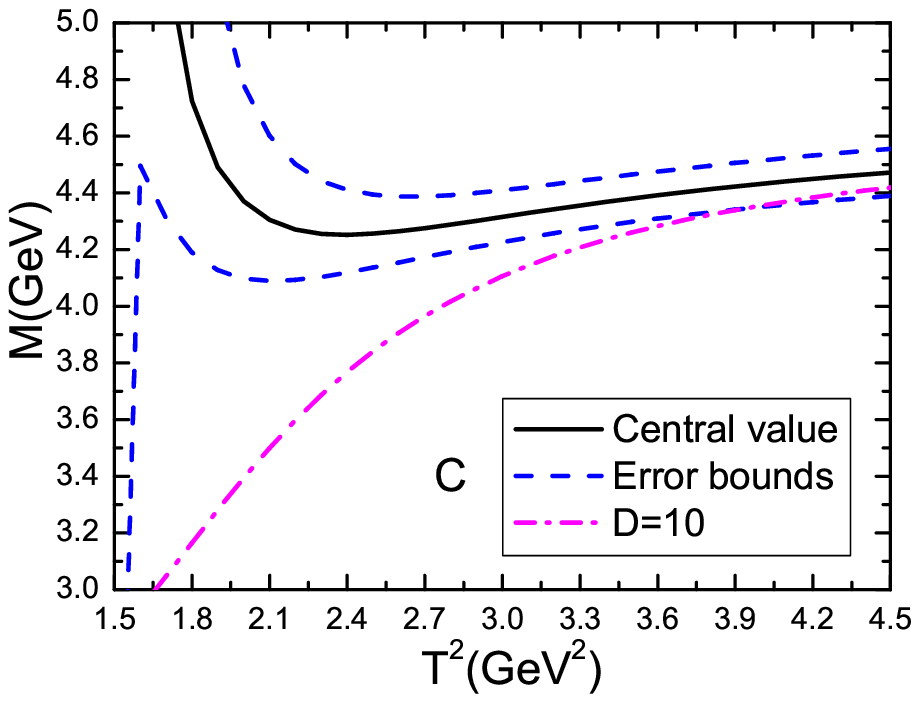}
 \includegraphics[totalheight=5cm,width=7cm]{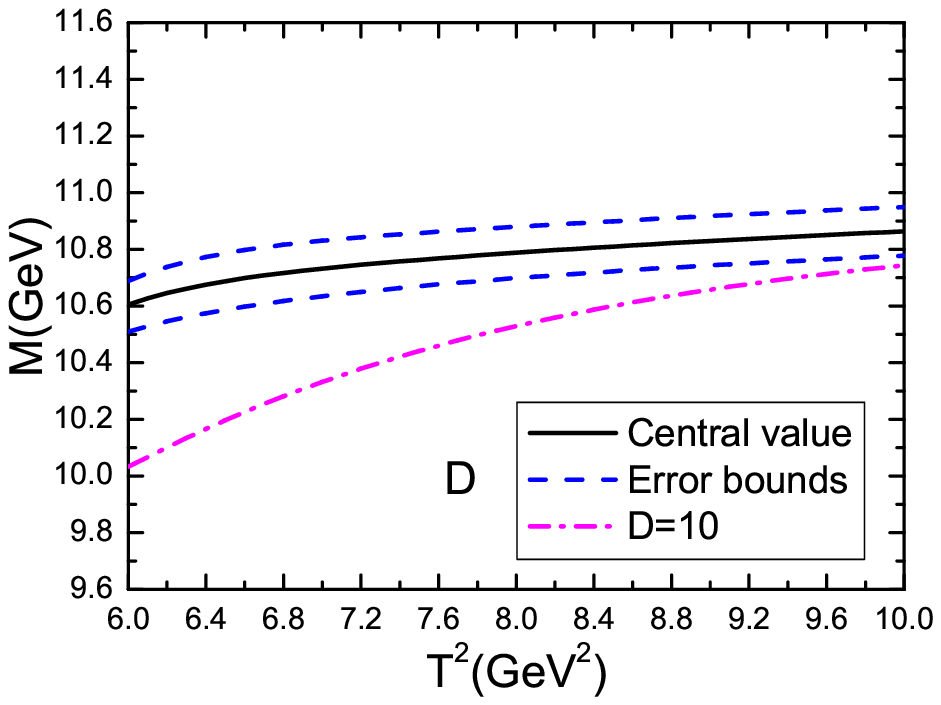}
 \includegraphics[totalheight=5cm,width=7cm]{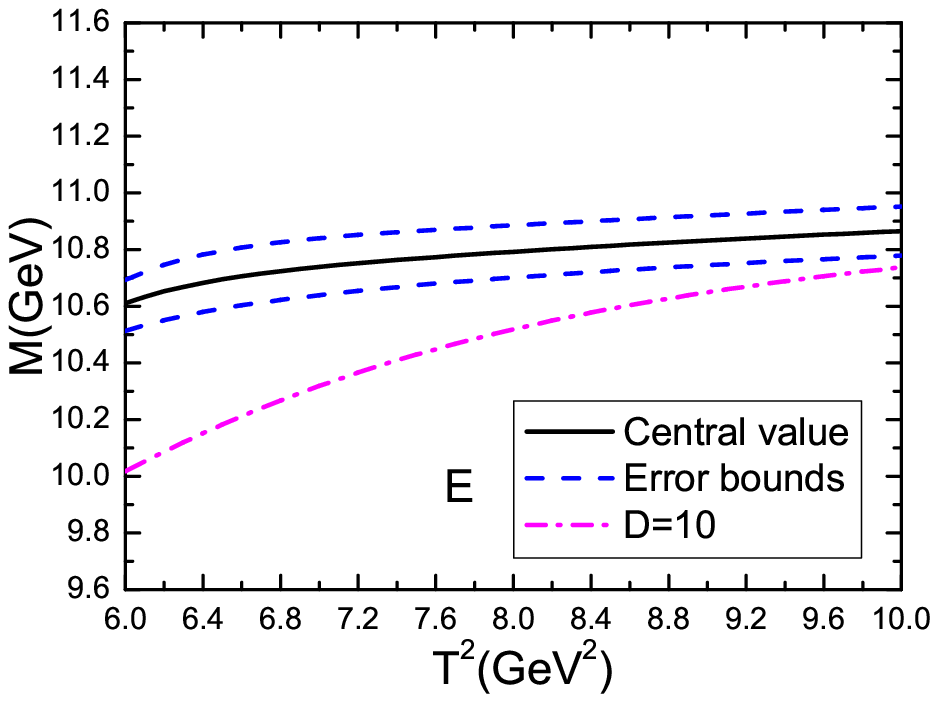}
 \includegraphics[totalheight=5cm,width=7cm]{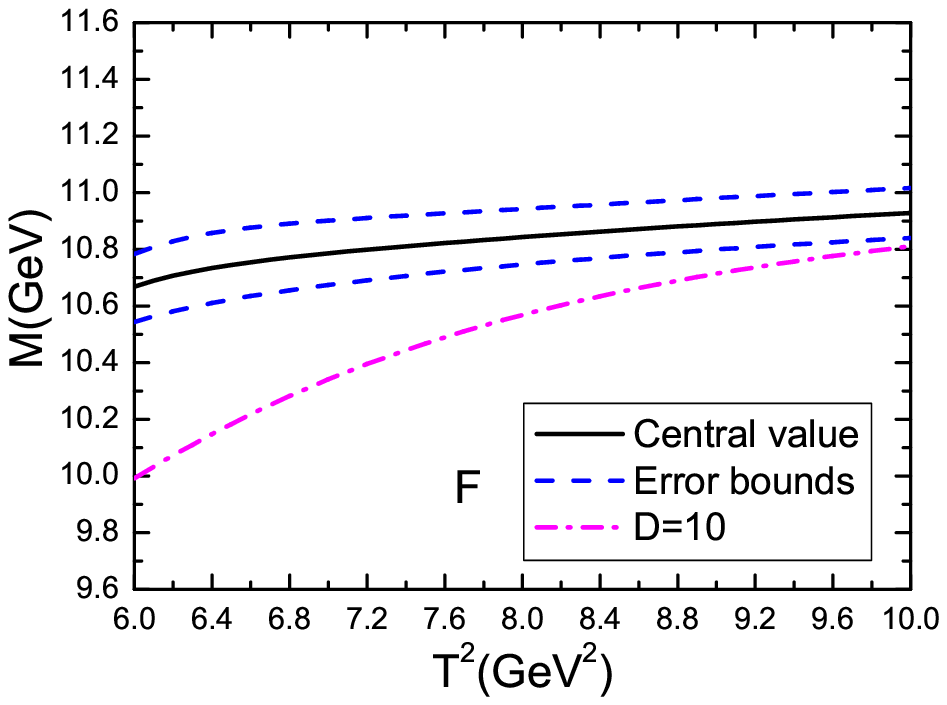}
 \caption{ The masses  of the doubly heavy pentaquark states  with variations of the Borel parameters $T^2$, where the $A$, $B$, $C$, $D$, $E$ and $F$ denote
  the $P_{cc,\frac{1}{2}}$, $P_{cc,\frac{3}{2}}$, $P_{cc,\frac{5}{2}}$, $P_{bb,\frac{1}{2}}$, $P_{bb,\frac{3}{2}}$ and $P_{bb,\frac{5}{2}}$,  respectively, the $D=10$ denotes the predictions based on the truncations  of the operator product expansion up to the vacuum condensates of dimension $10$.  }
\end{figure}

\begin{figure}
 \centering
 \includegraphics[totalheight=5cm,width=7cm]{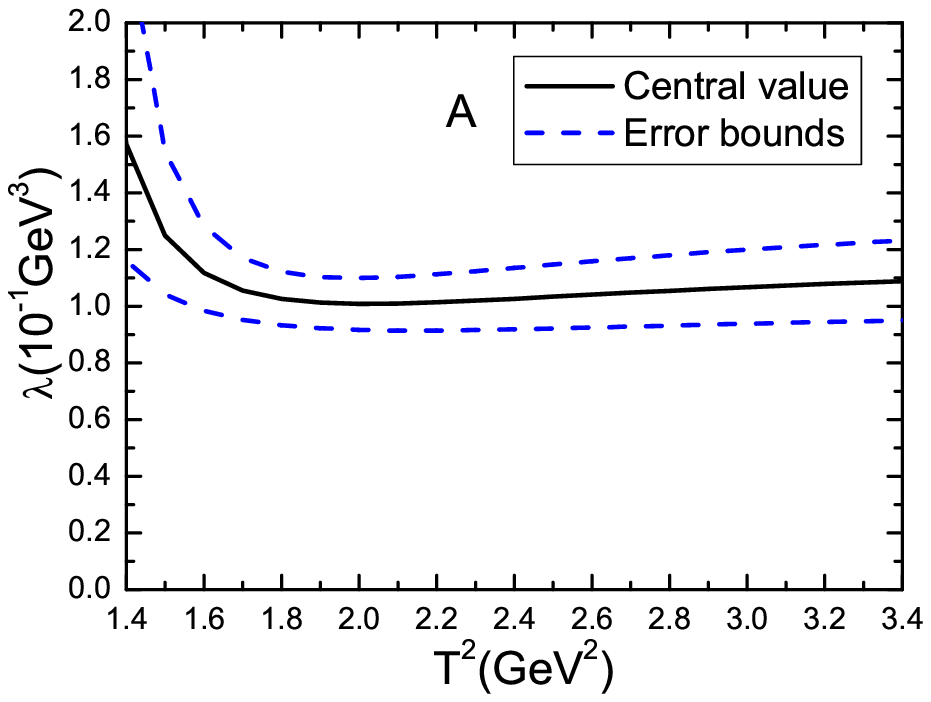}
 \includegraphics[totalheight=5cm,width=7cm]{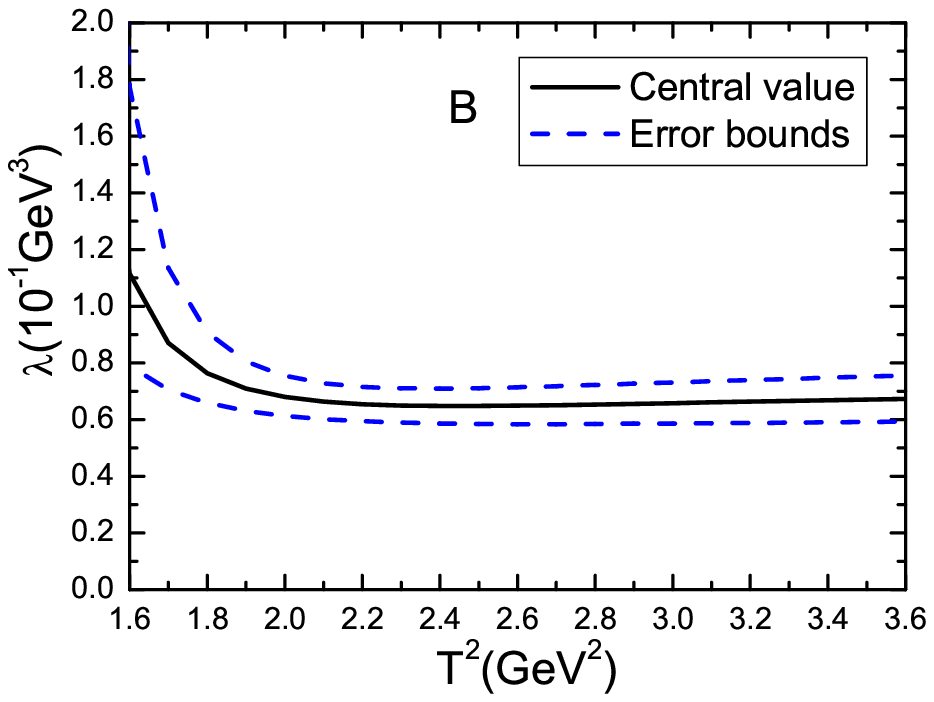}
 \includegraphics[totalheight=5cm,width=7cm]{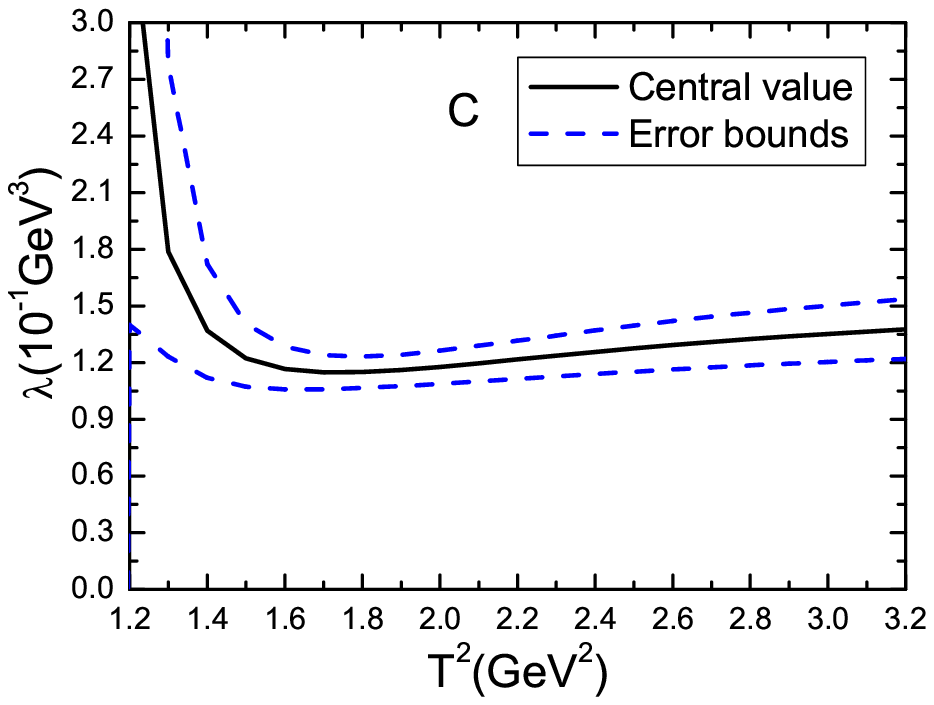}
 \includegraphics[totalheight=5cm,width=7cm]{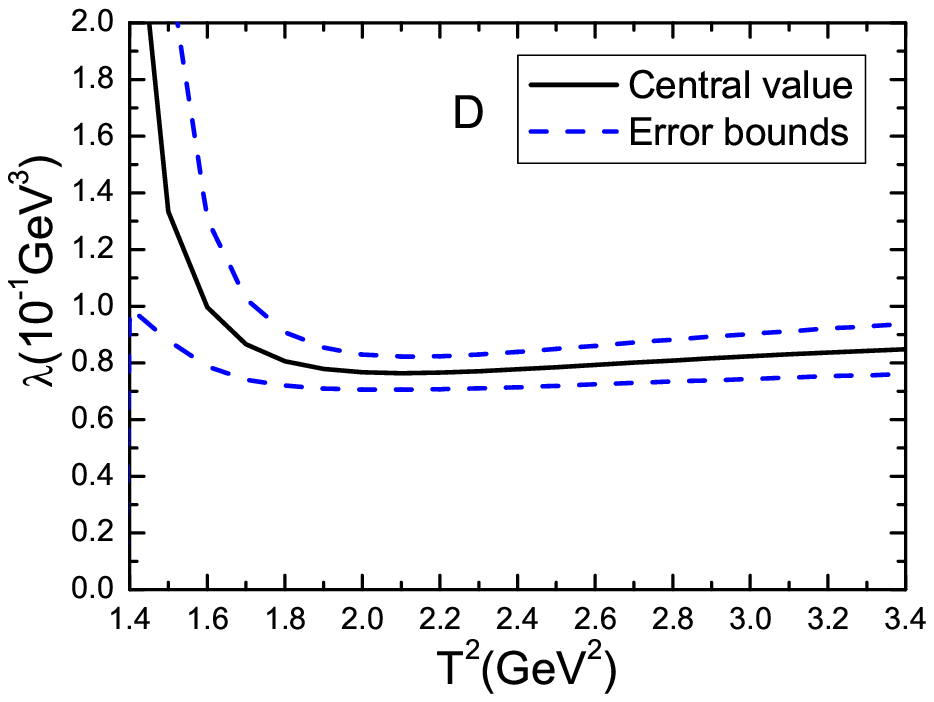}
 \includegraphics[totalheight=5cm,width=7cm]{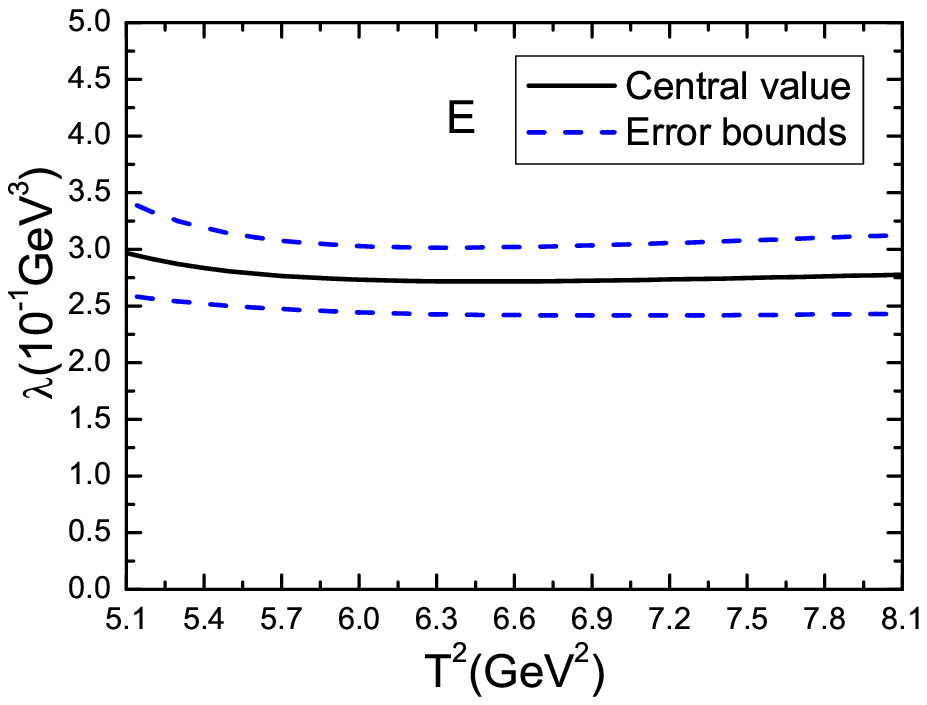}
 \includegraphics[totalheight=5cm,width=7cm]{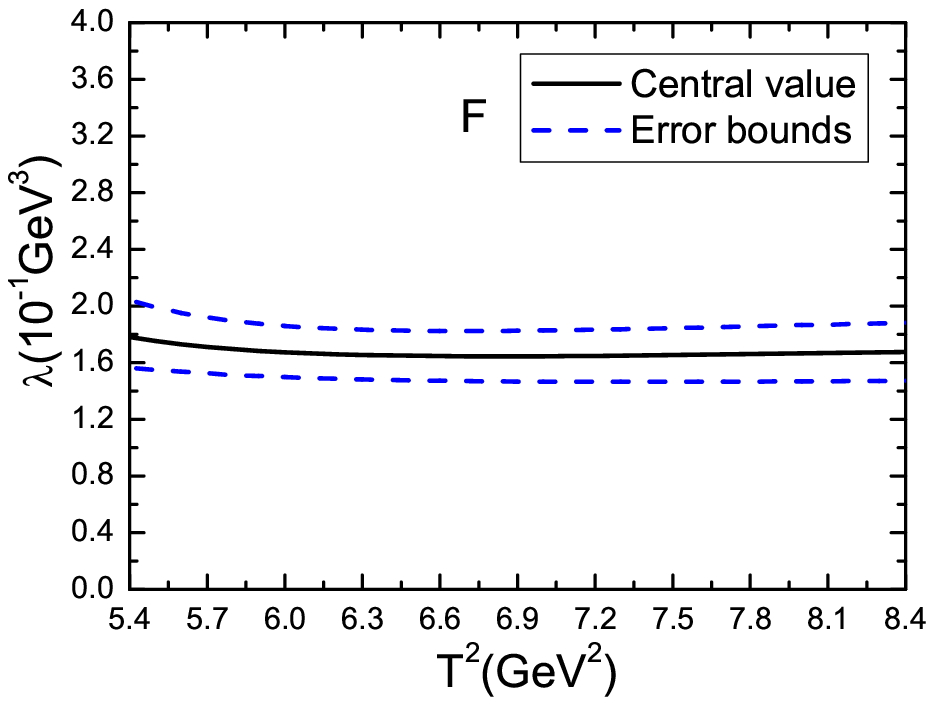}
 \includegraphics[totalheight=5cm,width=7cm]{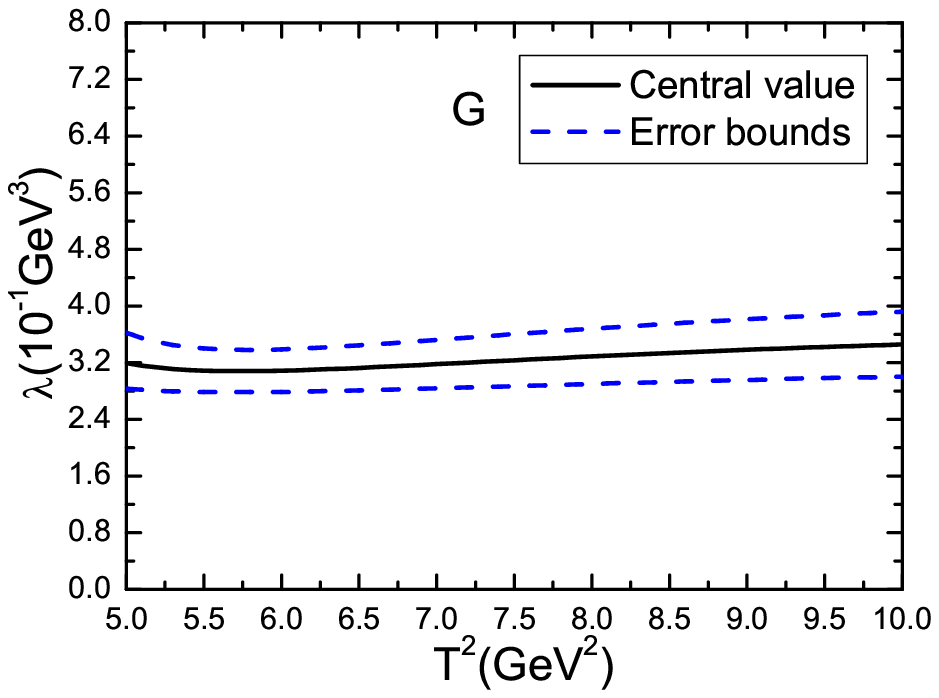}
 \includegraphics[totalheight=5cm,width=7cm]{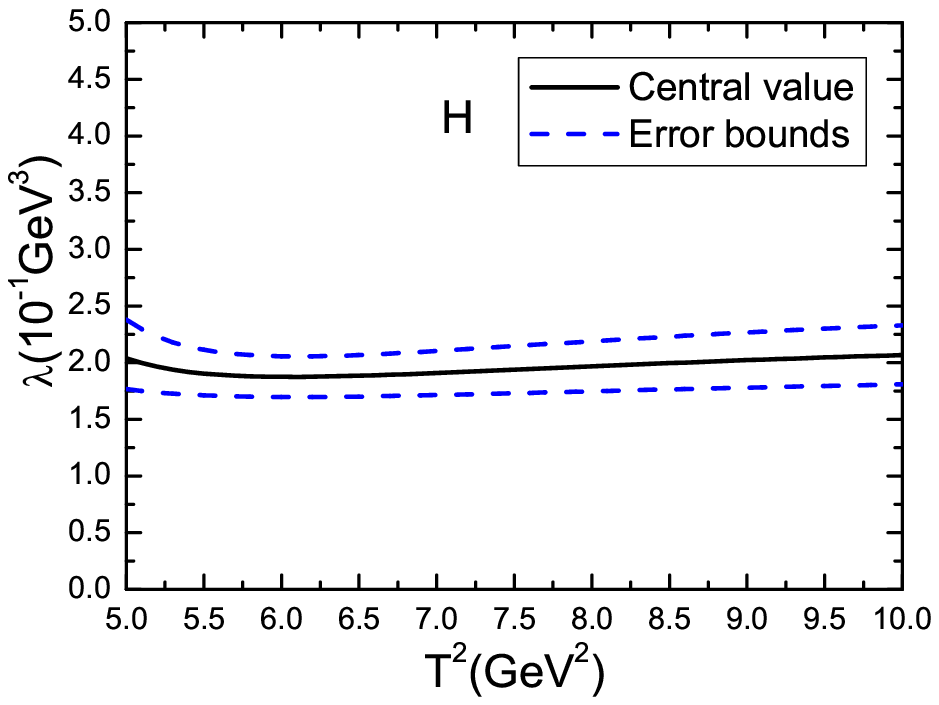}
 \caption{ The pole residues  of the doubly heavy baryon states  with variations of the Borel parameters $T^2$, where the $A$, $B$, $C$, $D$, $E$, $F$, $G$ and $H$ denote the $\Xi_{cc}$, $\Xi_{cc}^*$, $\Omega_{cc}$, $\Omega_{cc}^*$, $\Xi_{bb}$, $\Xi_{bb}^*$, $\Omega_{bb}$ and $\Omega_{bb}^*$,  respectively.  }
\end{figure}

\begin{figure}
 \centering
 \includegraphics[totalheight=5cm,width=7cm]{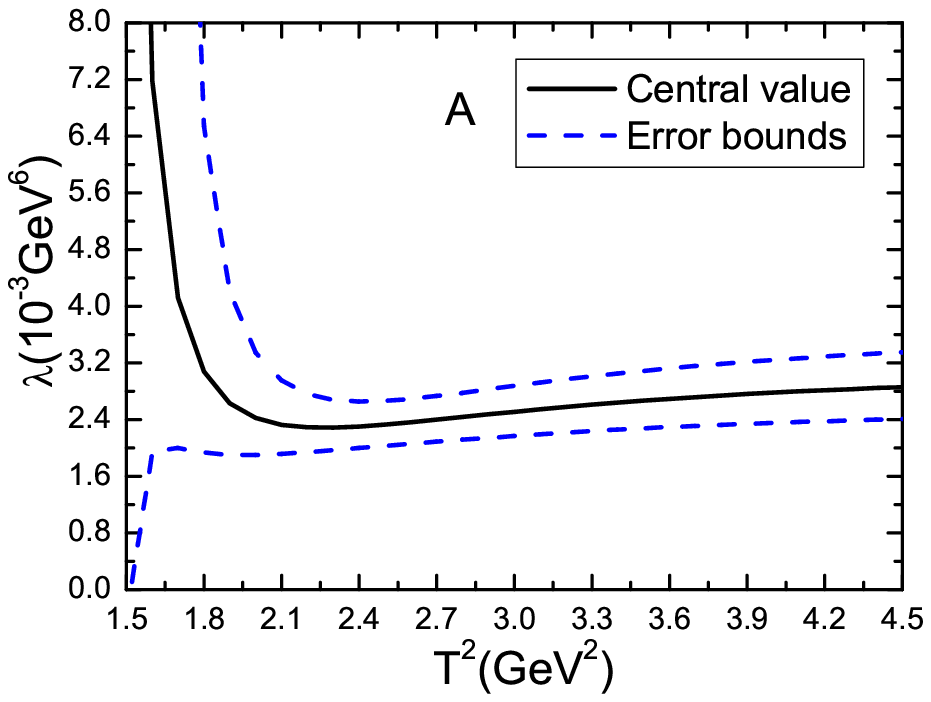}
 \includegraphics[totalheight=5cm,width=7cm]{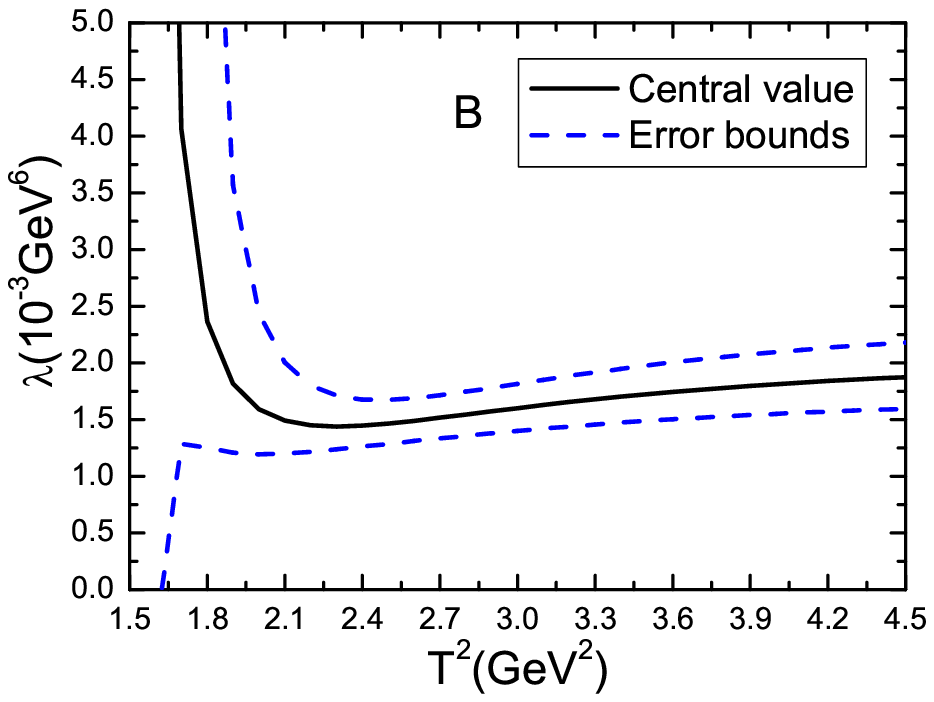}
 \includegraphics[totalheight=5cm,width=7cm]{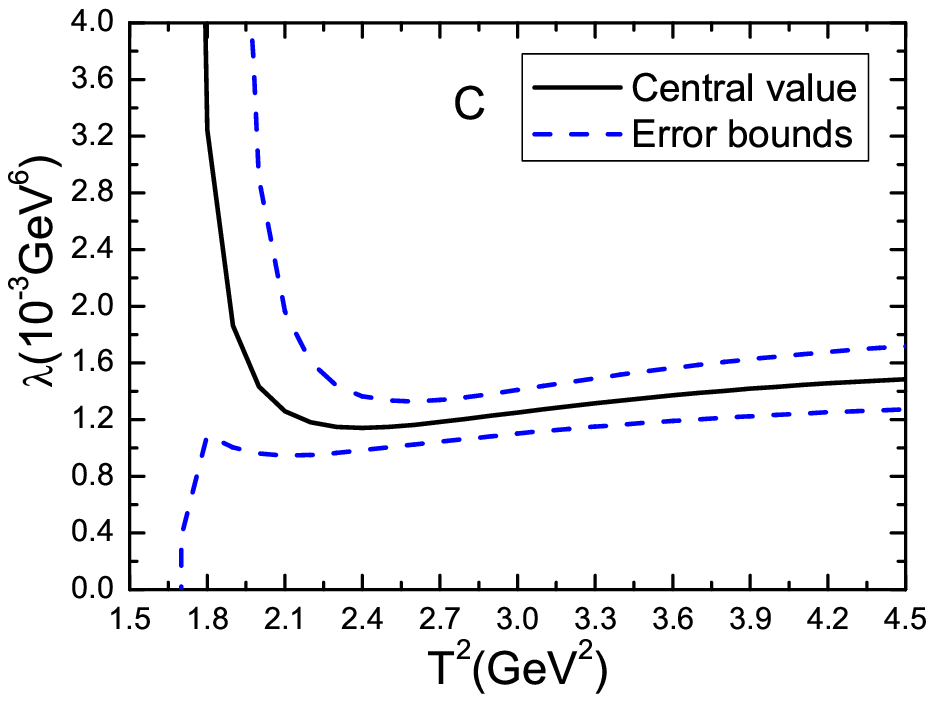}
 \includegraphics[totalheight=5cm,width=7cm]{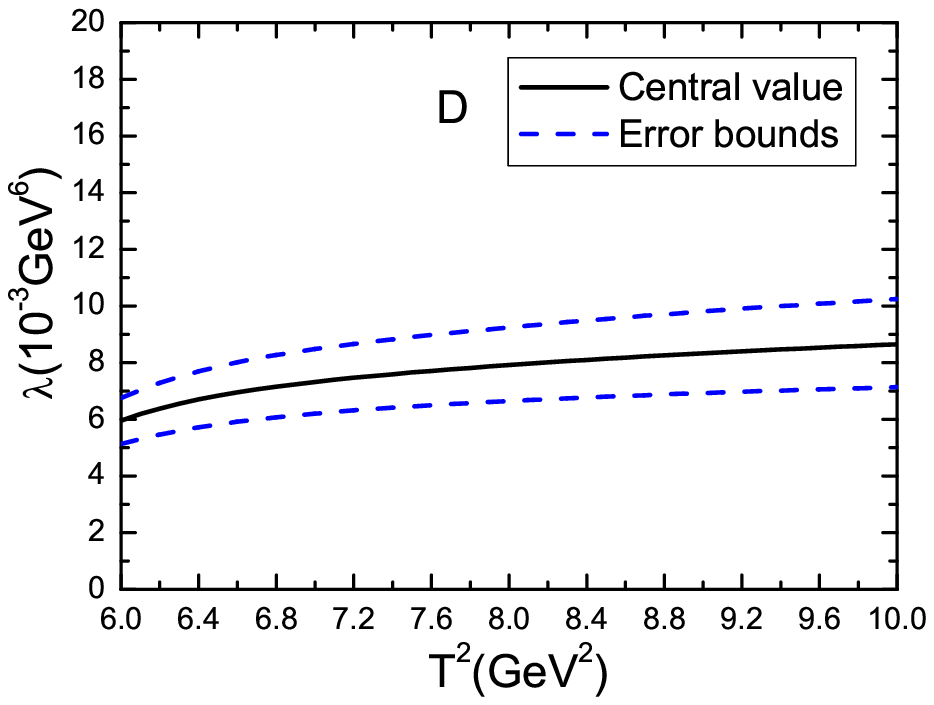}
 \includegraphics[totalheight=5cm,width=7cm]{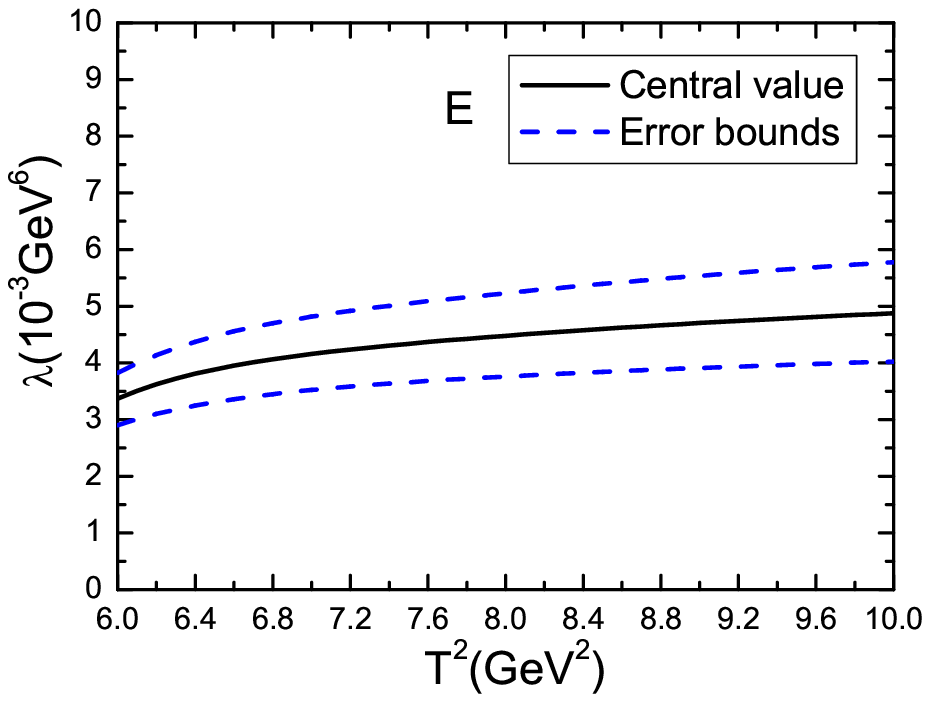}
 \includegraphics[totalheight=5cm,width=7cm]{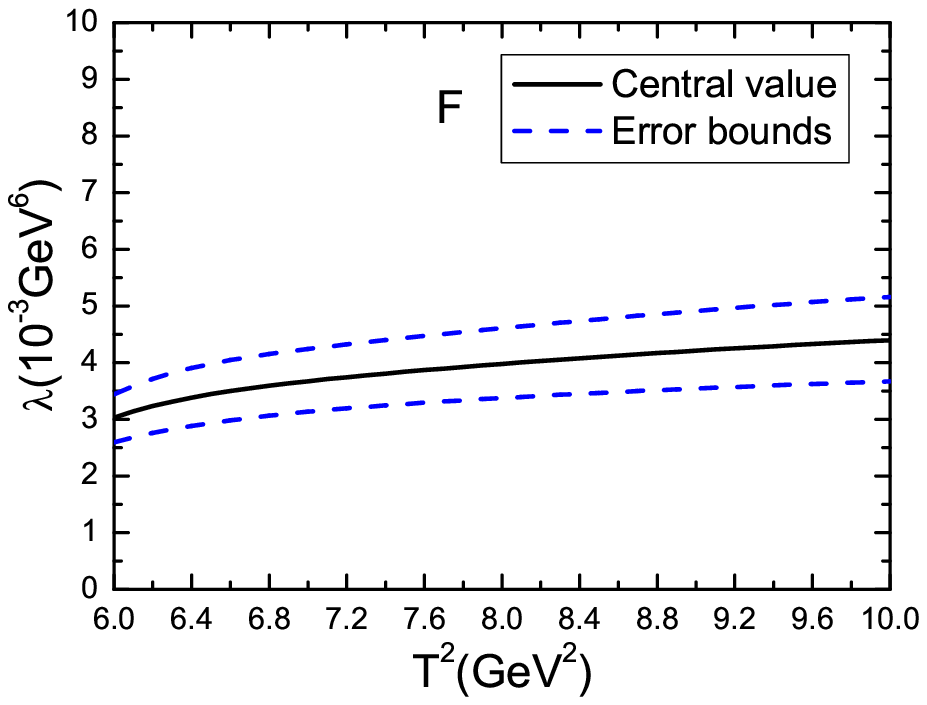}
 \caption{ The pole residues  of the doubly heavy pentaquark states  with variations of the Borel parameters $T^2$, where the $A$, $B$, $C$, $D$, $E$ and $F$ denote
  the $P_{cc,\frac{1}{2}}$, $P_{cc,\frac{3}{2}}$, $P_{cc,\frac{5}{2}}$, $P_{bb,\frac{1}{2}}$, $P_{bb,\frac{3}{2}}$ and $P_{bb,\frac{5}{2}}$,  respectively.  }
\end{figure}

We take into account  all uncertainties  of the input   parameters,
and obtain  the masses and pole residues of
 the doubly heavy baryon states and  pentaquark  states, which are shown explicitly in Table 3 and Figs.4-7. From Table 3, we can see that the  criterion  $\bf C4$ is  satisfied for the doubly heavy pentaquark states. In this article, we choose the effective heavy quark masses ${\mathbb{M}}_c=1.84\,\rm{GeV}$ and  ${\mathbb{M}}_b=5.17\,\rm{GeV}$ for the doubly heavy tetraquark states \cite{Wang-QQ-tetraquark}, if we choose slightly larger mass ${\mathbb{M}}_b=5.18\,\rm{GeV}$, the energy scale formula is satisfied even better.

 In Figs.4-7, we plot the masses and pole residues at much larger ranges of the Borel parameters than the Borel windows.
 From the figures, we can see that there appear Borel platforms  in the Borel windows, the criterion  $\bf C3$ is also  satisfied. In Figs.4-5, we also plot the masses
 with the truncations of the  operator product expansion up to  the vacuum condensates of dimension $4$ for the doubly heavy baryon states and of
 dimension $10$ for the doubly heavy pentaquark states. From  the figures, we can see that without including the vacuum condensates of dimensions $5$, $7$ and $11$, $13$ for the doubly heavy baryon states and  pentaquark  states, respectively, we cannot obtain very stable QCD sum rules with respect to  variations of the Borel parameters, the higher dimensional vacuum condensates play an important role in determining the Borel platforms.

For the doubly heavy baryon states, the criteria $\bf C1$, $\bf C2$ and $\bf C3$ are satisfied, for the doubly heavy pentaquark states, the  criteria
$\bf C1$, $\bf C2$, $\bf C3$ and $\bf C4$ are satisfied, we expect to make reliable predictions, which can be confronted to  the experimental data in the future.

In the present work, we obtain the mass $M=4.21^{+0.10}_{-0.11}\,\rm{GeV}$ for the doubly charmed pentaquark state  $ccud\bar{q}$ with  $J^P={\frac{1}{2}}^-$. While in Ref.\cite{Yan-GuoFK}, Yan et al obtain the masses of the meson-baryon type doubly charmed  pentaquark states with $J^P={\frac{1}{2}}^-$  below $4.2\,\rm{GeV}$ based on  the unitarized coupled-channel approach, which are in qualitative agreement with the present predictions. In Ref.\cite{Azizi-Penta}, Azizi,  Sarac and Sundu study the meson-baryon type hidden-charm (hidden-bottom)  pentaquark states with $J^P={\frac{3}{2}}^\pm$ and ${\frac{5}{2}}^\pm$ based on the QCD sum rules,
the predicted masses $4.30\pm0.10\,\rm{GeV}$ and $4.20\pm 0.15\,\rm{GeV}$ ($10.96^{+0.84}_{-0.88}\,\rm{GeV}$ and $10.98^{+0.82}_{-0.82}\,\rm{GeV}$) for the hidden-charm (hidden-bottom) pentaquark states with $J^P={\frac{3}{2}}^-$ and ${\frac{5}{2}}^-$ respectively  are compatible with the present calculations  in magnitude, but differ from the present calculations quantitatively. We should bear in mind that they are quite different pentaquark states.

\begin{table}
\begin{center}
\begin{tabular}{|c|c|c|c|c|c|c|c|}\hline\hline
                &$J^P$                    &$\mu(\rm GeV)$    &$T^2 (\rm{GeV}^2)$  &$\sqrt{s_0} (\rm{GeV})$      &pole          &$|D(n=7/13)|$       \\  \hline
$ccq$           &${\frac{1}{2}}^+$        &1.0               &$2.0-2.6$           &$4.15\pm0.10$                &$(60-86)\%$   &$\ll 1\%$   \\ \hline
$ccq$           &${\frac{3}{2}}^+$        &1.0               &$2.2-2.8$           &$4.25\pm0.10$                &$(60-85)\%$   &$\ll 1\%$   \\  \hline
$ccs$           &${\frac{1}{2}}^+$        &1.0               &$2.2-2.8$           &$4.35\pm0.10$                &$(64-87)\%$   &$\ll 1\%$    \\ \hline
$ccs$           &${\frac{3}{2}}^+$        &1.0               &$2.4-3.0$           &$4.45\pm0.10$                &$(65-87)\%$   &$\ll 1\%$   \\   \hline

$bbq$           &${\frac{1}{2}}^+$        &2.2               &$6.8-7.6$           &$10.75\pm0.10$               &$(55-73)\%$   &$\ll 1\%$   \\ \hline
$bbq$           &${\frac{3}{2}}^+$        &2.2               &$7.1-7.9$           &$10.80\pm0.10$               &$(55-73)\%$   &$\ll 1\%$   \\  \hline
$bbs$           &${\frac{1}{2}}^+$        &2.2               &$7.4-8.2$           &$10.90\pm0.10$               &$(55-72)\%$   &$\ll 1\%$    \\ \hline
$bbs$           &${\frac{3}{2}}^+$        &2.2               &$7.7-8.5$           &$10.95\pm0.10$               &$(55-72)\%$   &$\ll 1\%$   \\ \hline

$ccud\bar{q}$   &${\frac{1}{2}}^-$        &2.0               &$2.8-3.2$           &$4.80\pm0.10$                &$(41-64)\%$   &$(1-2)\%$  \\   \hline
$ccud\bar{q}$   &${\frac{3}{2}}^-$        &2.2               &$3.0-3.4$           &$4.90\pm0.10$                &$(41-63)\%$   &$\sim 1\%$  \\   \hline
$ccud\bar{q}$   &${\frac{5}{2}}^-$        &2.4               &$3.2-3.6$           &$5.00\pm0.10$                &$(40-61)\%$   &$\leq 1\%$  \\   \hline

$bbud\bar{q}$   &${\frac{1}{2}}^-$        &2.9               &$6.9-7.7$           &$11.40\pm0.10$               &$(40-60)\%$   &$(6-14)\%$ \\   \hline
$bbud\bar{q}$   &${\frac{3}{2}}^-$        &2.9               &$6.9-7.7$           &$11.40\pm0.10$               &$(40-60)\%$   &$(7-16)\%$ \\   \hline
$bbud\bar{q}$   &${\frac{5}{2}}^-$        &3.1               &$7.2-8.0$           &$11.50\pm0.10$               &$(41-60)\%$   &$(5-10)\%$ \\   \hline
  \hline
\end{tabular}
\end{center}
\caption{ The energy scales $\mu$, Borel parameters $T^2$, continuum threshold parameters $s_0$,
 pole contributions (pole) and contributions of the highest vacuum condensates  for the doubly heavy baryon states and  pentaquark states.}
\end{table}

\begin{table}
\begin{center}
\begin{tabular}{|c|c|c|c|c|c|c|c|}\hline\hline
                &$J^P$                    &$M(\rm GeV)$              &$\lambda (10^{-1}\rm{GeV}^3)$     &$\lambda (10^{-3}\rm{GeV}^6)$             \\  \hline
$ccq$           &${\frac{1}{2}}^+$        &$3.63^{+0.08}_{-0.07}$    &$1.02^{+0.14}_{-0.10}$            &  \\ \hline
$ccq$           &${\frac{3}{2}}^+$        &$3.75^{+0.07}_{-0.07}$    &$0.65^{+0.07}_{-0.07}$            &  \\ \hline
$ccs$           &${\frac{1}{2}}^+$        &$3.75^{+0.08}_{-0.09}$    &$1.28^{+0.18}_{-0.17}$            &  \\ \hline
$ccs$           &${\frac{3}{2}}^+$        &$3.85^{+0.08}_{-0.08}$    &$0.81^{+0.09}_{-0.09}$            &  \\ \hline

$bbq$           &${\frac{1}{2}}^+$        &$10.22^{+0.07}_{-0.07}$   &$2.73^{+0.36}_{-0.31}$            &  \\ \hline
$bbq$           &${\frac{3}{2}}^+$        &$10.27^{+0.07}_{-0.07}$   &$1.65^{+0.21}_{-0.19}$            &  \\ \hline
$bbs$           &${\frac{1}{2}}^+$        &$10.33^{+0.07}_{-0.08}$   &$3.27^{+0.44}_{-0.41}$            &  \\ \hline
$bbs$           &${\frac{3}{2}}^+$        &$10.37^{+0.07}_{-0.08}$   &$1.97^{+0.26}_{-0.23}$            &  \\ \hline

$ccud\bar{q}$   &${\frac{1}{2}}^-$        &$4.21^{+0.10}_{-0.11}$    &                                   &$2.51^{+0.46}_{-0.39}$   \\ \hline
$ccud\bar{q}$   &${\frac{3}{2}}^-$        &$4.27^{+0.11}_{-0.10}$    &                                   &$1.65^{+0.30}_{-0.25}$   \\ \hline
$ccud\bar{q}$   &${\frac{5}{2}}^-$        &$4.37^{+0.11}_{-0.11}$    &                                   &$1.34^{+0.22}_{-0.20}$   \\ \hline

$bbud\bar{q}$   &${\frac{1}{2}}^-$        &$10.75^{+0.12}_{-0.12}$   &                                   &$7.53^{+1.52}_{-1.39}$   \\ \hline
$bbud\bar{q}$   &${\frac{3}{2}}^-$        &$10.76^{+0.11}_{-0.13}$   &                                   &$4.27^{+0.85}_{-0.78}$   \\ \hline
$bbud\bar{q}$   &${\frac{5}{2}}^-$        &$10.82^{+0.12}_{-0.13}$   &                                   &$3.87^{+0.74}_{-0.68}$   \\ \hline

   \hline
\end{tabular}
\end{center}
\caption{ The masses and pole residues of the doubly heavy baryon states and pentaquark states.}
\end{table}

\section{Conclusion}
In this article, we  study  the doubly heavy baryon states and  pentaquark states
with the QCD sum rules by carrying out the operator product expansion   up to   the vacuum condensates of dimension $7$ and $13$ respectively  in a consistent way.
In calculations,   we separate  the contributions of the negative parity and positive parity hadron  states  unambiguously,
and  study the masses and pole residues of the doubly heavy baryon states and  pentaquark states
  in details, and obtain very stable QCD sum rules in the Borel windows. The prediction $M_{\Xi_{cc}}=3.63^{+0.08}_{-0.07}\,\rm{GeV}$ is in excellent agrement with the LHCb data  $M_{\Xi_{cc}^{++}}=3621.40 \pm 0.72 \pm 0.27 \pm 0.14\, \rm{MeV}$, other predictions can be confronted to the experimental data in the future.

\section*{Appendix}
The explicit expressions of the QCD spectral densities: \\ \\
{\bf  For the $\Omega_{QQ}$  states}, \\

\begin{eqnarray}
\rho_{0}^1(s)&=& \frac{3}{8\pi^4}\int dydz\, yz\left(1-y-z\right) \left(s-\overline{m}_Q^2\right)\left(5s-3\overline{m}_Q^2\right) \nonumber\\
&&+\frac{3m_Q^2 }{8\pi^4} \int dydz \left(1-y-z\right)\left(s-\overline{m}_Q^2\right) \, , \nonumber
\end{eqnarray}

\begin{eqnarray}
\rho_{3}^1(s)&=&\frac{3m_s\langle\bar{s}s\rangle}{2\pi^2} \int dy \, y\left(1-y\right)
\left[1+\frac{s}{2}\delta\left(s-\widetilde{m}_Q^2\right)\right] \, , \nonumber
\end{eqnarray}

\begin{eqnarray}
\rho_{4}^1(s)&=&-\frac{m_Q^2}{6\pi^2}\langle\frac{\alpha_{s}GG}{\pi}\rangle \int dydz\,
\frac{z\left(1-y-z\right)}{y^2}\left(1+\frac{s}{2T^2}\right)\delta\left(s-\overline{m}_Q^2\right) \nonumber\\
&&-\frac{m_Q^4}{24\pi^2T^2}\langle\frac{\alpha_{s}GG}{\pi}\rangle \int dydz\, \frac{1-y-z}{y^3}   \delta\left(s-\overline{m}_Q^2\right) \nonumber\\
&&+\frac{m_Q^2}{8\pi^2}\langle\frac{\alpha_{s}GG}{\pi}\rangle \int dydz\, \frac{1-y-z}{y^2}  \delta\left(s-\overline{m}_Q^2\right) \nonumber\\
&&+\frac{3}{16\pi^2}\langle\frac{\alpha_{s}GG}{\pi}\rangle  \int dydz\,z\, \left[1+\frac{s}{3}\delta\left(s-\overline{m}_Q^2\right)\right] \nonumber\\
&&+\frac{m_Q^2 }{16\pi^2}\langle\frac{\alpha_{s}GG}{\pi}\rangle  \int dydz\,\frac{1}{y}\,\delta\left(s-\overline{m}_Q^2\right)\, , \nonumber
\end{eqnarray}

\begin{eqnarray}
\rho_{5}^1(s)&=&-\frac{m_s\langle\bar{s}g_{s}\sigma Gs\rangle}{2\pi^2} \int dy\, y\left(1-y\right)
\left(1+\frac{3s}{4 T^2}+\frac{s^2}{4T^4}\right)\delta\left(s-\widetilde{m}_Q^2\right) \, , \nonumber
\end{eqnarray}

\begin{eqnarray}
\rho_{7}^1(s)&=& -\frac{m_s m_Q^2\langle\bar{s}s\rangle }{18T^6} \langle\frac{\alpha_{s}GG}{\pi}\rangle \int dy\,\frac{1-y}{y^2}\,s\,\delta\left(s-\widetilde{m}_Q^2\right) \nonumber\\
&& -\frac{m_s m_Q^4\langle\bar{s}s\rangle}{36T^6}\langle\frac{\alpha_{s}GG}{\pi}\rangle\int dy\,\frac{1}{y^3}\,\delta\left(s-\widetilde{m}_Q^2\right) \nonumber\\
&&+\frac{m_s m_Q^2\langle\bar{s}s\rangle}{12T^4} \langle\frac{\alpha_{s}GG}{\pi}\rangle \int dy\,\frac{1}{y^2}\, \delta\left(s-\widetilde{m}_Q^2\right) \, ,\nonumber
\end{eqnarray}

\begin{eqnarray}
\rho_{0}^0(s)&=&\frac{3m_s}{8\pi^4}\int dydz\,  yz\, \left(s-\overline{m}_Q^2\right) \left(2s-\overline{m}_Q^2\right) +  \frac{3m_s m_Q^2}{4\pi^4}\int dydz\, \left(s-\overline{m}_Q^2\right) \, ,\nonumber
\end{eqnarray}

\begin{eqnarray}
\rho_{3}^0(s)&=&-\frac{3\langle\bar{s}s\rangle}{2\pi^2} \int dy \,  y\left(1-y\right) \,s  \, ,\nonumber
\end{eqnarray}

\begin{eqnarray}
\rho_{4}^0(s)&=&-\frac{m_s m_Q^2}{24\pi^2}\langle\frac{\alpha_{s}GG}{\pi}\rangle \int dydz\,
\frac{z}{y^2}\left(1+\frac{s}{T^2}\right)\delta\left(s-\overline{m}_Q^2\right) \nonumber\\
&&-\frac{m_s m_Q^4}{12\pi^2T^2}\langle\frac{\alpha_{s}GG}{\pi}\rangle \int dydz\,\frac{1}{y^3}\delta\left(s-\overline{m}_Q^2\right) \nonumber\\
&&+\frac{m_s m_Q^2}{4\pi^2}\langle\frac{\alpha_{s}GG}{\pi}\rangle \int dydz\,\frac{1}{y^2}\delta\left(s-\overline{m}_Q^2\right) \nonumber\\
&&-\frac{m_s}{16\pi^2}\langle\frac{\alpha_{s}GG}{\pi}\rangle  \int dydz\,\left[1+\frac{s}{2}\,\delta\left(s-\overline{m}_Q^2\right)\right] \, ,\nonumber
\end{eqnarray}

\begin{eqnarray}
\rho_{5}^0(s)&=&\frac{3\langle\bar{s}g_{s}\sigma Gs\rangle}{4\pi^2} \int dy \,y\left(1-y\right)
\left[1+\left(s+\frac{s^2}{2T^2}\right)\delta\left(s-\widetilde{m}_Q^2\right)\right]\nonumber\\
&& -\frac{\langle\bar{s}g_{s}\sigma Gs\rangle}{8\pi^2} \int dy \,\left[1+\frac{3s}{2}\delta\left(s-\widetilde{m}_Q^2\right)\right] \, ,\nonumber
\end{eqnarray}

\begin{eqnarray}
\rho_{7}^0(s)&=& \frac{m_Q^2\langle\bar{s}s\rangle}{18T^4} \langle\frac{\alpha_{s}GG}{\pi}\rangle \int dy \, \frac{1-y}{y^2}\, s\,\delta\left(s-\widetilde{m}_Q^2\right) \nonumber\\
&&+ \frac{m_Q^4\langle\bar{s}s\rangle}{9T^4} \langle\frac{\alpha_{s}GG}{\pi}\rangle \int dy \, \frac{1}{y^3}  \delta\left(s-\widetilde{m}_Q^2\right) \nonumber\\
&& -\frac{m_Q^2\langle\bar{s}s\rangle}{3T^2}\langle\frac{\alpha_{s}GG}{\pi}\rangle \int dy \,\frac{1}{y^2}\delta\left(s-\widetilde{m}_Q^2\right) \nonumber\\
&&+ \frac{\langle\bar{s}s\rangle}{24}\langle\frac{\alpha_{s}GG}{\pi}\rangle \int dy \,\left(1+\frac{s}{T^2}\right)\delta\left(s-\widetilde{m}_Q^2\right)\nonumber\\
&& -\frac{\langle\bar{s}s\rangle}{12}\langle\frac{\alpha_{s}GG}{\pi}\rangle \int dy \,
y\left(1-y\right)\left(1+\frac{s}{T^2}+\frac{s^2}{2T^4}+\frac{s^3}{2T^6}\right)\delta\left(s-\widetilde{m}_Q^2\right) \, . \nonumber
\end{eqnarray}

{\bf  For the $\Omega_{QQ}^*$  states}, \\

\begin{eqnarray}
\rho_{0}^1(s)&=&\frac{3}{16\pi^4}\int dydz\, yz \left(1-y-z\right)\left(s-\overline{m}_Q^2\right)\left(2s-\overline{m}_Q^2\right)  \nonumber\\
&&+\frac{3m_Q^2 }{16\pi^4}\int dydz\,\left(1-y-z\right)\left(s-\overline{m}_Q^2\right) \, ,\nonumber
\end{eqnarray}

\begin{eqnarray}
\rho_{3}^1(s)&=& \frac{m_s\langle\bar{s}s\rangle}{4\pi^2} \int dy\,y\left(1-y\right)
\Big[1+s\,\delta\left(s-\widetilde{m}_Q^2\right)\Big] \, , \nonumber
\end{eqnarray}

\begin{eqnarray}
\rho_{4}^1(s)&=&-\frac{m_Q^2}{48\pi^2}\langle\frac{\alpha_{s}GG}{\pi}\rangle \int dydz\,
\frac{z\left(1-y-z\right)}{y^2}\left(1+\frac{s}{T^2}\right)\delta\left(s-\overline{m}_Q^2\right) \nonumber\\
&&-\frac{m_Q^4}{48\pi^2T^2}\langle\frac{\alpha_{s}GG}{\pi}\rangle \int dydz\,\frac{1-y-z}{y^3}\delta\left(s-\overline{m}_Q^2\right) \nonumber\\
&&+\frac{m_Q^2}{16\pi^2}\langle\frac{\alpha_{s}GG}{\pi}\rangle \int dydz\,\frac{1-y-z}{y^2}\delta\left(s-\overline{m}_Q^2\right) \nonumber\\
&&-\frac{1}{48\pi^2}\langle\frac{\alpha_{s}GG}{\pi}\rangle \int dydz\,\left(1-y-z\right)\left[1+\frac{s}{4}\delta\left(s-\overline{m}_Q^2\right)\right] \, ,\nonumber
\end{eqnarray}

\begin{eqnarray}
\rho_{5}^1(s)&=&-\frac{ m_s\langle\bar{s}g_{s}\sigma Gs\rangle}{16\pi^2} \int dy\, y\left(1-y\right)
\left(1+\frac{4s}{3}+\frac{2s^2}{3T^4}\right)\delta\left(s-\widetilde{m}_Q^2\right) \, , \nonumber
\end{eqnarray}

\begin{eqnarray}
\rho_{7}^1(s)&=& \frac{m_s m_Q^2\langle\bar{s}s\rangle}{72T^4}\langle\frac{\alpha_{s}GG}{\pi}\rangle \int dy\,
\frac{1-y}{y^2}\left(1-\frac{s}{T^2}\right)\delta\left(s-\widetilde{m}_Q^2\right) \nonumber\\
&& -\frac{m_s m_Q^4\langle\bar{s}s\rangle}{72T^6}\langle\frac{\alpha_{s}GG}{\pi}\rangle \int dy\,\frac{1}{y^3}\delta\left(s-\widetilde{m}_Q^2\right) \nonumber\\
&&+\frac{m_s m_Q^2\langle\bar{s}s\rangle}{24T^4} \langle\frac{\alpha_{s}GG}{\pi}\rangle \int dy\,\frac{1}{y^2}\delta\left(s-\widetilde{m}_Q^2\right)\nonumber\\
&& -\frac{m_s \langle\bar{s}s\rangle}{144T^2} \langle\frac{\alpha_{s}GG}{\pi}\rangle \int dy\,\left(1+\frac{s}{2T^2}\right)\delta\left(s-\widetilde{m}_Q^2\right)\, ,\nonumber
\end{eqnarray}

\begin{eqnarray}
\rho_{0}^0(s)&=&\frac{3m_s}{32\pi^4}\int dydz\, yz \left(s-\overline{m}_Q^2\right)\left(3s-\overline{m}_Q^2\right)  + \frac{3m_s m_Q^2}{16\pi^4} \int dydz\,\left(s-\overline{m}_Q^2\right) \, , \nonumber
\end{eqnarray}

\begin{eqnarray}
\rho_{3}^0(s)&=& -\frac{\langle\bar{s}s\rangle}{2\pi^2} \int dy\,y\left(1-y\right)\,s\, ,\nonumber
\end{eqnarray}

\begin{eqnarray}
\rho_{4}^0(s)&=&-\frac{m_s m_Q^2}{48\pi^2T^2}\langle\frac{\alpha_{s}GG}{\pi}\rangle \int dydz\,\frac{z}{y^2}\,s\,\delta\left(s-\overline{m}_Q^2\right) \nonumber\\
&&-\frac{m_s m_Q^4}{48\pi^2T^2}\langle\frac{\alpha_{s}GG}{\pi}\rangle \int dydz\,\frac{1}{y^3}\delta\left(s-\overline{m}_Q^2\right) \nonumber\\
&&+\frac{m_s m_Q^2}{16\pi^2}\langle\frac{\alpha_{s}GG}{\pi}\rangle \int dydz\,\frac{1}{y^2}\delta\left(s-\overline{m}_Q^2\right) \nonumber\\
&&-\frac{m_s}{64\pi^2}\langle\frac{\alpha_{s}GG}{\pi}\rangle \int dydz\,\left[1+\frac{s}{3}\delta\left(s-\overline{m}_Q^2\right)\right]\, ,\nonumber
\end{eqnarray}

\begin{eqnarray}
\rho_{5}^0(s)&=& \frac{\langle\bar{s}g_{s}\sigma Gs\rangle}{16\pi^2} \int dy\,y\left(1-y\right)
\left[3+\left(4s+\frac{2s^2}{T^2}\right)\delta\left(s-\widetilde{m}_Q^2\right)\right]\, ,\nonumber
\end{eqnarray}

\begin{eqnarray}
\rho_{7}^0(s)&=&-\frac{m_Q^2\langle\bar{s}s\rangle}{36T^2} \langle\frac{\alpha_{s}GG}{\pi}\rangle \int dy\,
 \frac{1-y}{y^2}\left(1-\frac{s}{T^2}\right)\delta\left(s-\widetilde{m}_Q^2\right) \nonumber\\
&&+ \frac{m_Q^4\langle\bar{s}s\rangle}{36T^4} \langle\frac{\alpha_{s}GG}{\pi}\rangle \int dy\,\frac{1}{y^3}\delta\left(s-\widetilde{m}_Q^2\right) \nonumber\\
&& -\frac{m_Q^2\langle\bar{s}s\rangle}{12T^2}\langle\frac{\alpha_{s}GG}{\pi}\rangle \int dy\,\frac{1}{y^2}\delta\left(s-\widetilde{m}_Q^2\right)\nonumber\\
&&+ \frac{\langle\bar{s}s\rangle}{72} \langle\frac{\alpha_{s}GG}{\pi}\rangle \int dy\,\left(1+\frac{s}{2T^2}\right)\delta\left(s-\widetilde{m}_Q^2\right)\nonumber\\
&&-\frac{\langle\bar{s}s\rangle}{72T^2} \langle\frac{\alpha_{s}GG}{\pi}\rangle \int dy\,
y\left(1-y\right)\left(s+\frac{s^2}{T^2}+\frac{s^3}{T^4}\right)\delta\left(s-\widetilde{m}_Q^2\right)\, . \nonumber
\end{eqnarray}

{\bf  For the $QQcd\bar{q}$  states with $J^P={\frac{1}{2}}^-$}, \\

\begin{eqnarray}
\rho_{0}^1(s)&=&\frac{1}{61440\pi^8}\int dydz\,yz\left(1-y-z\right)^4 \left(s-\overline{m}_Q^2\right)^4\left(8s-3\overline{m}_Q^2\right) \nonumber\\
&&+ \frac{m_Q^2}{24576\pi^8} \int dydz\,\left(1-y-z\right)^4\left(s-\overline{m}_Q^2\right)^4 \, ,\nonumber
\end{eqnarray}

\begin{eqnarray}
\rho_{4}^1(s)&=&-\frac{ m_Q^2}{9216\pi^6}\langle\frac{\alpha_{s}GG}{\pi}\rangle \int dydz\, \frac{z\left(1-y-z\right)^4}{y^2}\left(s-\overline{m}_Q^2\right)\left(5s-3\overline{m}_Q^2\right) \nonumber\\
&&-\frac{m_Q^4}{9216\pi^6}\langle\frac{\alpha_{s}GG}{\pi}\rangle \int dydz\,\frac{\left(1-y-z\right)^4}{y^3}\left(s-\overline{m}_Q^2\right)  \nonumber\\
&&+\frac{m_Q^2}{6144\pi^6}\langle\frac{\alpha_{s}GG}{\pi}\rangle \int dydz\,\frac{\left(1-y-z\right)^4}{y^2}\left(s-\overline{m}_Q^2\right)^2 \nonumber\\
&&+\frac{1}{1024\pi^6}\langle\frac{\alpha_{s}GG}{\pi}\rangle \int dydz\, yz\left(1-y-z\right)^2\left(s-\overline{m}_Q^2\right)^2\left(2s-\overline{m}_Q^2\right)  \nonumber\\
&&+\frac{m_Q^2}{2048\pi^6} \langle\frac{\alpha_{s}GG}{\pi}\rangle \int dydz\, \left(1-y-z\right)^2\left(s-\overline{m}_Q^2\right)^2 \nonumber\\
&&+\frac{1}{6144\pi^6}\langle\frac{\alpha_{s}GG}{\pi}\rangle \int dydz\,
z\left(1-y-z\right)^3\left(s-\overline{m}_Q^2\right)^2\left(2s-\overline{m}_Q^2\right)   \nonumber\\
&&+\frac{m_Q^2}{6144\pi^6}\langle\frac{\alpha_{s}GG}{\pi}\rangle \int dydz\,\frac{\left(1-y-z\right)^3}{y}\left(s-\overline{m}_Q^2\right)^2 \, ,\nonumber
\end{eqnarray}

\begin{eqnarray}
\rho_{6}^1(s)&=&\frac{\langle\bar{q}q\rangle^2}{24\pi^4} \int dydz\,yz\left(1-y-z\right)\left(s-\overline{m}_Q^2\right)\left(5s-3\overline{m}_Q^2\right) \nonumber\\
&&+  \frac{m_Q^2\langle\bar{q}q\rangle^2}{24\pi^4}  \int dydz\,
\left(1-y-z\right)\left(s-\overline{m}_Q^2\right)\, , \nonumber
\end{eqnarray}

\begin{eqnarray}
\rho_{8}^1(s)&=& -\frac{\langle\bar{q}q\rangle \langle\bar{q}g_{s}\sigma Gq\rangle}{24\pi^4} \int dydz\,yz\left(4s-3\overline{m}_Q^2\right)
-\frac{m_Q^2\langle\bar{q}q\rangle\langle\bar{q}g_{s}\sigma Gq\rangle }{48\pi^4}  \int dydz\,  , \nonumber
\end{eqnarray}

\begin{eqnarray}
\rho_{10}^1(s)&=&\left[\frac{\langle\bar{q}g_{s}\sigma Gq\rangle^2}{64\pi^4}+\frac{\langle\bar{q}q\rangle^2}{72\pi^2}\langle\frac{\alpha_{s}GG}{\pi}\rangle\right] \int dy\,y\left(1-y\right)\left[1+\frac{s}{2}\delta\left(s-\widetilde{m}_Q^2\right)\right]\nonumber\\
&&-\frac{m_Q^2\langle\bar{q}q\rangle^2}{54\pi^2}  \langle\frac{\alpha_{s}GG}{\pi}\rangle \int dydz\,
\frac{z\left(1-y-z\right)}{y^2}\left(1+\frac{s}{2T^2}\right)\delta\left(s-\overline{m}_Q^2\right)  \nonumber\\
&& -\frac{m_Q^4\langle\bar{q}q\rangle^2}{216\pi^2T^2} \langle\frac{\alpha_{s}GG}{\pi}\rangle \int dydz\,\frac{1-y-z}{y^3}\delta\left(s-\overline{m}_Q^2\right)  \nonumber\\
&&+ \frac{m_Q^2\langle\bar{q}q\rangle^2}{72\pi^2}\langle\frac{\alpha_{s}GG}{\pi}\rangle \int dydz\,\frac{1-y-z}{y^2}\delta\left(s-\overline{m}_Q^2\right)  \nonumber\\
&&+\frac{1}{96\pi^2}\left[\langle\bar{q}q\rangle^2 \langle\frac{\alpha_{s}GG}{\pi}\rangle+\frac{11  \langle\bar{q}g_{s}\sigma Gq\rangle^2}{64\pi^2}\right]\int dydz\,z\, \left[1+\frac{s}{3}\delta\left(s-\overline{m}_Q^2\right)\right] \nonumber\\
&&+ \frac{m_Q^2}{288\pi^2}\left[\langle\bar{q}q\rangle^2 \langle\frac{\alpha_{s}GG}{\pi}\rangle+\frac{11  \langle\bar{q}g_{s}\sigma Gq\rangle^2}{64\pi^2}\right]\int dydz\,\frac{1}{y}\delta\left(s-\overline{m}_Q^2\right) \, , \nonumber
\end{eqnarray}

\begin{eqnarray}
\rho_{3}^0(s)&=&-\frac{\langle\bar{q}q\rangle}{768\pi^6} \int dydz\,yz\left(1-y-z\right)^2\left(s-\overline{m}_Q^2\right)^3\left(3s-\overline{m}_Q^2\right) \nonumber\\
&& -\frac{m_Q^2 \langle\bar{q}q\rangle}{192\pi^6} \int dydz\,\left(1-y-z\right)^2\left(s-\overline{m}_Q^2\right)^3 \, ,\nonumber
\end{eqnarray}

\begin{eqnarray}
\rho_{5}^0(s)&=&\frac{\langle\bar{q}g_{s}\sigma Gq\rangle}{768\pi^6} \int dydz\, yz\left(1-y-z\right)\left(s-\overline{m}_Q^2\right)^2\left(5s-2\overline{m}_Q^2\right)\nonumber\\
&&-\frac{\langle\bar{q}g_{s}\sigma Gq\rangle}{3072\pi^6} \int dydz\, z\left(1-y-z\right)^2\left(s-\overline{m}_Q^2\right)^2\left(5s-2\overline{m}_Q^2\right) \nonumber\\
&&+ \frac{m_Q^2\langle\bar{q}g_{s}\sigma Gq\rangle}{128\pi^6} \int dydz\,\left(1-y-z\right)\left(s-\overline{m}_Q^2\right)^2 \nonumber\\
&& -\frac{m_Q^2 \langle\bar{q}g_{s}\sigma Gq\rangle}{512\pi^6} \int dydz\,\frac{\left(1-y-z\right)^2}{y}\left(s-\overline{m}_Q^2\right)^2 \, ,\nonumber
\end{eqnarray}

\begin{eqnarray}
\rho_{7}^0(s)&=&\frac{m_Q^2\langle\bar{q}q\rangle}{576\pi^4}  \langle\frac{\alpha_{s}GG}{\pi}\rangle  \int dydz\, \frac{z\left(1-y-z\right)^2}{y^2}\left(3s-2\overline{m}_Q^2\right) \nonumber\\
&&+ \frac{m_Q^4\langle\bar{q}q\rangle}{288\pi^4} \langle\frac{\alpha_{s}GG}{\pi}\rangle  \int dydz\,\frac{\left(1-y-z\right)^2}{y^3}  \nonumber\\
&& -\frac{m_Q^2\langle\bar{q}q\rangle}{96\pi^4} \langle\frac{\alpha_{s}GG}{\pi}\rangle  \int dydz\,\frac{\left(1-y-z\right)^2}{y^2}\left(s-\overline{m}_Q^2\right) \nonumber\\
&&+\frac{\langle\bar{q}q\rangle }{768\pi^4}\langle\frac{\alpha_{s}GG}{\pi}\rangle \int dydz\, \left(1-y-z\right)^2\left(s-\overline{m}_Q^2\right)\left(2s-\overline{m}_Q^2\right)  \nonumber\\
&&-\frac{\langle\bar{q}q\rangle }{288\pi^4}\langle\frac{\alpha_{s}GG}{\pi}\rangle \int dydz\, yz\left(s-\overline{m}_Q^2\right)\left(2s-\overline{m}_Q^2\right)  \nonumber\\
&&-\frac{m_Q^2\langle\bar{q}q\rangle }{144\pi^4}\langle\frac{\alpha_{s}GG}{\pi}\rangle \int dydz\, \left(s-\overline{m}_Q^2\right) \, ,\nonumber
\end{eqnarray}

\begin{eqnarray}
\rho_{9}^0(s)&=&-\frac{\langle\bar{q}q\rangle^3 }{6\pi^2} \int dy\,y\left(1-y\right)\,s  \, ,\nonumber
\end{eqnarray}

\begin{eqnarray}
\rho_{11}^0(s)&=& \frac{\langle\bar{q}q\rangle^2 \langle\bar{q}g_{s}\sigma Gq\rangle}{4\pi^2}\int dy\,y\left(1-y\right)\left[1
+\left(s+\frac{s^2}{2T^2}\right)\delta\left(s-\widetilde{m}_Q^2\right)\right] \nonumber\\
&& -\frac{\langle\bar{q}q\rangle^2 \langle\bar{q}g_{s}\sigma Gq\rangle}{144\pi^2}\int dy\,\left[1+\frac{3s}{2}\delta\left(s-\widetilde{m}_Q^2\right)\right] \, ,\nonumber
\end{eqnarray}

\begin{eqnarray}
\rho_{13}^0(s)&=& -\frac{\langle\bar{q}q\rangle \langle\bar{q}g_{s}\sigma Gq\rangle^2}{16\pi^2} \int dy\,
y\left(1-y\right)\left(1+\frac{s}{T^2}+\frac{s^2}{2T^4}+\frac{s^3}{2T^6}\right)\delta\left(s-\widetilde{m}_Q^2\right) \nonumber\\
&&+\frac{m_Q^2\langle\bar{q}q\rangle^3}{162T^4} \langle\frac{\alpha_{s}GG}{\pi}\rangle \int dy\,\frac{1-y}{y^2}\, s\, \delta\left(s-\widetilde{m}_Q^2\right)  \nonumber\\
&&+\frac{m_Q^4\langle\bar{q}q\rangle^3}{81T^4} \langle\frac{\alpha_{s}GG}{\pi}\rangle \int dy\,\frac{1}{y^3}\delta\left(s-\widetilde{m}_Q^2\right)  \nonumber\\
&&-\frac{m_Q^2\langle\bar{q}q\rangle^3}{27T^2} \langle\frac{\alpha_{s}GG}{\pi}\rangle \int dy\,\frac{1}{y^2}\delta\left(s-\widetilde{m}_Q^2\right)\nonumber\\
&&+\frac{\langle\bar{q}q\rangle}{216}\left[\langle\bar{q}q\rangle^2 \langle\frac{\alpha_{s}GG}{\pi}\rangle  -\frac{ \langle\bar{q}g_{s}\sigma Gq\rangle^2}{16\pi^2}\right]\int dy\,\left(1+\frac{s}{T^2}\right)\delta\left(s-\widetilde{m}_Q^2\right)\nonumber\\
&&+ \frac{\langle\bar{q}q\rangle \langle\bar{q}g_{s}\sigma Gq\rangle^2}{288\pi^2}\int dy\, \left(1+\frac{s}{T^2}+\frac{3s^2}{2T^4}\right)\delta\left(s-\widetilde{m}_Q^2\right) \nonumber\\
&&-\frac{\langle\bar{q}q\rangle^3}{36} \langle\frac{\alpha_{s}GG}{\pi}\rangle \int dy\,
y\left(1-y\right)\left(1+\frac{s}{T^2}+\frac{s^2}{2T^4}+\frac{s^3}{2T^6}\right)\delta\left(s-\widetilde{m}_Q^2\right) \, .\nonumber
\end{eqnarray}

{\bf  For the $QQcd\bar{q}$  states with $J^P={\frac{3}{2}}^-$}, \\

\begin{eqnarray}
\rho_{0}^1(s)&=&\frac{1}{245760\pi^8}\int dydz\,yz\left(1-y-z\right)^4\left(s-\overline{m}_Q^2\right)^4\left(7s-2\overline{m}_Q^2\right)  \nonumber\\
&&+ \frac{m_Q^2}{49152\pi^8}\int dydz\,\left(1-y-z\right)^4\left(s-\overline{m}_Q^2\right)^4 \, ,\nonumber
\end{eqnarray}

\begin{eqnarray}
\rho_{4}^1(s)&=&-\frac{m_Q^2}{18432\pi^6}\langle\frac{\alpha_{s}GG}{\pi}\rangle \int dydz\, \frac{z\left(1-y-z\right)^4}{y^2} \left(s-\overline{m}_Q^2\right)\left(2s-\overline{m}_Q^2\right)  \nonumber\\
&&-\frac{m_Q^4}{18432\pi^6}\langle\frac{\alpha_{s}GG}{\pi}\rangle \int dydz\,\frac{\left(1-y-z\right)^4}{y^3}\left(s-\overline{m}_Q^2\right)  \nonumber\\
&&+\frac{m_Q^2}{12288\pi^6}\langle\frac{\alpha_{s}GG}{\pi}\rangle \int dydz\,\frac{\left(1-y-z\right)^4}{y^2}\left(s-\overline{m}_Q^2\right)^2 \nonumber\\
&&+\frac{1}{12288\pi^6}\langle\frac{\alpha_{s}GG}{\pi}\rangle \int dydz\,yz\left(1-y-z\right)^2\left(s-\overline{m}_Q^2\right)^2\left(5s-2\overline{m}_Q^2\right) \nonumber\\
&&+ \frac{m_Q^2}{4096\pi^6}\langle\frac{\alpha_{s}GG}{\pi}\rangle \int dydz\,\left(1-y-z\right)^2\left(s-\overline{m}_Q^2\right)^2  \nonumber\\
&&-\frac{1}{442368\pi^6}\langle\frac{\alpha_{s}GG}{\pi}\rangle \int dydz\,\left(1-y-z\right)^4\left(s-\overline{m}_Q^2\right)^2\left(7s-4\overline{m}_Q^2\right) \, ,\nonumber
\end{eqnarray}

\begin{eqnarray}
\rho_{6}^1(s)&=&\frac{\langle\bar{q}q\rangle^2}{48\pi^4} \int dydz\, yz\left(1-y-z\right)
\left(s-\overline{m}_Q^2\right)\left(2s-\overline{m}_Q^2\right)\nonumber\\
&&+ \frac{m_Q^2\langle\bar{q}q\rangle^2}{48\pi^4}  \int dydz\,\left(1-y-z\right)\left(s-\overline{m}_Q^2\right)\, ,\nonumber
\end{eqnarray}

\begin{eqnarray}
\rho_{8}^1(s)&=& -\frac{\langle\bar{q}q\rangle \langle\bar{q}g_{s}\sigma Gq\rangle}{96\pi^4}\int dydz\,
yz\left(3s-2\overline{m}_Q^2\right)   -\frac{m_Q^2\langle\bar{q}q\rangle \langle\bar{q}g_{s}\sigma Gq\rangle}{96\pi^4} \int dydz\, ,\nonumber
\end{eqnarray}

\begin{eqnarray}
\rho_{10}^1(s)&=& \left[\frac{\langle\bar{q}g_{s}\sigma Gq\rangle^2}{384\pi^4} +\frac{\langle\bar{q}q\rangle^2}{432\pi^2}\langle\frac{\alpha_{s}GG}{\pi}\rangle\right] \int dy\, y\left(1-y\right) \Big[1+s\,\delta\left(s-\widetilde{m}_Q^2\right)\Big] \nonumber\\
&&-\frac{m_Q^2\langle\bar{q}q\rangle^2}{432\pi^2} \langle\frac{\alpha_{s}GG}{\pi}\rangle \int dydz\,
\frac{z\left(1-y-z\right)}{y^2}
\left(1+\frac{s}{T^2}\right)\delta\left(s-\overline{m}_Q^2\right)  \nonumber\\
&& -\frac{m_Q^4\langle\bar{q}q\rangle^2}{432\pi^2T^2}\langle\frac{\alpha_{s}GG}{\pi}\rangle \int dydz\,\frac{1-y-z}{y^3}\delta\left(s-\overline{m}_Q^2\right)  \nonumber\\
&&+ \frac{m_Q^2\langle\bar{q}q\rangle^2}{144\pi^2}\langle\frac{\alpha_{s}GG}{\pi}\rangle \int dydz\, \frac{1-y-z}{y^2}\delta\left(s-\overline{m}_Q^2\right)  \nonumber\\
&&-\frac{\langle\bar{q}q\rangle^2}{432\pi^2} \langle\frac{\alpha_{s}GG}{\pi}\rangle \int dydz\, \left(1-y-z\right)\left[1+\frac{s}{4}\delta\left(s-\overline{m}_Q^2\right)\right] \nonumber\\
&&+\frac{\langle\bar{q}g_{s}\sigma Gq\rangle^2}{6912\pi^4} \int dydz\,
\left(1-y-z\right)\left[1+\frac{s}{4}\delta\left(s-\overline{m}_Q^2\right)\right]  \, , \nonumber
\end{eqnarray}

\begin{eqnarray}
\rho_{3}^0(s)&=&-\frac{\langle\bar{q}q\rangle}{3072\pi^6} \int dydz\,yz\left(1-y-z\right)^2\left(s-\overline{m}_Q^2\right)^3\left(5s-\overline{m}_Q^2\right) \nonumber\\
&& -\frac{m_Q^2 \langle\bar{q}q\rangle}{768\pi^6}\int dydz\,\left(1-y-z\right)^2\left(s-\overline{m}_Q^2\right)^3 \, ,\nonumber
\end{eqnarray}

\begin{eqnarray}
\rho_{5}^0(s)&=&\frac{\langle\bar{q}g_{s}\sigma Gq\rangle}{1536\pi^6} \int dydz\,yz\left(1-y-z\right)\left(s-\overline{m}_Q^2\right)^2\left(4s-\overline{m}_Q^2\right) \nonumber\\
&&+ \frac{m_Q^2\langle\bar{q}g_{s}\sigma Gq\rangle}{512\pi^6} \int dydz\,\left(1-y-z\right)\left(s-\overline{m}_Q^2\right)^2 \, ,\nonumber
\end{eqnarray}

\begin{eqnarray}
\rho_{7}^0(s)&=& \frac{m_Q^2\langle\bar{q}q\rangle}{1152\pi^4} \langle\frac{\alpha_{s}GG}{\pi}\rangle \int dydz\,
\frac{z\left(1-y-z\right)^2}{y^2}\left(2s-\overline{m}_Q^2\right) \nonumber\\
&&+ \frac{m_Q^4\langle\bar{q}q\rangle}{1152\pi^4} \langle\frac{\alpha_{s}GG}{\pi}\rangle \int dydz\, \frac{\left(1-y-z\right)^2}{y^3} \nonumber\\
&& -\frac{m_Q^2\langle\bar{q}q\rangle}{384\pi^4}\langle\frac{\alpha_{s}GG}{\pi}\rangle \int dydz\,\frac{\left(1-y-z\right)^2}{y^2}\left(s-\overline{m}_Q^2\right) \nonumber\\
&&+\frac{\langle\bar{q}q\rangle}{9216\pi^4} \langle\frac{\alpha_{s}GG}{\pi}\rangle \int dydz\,
\left(1-y-z\right)^2\left(s-\overline{m}_Q^2\right)\left(5s-3\overline{m}_Q^2\right)\nonumber\\
&& -\frac{\langle\bar{q}q\rangle}{1152\pi^4}\langle\frac{\alpha_{s}GG}{\pi}\rangle \int dydz\,yz\left(s-\overline{m}_Q^2\right)\left(3s-\overline{m}_Q^2\right) \nonumber\\
&& -\frac{m_Q^2\langle\bar{q}q\rangle}{576\pi^4} \langle\frac{\alpha_{s}GG}{\pi}\rangle \int dydz\,\left(s-\overline{m}_Q^2\right) \, ,\nonumber
\end{eqnarray}

\begin{eqnarray}
\rho_{9}^0(s)&=&  -\frac{\langle\bar{q}q\rangle^3}{18\pi^2}\int dy\, y\left(1-y\right)\, s\, ,\nonumber
\end{eqnarray}

\begin{eqnarray}
\rho_{11}^0(s)&=&\frac{\langle\bar{q}q\rangle^2 \langle\bar{q}g_{s}\sigma Gq\rangle}{16\pi^2} \int dy\,
y\left(1-y\right)\left[1+\left(\frac{4s}{3}+\frac{2s^2}{3T^2}\right)\delta\left(s-\widetilde{m}_Q^2\right)\right]\, ,\nonumber
\end{eqnarray}

\begin{eqnarray}
\rho_{13}^0(s)&=& -\frac{ \langle\bar{q}q\rangle\langle\bar{q}g_{s}\sigma Gq\rangle^2}{96\pi^2T^2}\int dy\, y\left(1-y\right)\left(s+\frac{s^2}{T^2}+\frac{s^3}{T^4}\right)\delta\left(s-\widetilde{m}_Q^2\right) \nonumber\\
&& -\frac{m_Q^2\langle\bar{q}q\rangle^3}{324T^2}\langle\frac{\alpha_{s}GG}{\pi}\rangle \int dy\,
\frac{1-y}{y^2}\left(1-\frac{s}{T^2}\right)\delta\left(s-\widetilde{m}_Q^2\right)  \nonumber\\
&&+ \frac{m_Q^4\langle\bar{q}q\rangle^3}{324T^4}\langle\frac{\alpha_{s}GG}{\pi}\rangle \int dy\,\frac{1}{y^3}\delta\left(s-\widetilde{m}_Q^2\right) \nonumber\\
&&-\frac{m_Q^2\langle\bar{q}q\rangle^3}{108T^2} \langle\frac{\alpha_{s}GG}{\pi}\rangle \int dy\,\frac{1}{y^2}\delta\left(s-\widetilde{m}_Q^2\right)\nonumber\\
&&+ \frac{\langle\bar{q}q\rangle}{648}\left[\langle\bar{q}q\rangle^2 \langle\frac{\alpha_{s}GG}{\pi}\rangle  -\frac{ \langle\bar{q}g_{s}\sigma Gq\rangle^2}{16\pi^2}\right]\int dy\, \left(1+\frac{s}{2T^2}\right)\delta\left(s-\widetilde{m}_Q^2\right) \nonumber \\
&&-\frac{\langle\bar{q}q\rangle^3}{216T^2}\langle\frac{\alpha_{s}GG}{\pi}\rangle\int dy\, y\left(1-y\right)\left(s+\frac{s^2}{T^2}+\frac{s^3}{T^4}\right)\delta\left(s-\widetilde{m}_Q^2\right) \, .\nonumber
\end{eqnarray}

{\bf  For the $QQcd\bar{q}$  states with $J^P={\frac{5}{2}}^-$}, \\

\begin{eqnarray}
\rho_{0}^1(s)&=&\frac{1}{2457600\pi^8}\int dydz\,yz\left(1-y-z\right)^4 \left(4+y+z\right)\left(s-\overline{m}_Q^2\right)^4 \left(7s-2\overline{m}_Q^2\right) \nonumber\\
&&+\frac{m_Q^2}{491520\pi^8}\int dydz\,\left(1-y-z\right)^4\left(4+y+z\right)\left(s-\overline{m}_Q^2\right)^4 \, ,\nonumber
\end{eqnarray}

\begin{eqnarray}
\rho_{4}^1(s)&=&-\frac{m_Q^2}{184320\pi^6}\langle\frac{\alpha_{s}GG}{\pi}\rangle \int dydz\,
\frac{z\left(1-y-z\right)^4 \left(4+y+z\right)}{y^2}\left(s-\overline{m}_Q^2\right)\left(2s-\overline{m}_Q^2\right)  \nonumber\\
&&-\frac{m_Q^4}{184320\pi^6} \langle\frac{\alpha_{s}GG}{\pi}\rangle \int dydz\,\frac{\left(1-y-z\right)^4\left(4+y+z\right)}{y^3}\left(s-\overline{m}_Q^2\right) \nonumber\\
&&+\frac{m_Q^2}{122880\pi^6}\langle\frac{\alpha_{s}GG}{\pi}\rangle \int dydz\,\frac{\left(1-y-z\right)^4\left(4+y+z\right)}{y^2}\left(s-\overline{m}_Q^2\right)^2 \nonumber\\
&&-\frac{1}{221184\pi^6}\langle\frac{\alpha_{s}GG}{\pi}\rangle \int dydz\,yz\left(1-y-z\right)^2 \left(4-y-z\right)\left(s-\overline{m}_Q^2\right)^2 \left(5s-2\overline{m}_Q^2\right)   \nonumber\\
&&-\frac{m_Q^2}{73728\pi^6} \langle\frac{\alpha_{s}GG}{\pi}\rangle \int dydz\,\left(1-y-z\right)^2\left(4-y-z\right)\left(s-\overline{m}_Q^2\right)^2  \nonumber\\
&&-\frac{1}{221184\pi^6}\langle\frac{\alpha_{s}GG}{\pi}\rangle \int dydz\,z\left(1-y-z\right)^3\left(s-\overline{m}_Q^2\right)^2\left(5s-2\overline{m}_Q^2\right)   \nonumber\\
&&-\frac{m_Q^2}{294912\pi^6}\langle\frac{\alpha_{s}GG}{\pi}\rangle \int dydz\,\frac{\left(1-y-z\right)^3 \left(3+y+z\right)}{y}\left(s-\overline{m}_Q^2\right)^2 \nonumber\\
&&-\frac{1}{884736\pi^6}\langle\frac{\alpha_{s}GG}{\pi}\rangle \int dydz\,\left(1-y-z\right)^4\left(s-\overline{m}_Q^2\right)^2\left(7s-4\overline{m}_Q^2\right)  \nonumber\\
&&+\frac{1}{1474560\pi^6}\langle\frac{\alpha_{s}GG}{\pi}\rangle \int dydz\,\left(1-y-z\right)^5\left(s-\overline{m}_Q^2\right)^2\left(3s-2\overline{m}_Q^2\right) \, ,\nonumber
\end{eqnarray}

\begin{eqnarray}
\rho_{6}^1(s)&=& \frac{\langle\bar{q}q\rangle^2}{96\pi^4} \int dydz\,yz\left(1-y-z\right)\left(s-\overline{m}_Q^2\right)\left(2s-\overline{m}_Q^2\right)
 \nonumber\\
&&+ \frac{m_Q^2\langle\bar{q}q\rangle^2}{96\pi^4}\int dydz\,\left(1-y-z\right)\left(s-\overline{m}_Q^2\right)\, , \nonumber
\end{eqnarray}

\begin{eqnarray}
\rho_{8}^1(s)&=& -\frac{\langle\bar{q}q\rangle \langle\bar{q}g_{s}\sigma Gq\rangle}{192\pi^4}\int dydz\,yz \, \left(3s-2\overline{m}_Q^2\right)   -\frac{m_Q^2\langle\bar{q}q\rangle \langle\bar{q}g_{s}\sigma Gq\rangle}{192\pi^4}\int dydz\,  \nonumber\\
&& -\frac{\langle\bar{q}q\rangle \langle\bar{q}g_{s}\sigma Gq\rangle}{2304\pi^4}\int dydz\,z\left(1-y-z\right)\left(3s-2\overline{m}_Q^2\right) \nonumber\\
&& -\frac{m_Q^2 \langle\bar{q}q\rangle \langle\bar{q}g_{s}\sigma Gq\rangle}{2304\pi^4}\int dydz\,\frac{1-y-z}{y} \, ,\nonumber
\end{eqnarray}

\begin{eqnarray}
\rho_{10}^1(s)&=& \left[\frac{\langle\bar{q}g_{s}\sigma Gq\rangle^2}{768\pi^4}+ \frac{\langle\bar{q}q\rangle^2}{864\pi^2} \langle\frac{\alpha_{s}GG}{\pi}\rangle\right] \int dy\,y\left(1-y\right)\Big[1+s\,\delta\left(s-\widetilde{m}_Q^2\right)\Big] \nonumber\\
&&-\frac{m_Q^2\langle\bar{q}q\rangle^2 }{864\pi^2} \langle\frac{\alpha_{s}GG}{\pi}\rangle \int dydz\,
\frac{z\left(1-y-z\right)}{y^2}\left(1+\frac{s}{T^2}\right)\delta\left(s-\overline{m}_Q^2\right)  \nonumber\\
&&-\frac{m_Q^4\langle\bar{q}q\rangle^2}{864\pi^2T^2} \langle\frac{\alpha_{s}GG}{\pi}\rangle \int dydz\, \frac{1-y-z}{y^3} \delta\left(s-\overline{m}_Q^2\right)   \nonumber\\
&&+ \frac{m_Q^2\langle\bar{q}q\rangle^2}{288\pi^2}\langle\frac{\alpha_{s}GG}{\pi}\rangle \int dydz\,
 \frac{1-y-z}{y^2}\delta\left(s-\overline{m}_Q^2\right) \nonumber\\
&& -\frac{\langle\bar{q}q\rangle^2}{864\pi^2} \langle\frac{\alpha_{s}GG}{\pi}\rangle \int dydz\,
\left(1-y-z\right)\left[1+\frac{s}{4}\delta\left(s-\overline{m}_Q^2\right)\right]\nonumber\\
&&+  \frac{m_Q^2 \langle\bar{q}g_{s}\sigma Gq\rangle^2}{9216\pi^4} \int dydz\,\frac{1}{y}\delta\left(s-\overline{m}_Q^2\right) \nonumber\\
&& -\frac{\langle\bar{q}g_{s}\sigma Gq\rangle^2}{27648\pi^4}\int dydz\,\left(1-4y-4z\right)\left[1+\frac{s}{2}\delta\left(s-\overline{m}_Q^2\right)\right] \nonumber\\
&&-\frac{m_Q^2\langle\bar{q}g_{s}\sigma Gq\rangle^2}{27648\pi^4} \int dydz\,\frac{1-y-z}{yz}\delta\left(s-\overline{m}_Q^2\right) \, ,\nonumber
\end{eqnarray}

\begin{eqnarray}
\rho_{3}^0(s)&=& -\frac{\langle\bar{q}q\rangle}{18432\pi^6} \int dydz\,yz\left(1-y-z\right)^2\left(2+y+z\right)\left(s-\overline{m}_Q^2\right)^3\left(5s-\overline{m}_Q^2\right) \nonumber\\
&& -\frac{m_Q^2 \langle\bar{q}q\rangle}{4608\pi^6} \int dydz\,\left(1-y-z\right)^2\left(2+y+z\right)\left(s-\overline{m}_Q^2\right)^3  \, ,\nonumber
\end{eqnarray}

\begin{eqnarray}
\rho_{5}^0(s)&=&\frac{\langle\bar{q}g_{s}\sigma Gq\rangle}{6144\pi^6} \int dydz\,yz\left(1-y-z\right)\left(1+y+z\right)\left(s-\overline{m}_Q^2\right)^2\left(4s-\overline{m}_Q^2\right)
 \nonumber\\
&&+\frac{m_Q^2\langle\bar{q}g_{s}\sigma Gq\rangle }{2048\pi^6} \int dydz\,\left(1-y-z\right)\left(1+y+z\right)\left(s-\overline{m}_Q^2\right)^2 \, ,\nonumber
\end{eqnarray}

\begin{eqnarray}
\rho_{7}^0(s)&=&\frac{m_Q^2\langle\bar{q}q\rangle }{6912\pi^4} \langle\frac{\alpha_{s}GG}{\pi}\rangle \int dydz\, \frac{z\left(1-y-z\right)^2\left(2+y+z\right)}{y^2} \left(2s-\overline{m}_Q^2\right) \nonumber\\
&&+ \frac{m_Q^4\langle\bar{q}q\rangle}{6912\pi^4} \langle\frac{\alpha_{s}GG}{\pi}\rangle \int dydz\,\frac{\left(1-y-z\right)^2\left(2+y+z\right)}{y^3} \nonumber\\
&&-\frac{m_Q^2\langle\bar{q}q\rangle}{2304\pi^4} \langle\frac{\alpha_{s}GG}{\pi}\rangle \int dydz\,\frac{\left(1-y-z\right)^2\left(2+y+z\right)}{y^2}\left(s-\overline{m}_Q^2\right)  \nonumber\\
&&+\frac{\langle\bar{q}q\rangle}{9216\pi^4}\langle\frac{\alpha_{s}GG}{\pi}\rangle \int dydz\,\left(1+2y\right)z\left(1-y-z\right)\left(s-\overline{m}_Q^2\right)\left(3s-\overline{m}_Q^2\right)  \nonumber\\
&&+\frac{m_Q^2\langle\bar{q}q\rangle}{9216\pi^4}\langle\frac{\alpha_{s}GG}{\pi}\rangle \int dydz\,
\frac{\left(1-y-z\right)\left(1+5y+z\right)}{y}\left(s-\overline{m}_Q^2\right) \nonumber\\
&&+\frac{\langle\bar{q}q\rangle}{18432\pi^4} \langle\frac{\alpha_{s}GG}{\pi}\rangle \int dydz\,
\left(1-y-z\right)^2\left(s-\overline{m}_Q^2\right)\left(5s-3\overline{m}_Q^2\right) \nonumber\\
&& -\frac{\langle\bar{q}q\rangle}{55296\pi^4}\langle\frac{\alpha_{s}GG}{\pi}\rangle \int dydz\,
\left(1-y-z\right)^3\left(s-\overline{m}_Q^2\right)\left(7s-5\overline{m}_Q^2\right) \, ,\nonumber
\end{eqnarray}

\begin{eqnarray}
\rho_{9}^0(s)&=& -\frac{\langle\bar{q}q\rangle^3}{36\pi^2} \int dy\,y\left(1-y\right)\, s\,  ,\nonumber
\end{eqnarray}

\begin{eqnarray}
\rho_{11}^0(s)&=& \frac{\langle\bar{q}q\rangle^2 \langle\bar{q}g_{s}\sigma Gq\rangle }{32\pi^2} \int dy\,
 y\left(1-y\right) \left[1+\left(\frac{4s}{3}+\frac{2s^2}{3T^2}\right)\delta\left(s-\widetilde{m}_Q^2\right)\right] \nonumber\\
&&+\frac{\langle\bar{q}q\rangle^2 \langle\bar{q}g_{s}\sigma Gq\rangle }{3456\pi^2} \int dy\,\Big[1+2s\,\delta\left(s-\widetilde{m}_Q^2\right)\Big]\, ,\nonumber
\end{eqnarray}

\begin{eqnarray}
\rho_{13}^0(s)&=&-\frac{\langle\bar{q}q\rangle \langle\bar{q}g_{s}\sigma Gq\rangle^2}{192\pi^2T^2}\int dy\,
 y\left(1-y\right)\left(s+\frac{s^2}{T^2}+\frac{s^3}{T^4}\right)\delta\left(s-\widetilde{m}_Q^2\right) \nonumber\\
&&-\frac{m_Q^2\langle\bar{q}q\rangle^3}{648T^2} \langle\frac{\alpha_{s}GG}{\pi}\rangle  \int dy\, \frac{1-y}{y^2}\left(1-\frac{s}{T^2}\right)\delta\left(s-\widetilde{m}_Q^2\right)  \nonumber\\
&&+\frac{m_Q^4\langle\bar{q}q\rangle^3 }{648T^4}\langle\frac{\alpha_{s}GG}{\pi}\rangle  \int dy\,\frac{1}{y^3}\delta\left(s-\widetilde{m}_Q^2\right) \nonumber\\
&&-\frac{m_Q^2\langle\bar{q}q\rangle^3}{216T^2} \langle\frac{\alpha_{s}GG}{\pi}\rangle  \int dy\,\frac{1}{y^2}\delta\left(s-\widetilde{m}_Q^2\right)\nonumber\\
&&+ \frac{\langle\bar{q}q\rangle^3}{1296}\langle\frac{\alpha_{s}GG}{\pi}\rangle  \int dy\,\left(1+\frac{s}{2T^2}\right)\delta\left(s-\widetilde{m}_Q^2\right) \nonumber\\
&& -\frac{\langle\bar{q}q\rangle \langle\bar{q}g_{s}\sigma Gq\rangle^2}{13824\pi^2T^2} \int dy\,\left(s+\frac{4s^2}{T^2}\right)\delta\left(s-\widetilde{m}_Q^2\right) \nonumber\\
&&-\frac{\langle\bar{q}q\rangle^3}{432T^2} \langle\frac{\alpha_{s}GG}{\pi}\rangle   \int dy\,
 y\left(1-y\right)\left(s+\frac{s^2}{T^2}+\frac{s^3}{T^4}\right)\delta\left(s-\widetilde{m}_Q^2\right) \, , \nonumber
\end{eqnarray}
where $\int dydz =\int_{y_i}^{y_f}dy \int_{z_i}^{1-y}dz$, $\int dy=\int_{y_i}^{y_f}dy$, $y_{f}=\frac{1+\sqrt{1-4m_{Q}^{2}/s}}{2}$, $y_{i}=\frac{1-\sqrt{1-4m_{Q}^{2}/s}}{2}$, $z_{i}=\frac{ym_{Q}^{2}}{ys-m_{Q}^{2}}$, $\overline{m}_{Q}^{2}=\frac{(y+z)m_{Q}^{2}}{yz}$, $\widetilde{m}_{Q}^{2}=\frac{m_{Q}^{2}}{y(1-y)}$, $\int_{y_{i}}^{y_{f}}dy\rightarrow\int_{0}^{1}$, $\int_{z_{i}}^{1-y}dz\rightarrow\int_{0}^{1-y}dz$, when the $\delta$ functions $\delta(s-\overline{m}_{Q}^{2})$ and $\delta(s-\widetilde{m}_{Q}^{2})$ appear. We can obtain the QCD spectral densities of the $\Xi_{cc}$ and $\Xi^*_{cc}$ with a simple replacement $m_s \to 0$, $\langle\bar{s}s\rangle \to \langle\bar{q}q\rangle$, $\langle\bar{s}g_s\sigma Gs\rangle \to \langle\bar{q}g_s\sigma Gq\rangle$ for the QCD spectral densities of the $\Omega_{cc}$ and $\Omega^*_{cc}$, respectively.

\section*{Acknowledgements}
This  work is supported by National Natural Science Foundation,
Grant Number 11775079, and the Fundamental Research Funds for the
Central Universities, Grant Number 2016MS155.

\end{document}